\documentclass[epj]{svjour}
\usepackage{latexsym}
\usepackage{graphics}
\usepackage{bm}
\usepackage{amsmath}
\usepackage{multirow}
\usepackage[demo]{graphicx}
\usepackage{adjustbox}
\begin{document}
\title{Excited State Mass spectra of Singly Charmed Baryons}
\author{Zalak Shah\inst{1}, Kaushal Thakkar\inst{2,*}, Ajay Kumar Rai\inst{1,*}\and P. C. Vinodkumar\inst{3}
%
}                     
\offprints{raiajayk@gmail.com}          
\institute{Department of Applied Physics, Sardar Vallabhbhai National Institute of Technology, Surat, Gujarat, India-395007\and Department of Applied Sciences \& Humanities, GIDC Degree Engineering college, Abrama, Navsari, India-396406 \and Department of physics, Sardar Patel University, V.V. Nagar, Anand, India-388120\\
$^*$Corresponding Author: kaushal2physics@gmail.com,raiajayk@gmail.com}
\date{Received: date / Revised version: date}
%
\abstract{
Mass spectra of excited states of the singly charmed baryons are calculated using the hypercentral description of three body system. The baryon consist of a charm quark and light quarks (u, d and s) are studied in the framework of QCD motivated constituent quark model. The form  of confinement potential is hyper coloumb plus power potential with potential index $\nu$, varying from 0.5 to 2.0. The first order correction to the confinement potential is also incorporated in this approach. The radial as well as orbital excited state masses of $\Sigma_{c}^{++},  \Sigma_{c}^+, \Sigma_{c}^0, \Xi_c^+, \Xi_c^0, \Lambda_c^+, \Omega_c^0$ baryons, are reported in this paper. We have incorporated spin-spin, spin-orbit and tensor interactions perturbatively in the present study. The semi-electronic decay of  $\Omega_c$ and $\Xi_c$ are also calculated using the spectroscopic parameters of these baryons. The computed results are compared with other theoretical predictions as well as with the available experimental observations. We also construct the Regge trajectory in ($n_r$, $M^{2}$) and (J, $M^{2}$) plane for these baryons.
\PACS{
      {12.39.pn, 12.40.Yx, 14.20.-c}   \and
      {PACS-key}{discribing text of that key}
     } 
} 
\authorrunning {Z. Shah \textit{et al.}}
\titlerunning{Excited State Mass spectra of Singly Charmed Baryons}
\maketitle
\section{Introduction}
\label{intro}
Significant developments have been made in the field of heavy hadrons recently. A lot of new experimental results have been reported by different colliders like CLEO, Belle, BABAR, LHCb etc., on ground states and many new excited states of heavy flavour baryons \cite{CDF,cleo,2,3,belle}. Particle Data Group (PDG) listed 19 known charm baryons in their PDG-2014 \cite{olive}. Thus, a rich dynamical study on heavy flavor mesons and baryons with their properties become important from the point of view of understanding QCD at the different energy scale.\\

After the discovery of J/$\psi$, an evidence of the first single charmed baryon was found. $\Lambda_{c}^{+}$ has been discovered at BNL \cite{1} in 1975 . Shortly, $\Sigma_{c}^{++}$ mass was discovered at FNAL \cite{knapp} in 1976. Then $\Xi_{c}$ were announced by Belle \cite{2} and also by BABAR collaboration  \cite{3}. The WA62 collaboration announced the first ground state of $\Omega_{c}$ \cite{WA62}. Charmed baryons were mainly discovered at the B-factories, whereas bottom baryons are more recently studied at LHC and also in Lattice-QCD \cite{4,86}. Experimentally, only singly charmed baryons have been  perceived. The only  discovery of doubly charm baryon ($\Xi_{cc}^{++}$) is reported by SELEX collaboration \cite{selex}, whereas no triply heavy baryons have been observed yet. Recently, the spectra of triply heavy baryons are calculated by quark model \cite{89} as well as Lattice QCD \cite{88}. The mass spectra, width, lifetime and form factors have been often reported by numerous experimental groups but the spin and parity identification of some states are still missing. These problems will be taken care by the study of angular distribution of particle decays.
The recent review article by Hai-Yang Cheng gives 
details of the spectroscopy, strong decays, lifetimes, nonleptonic and semileptonic weak decays, and electromagnetic decays of these charmed baryons \cite{H}. Furthermore, the future experiments at J-PARC, PANDA \cite{panda}, LHCb, and Belle II are expected to give further information on charmed baryons in near future. 

\subsection{Singly Charmed Baryons}
As we know, Baryons are strongly interacting fermions made up of three quarks. Baryons having only u and d quark are called nucleons ($I_{3}=\frac{1}{2}$) or $\Delta$ ($I_{3}=\frac{3}{2}$) resonances. Combination of u, d and s quarks are called hyperons. The singly charmed baryon is composed of a charm quark and two light quarks belong to SU(4) multiplets. SU(4) group includes all of the baryons containing zero, one, two or three charm quarks. 
The multiplet numerology of the tensor product of three fundamental rendition:
\par  \textbf{$4 \otimes 4 \otimes 4 = 20 \oplus 20^{'}_{1} \oplus 20^{'}_{2} \oplus \bar{4}$}.\\
Representation shows totally symmetric 20-plet, the mixed symmetric 20$^{'}$-plet and the total anti symmetric $\bar{4}$ multiplet. Ground state baryons in 20 plet is equivalent of the SU(3) decuplet with positive parity $J^{P}$= $\frac{3}{2}^{+}$ and baryons in 20$^{'}$ plet is equivalent of SU(3) octet having positive parity $J^{P}$= $\frac{1}{2}^{+}$, while baryon $\Lambda$ in $\bar{4}$ is equivalent of the SU(3) singlet having a negative parity J=$\frac{1}{2}^{-}$ \cite{crede}.\\

\noindent According to the symmetry, the charmed baryons belong to two different SU(3) flavor representations: the symmetric sextet $6_{s}$ and anti symmetric anti triplet $\bar{3}_{A}$,
\begin{center}
$3 \otimes 3 = 6_{s} + \bar{3}_{A}$.
\end{center}Triplet baryons are anti symmetric under the interchange of the u (d) and s quarks; thus, the product of spin and spatial wave functions must be symmetric. In contrast, the product of spin and spatial wave functions of sextet baryons are required to be anti symmetric due to their symmetric flavor wave functions under the interchange of the two light quarks \cite{99}.

\noindent The symmetric flavor wave function for sextet can be written as\\
\[
    \phi_{\Sigma_{c}}=
\begin{cases}
uuc & \text{for}~~ \Sigma_{c}^{++} \\
        \frac{1}{\sqrt{2}} (ud +du)c & \text{for}~~ \Sigma_{c}^{+}\\
        ddc & \text{for}~~ \Sigma_{c}^{0}
\end{cases}
\]

\[
    \phi_{\Omega_{c}}=
\begin{cases}
        ssc & \text{for}~~ \Omega_{c}^{0}
\end{cases}
\]
while for a baryons belonging to $\bar{3}_{A}$, the antisymmetric flavor wave function is given by
\[
    \phi_{\Lambda_{c}}=
\begin{cases}
        \dfrac{1}{\sqrt{2}} (ud-du)c & \text{for}~~ \Lambda_{c}^{+}
\end{cases}
\]

\[
    \phi_{\Xi_{c}}=
\begin{cases}
        \dfrac{1}{\sqrt{2}} (us-su)c & \text{for}~~ \Xi_{c}^{+} \\
        \dfrac{1}{\sqrt{2}} (ds-sd)c & \text{for}~~ \Xi_{c}^{0}
\end{cases}
\]\\

\begin{table}
\caption{List of heavy singly charm baryons and their quark contents. In third and forth column we give multiplicity and spin-isospin values.}
\label{tab:1}
\begin{tabular}{llll}
\hline
Baryon & Quark Content& SU(3) multiplicity & (I, $I_{3}$) \\
\hline\noalign{\smallskip}
~$\Lambda_{c}^{+}$ & udc &  $\bar{3}$ & (0,0)\\
 $\Sigma_{c}^{++}$ & uuc & 6 &(1,1) \\
$\Sigma_{c}^{+}$ & udc & 6 &(1,0) \\
$\Sigma_{c}^{0}$ & ddc & 6 & (1,-1) \\
$\Xi_{c}^{+}$ & usc &$\bar{3}$ & ($\frac{1}{2}$,$\frac{1}{2}$)\\
$\Xi_{c}^{0}$ & dsc & $\bar{3}$ & ($\frac{1}{2}$,$-\frac{1}{2})$\\
$\Omega_{c}^{0}$ & ssc & 6 &(0,0) \\
\noalign{\smallskip}\hline
\end{tabular}
\end{table}

\par The constituent quark model can be considered as an intermediate phenomenological model which fits the experimental data. Various phenomenological models have been used to study heavy baryons by different approaches, starting from non-relativistic Isgur-Karl model \cite{Isgur}, then, relativized potential quark model \cite{Capstick}, relativistic quark model \cite{ebert2011}, variational approach \cite{Roberts2008}, the Fadeev approach \cite{13}, the algebraic approach \cite{9}, the Goldstone Boson Exchange Model \cite{6}, the Bonn model and Bethe-Salpter approach \cite{7}, the interacting quark-diquark model \cite{8}, the Feynman-Hellmann theorem \cite{28}, the Hypercentral Model  \cite{5}, the combined expansion in $1/m_Q$ and $1/N_c$ \cite{100}, QCD sum rules \cite{98}, a soliton model \cite{90}, chiral quark model \cite{chiral} and many more have been studied thenceforth. There are also many Lattice QCD studies which has examined the internal structure and quark dynamics of hadrons \cite{pacs,brown,alex,mathur,can}. An interesting and brief description of some models are given in M.M. Ginnani and E. Santopinto Review article \cite{ginnani2015}. The quark-di quark model could represent a partial solution to the problem of the missing resonance. Thus, it is useful especially if applied only to the excited states \cite{2}. The Capstick-Isgur model \cite{Capstick} is relativistic model and adjunct of relativised meson model to the three body. The U(7) model describes three quark system with the help of bosonic quantization \cite{9}, the elastic and inelastic form factors are explained decently. In Goldstone boson exchange model, as name suggests, boson interaction is considered in the chiral limit that QCD exhibits \cite{6}. The relativistic Interacting quark quark model is a relativistic version of Interacting model \cite{8}. In addition, they have much less resonances than a normal three quark model.\\
\begin{table}[b]
\caption{Quark mass parameters (in GeV) and constants used in the calculations.}
\label{tab:3}
\centering
\begin{tabular}{llllllll}
\hline\noalign{\smallskip}
$\bf{m_{u}}$ &$\bf{m_{d}}$& $\bf{m_{s}}$& $\bf{m_{c}}$ & $C_{F}$ & $C_{A}$ &$n_{f}$ & $\alpha_s(\mu_0=1GeV)$\\
\hline
0.338 & 0.350 &0.500 &1.275 & $\frac{2}{3}$ &3 &4 &0.6\\
\noalign{\smallskip}\hline
\end{tabular}
\end{table}
\par Our study here is based on Hypercentral Constituent Quark  Model (hCQM) with coulomb plus power potential (hCPP$_\nu$) \cite{bhavin,15,93,95,94}. The model has been used for a systematic calculation of various physical quantities. The hCQM scheme has been applied in case of baryons, which amounts to an average two two-body potential for the three quark system over the hyper angle  and works quite well. The Hypercentral Constituent Quark  Model (hCQM) is already used for ground state single charmed baryon mass calculations \cite{bhavin}.  The present study is useful to implant the model for excited states of single charmed baryon. We use the seven baryons ($\Sigma_{c}^{++}, \Sigma_{c}^+, \Sigma_{c}^0$, $\Xi_c^+$, $\Xi_c^0$,$\Lambda_c^+$, $\Omega_c^0$) to calculate mass spectra mentioned in Table 1. The quark content and SU(3) multiplicity of these baryons are also given in this table. We also construct the Regge trajectory for these baryon in ($n_r, M^2$) and ($J, M^2$) plane, where one can test their linearity, parallelism and equidistant. Here, $n_r$ is radial quantum number, $M$ is baryon mass and $J$ is the baryon spin.  As stated, we consider the masses of light quarks as unequal. According to our knowledge, only Ref. \cite{ebert2011} focused on the mass spectra of the radial as well as orbital excited states of heavy baryons ($\Sigma_{c}, \Xi_c,\Lambda_c$, $
\Omega_c$) precisely. In their work, a relativistic quark potential model was used in the quark-diquark picture. Recently, Ref. \cite{chen2015} calculated high excited states of $\Lambda_c$ and $\Xi_c$ within the relativistic flux tube (RFT) model. Moreover, a non-relativistic quark model with harmonic oscillator potential also showed excited mass spectra of $\Lambda_c$, $\Sigma_{c}$ and $\Omega_{c}$ baryons in terms of their $\lambda$ and $\rho$ excitation modes. Where domination of the modes depends on the heavy quark $M_{Q}$ and two light diquark masses $m_{q}$ \cite{yoshida}.  We anticipated the low lying states $J^{P}$, $\frac{1}{2}^{+}$ and $\frac{3}{2}^{+}$ and excited states, $\frac{5}{2}^{+}$, $\frac{7}{2}^{+}$,$\frac{1}{2}^{-}$,$\frac{3}{2}^{-}$,$\frac{5}{2}^{-}$,$\frac{7}{2}^{-}$ and $\frac{9}{2}^{-}$ of all (seven) baryons. Several S-wave, P-wave and D-wave single charm baryons are given in Table 3 with their known experimental masses.\\
\begin{table*}
\caption{Several S, P and D wave single charm baryons and their Exp. masses(in MeV) \cite{olive}.}
\label{tab:2}
\begin{tabular}{ll|ll|ll|ll}
\hline\noalign{\smallskip}
Names & Mass & Names & Mass &  Names & Mass &  Names & Mass \\
\noalign{\smallskip}\hline\noalign{\smallskip}
$\Lambda_{c}(2286)^{+}$ &2286.46$\pm$.014 & $\Sigma_{c}(2455)^{++}$ &2453.98$\pm$0.16 & $\Xi_{c}(2468)^{+}$ & 2467.8 & $\Omega_{c}(2695)^{0}$ & 2695.2 $\pm$ 1.7\\

$\Lambda_{c}(2595)^{+}$ &2592.25$\pm$.028 & $\Sigma_{c}(2455)^{+}$ &2452.9$\pm$0.4 & $\Xi_{c}(2468)^{0}$ & 2470.88 & $\Omega_{c}(2770)^{0}$ & 2765.9 $\pm$ 2.0\\

$\Lambda_{c}(2625)^{+}$ &2628.11$\pm$0.19 & $\Sigma_{c}(2455)^{0}$ & 2453.74$\pm$0.16 & $\Xi_{c}(2645)^{+}$ & $2645.9^{+0.5}_{-0.6}$ & • & • \\
$\Lambda_{c}(2880)^{+}$ &2881.53$\pm$.035 & $\Sigma_{c}(2520)^{++}$ & 2517.9$\pm$0.6 & $\Xi_{c}(2645)^{0}$ & 2645.9$\pm$0.5 & • & • \\

& & $\Sigma_{c}(2520)^{+}$ & 2517$\pm$2.3 & $\Xi_{c}(2790)^{+}$ & 2789.1$\pm$3.2 & • & • \\

& & $\Sigma_{c}(2520)^{0}$ & 2518.8$\pm$.06 & $\Xi_{c}(2790)^{0}$ & 2791.8$\pm$3.3 & • & • \\

• & • & & & $\Xi_{c}(2815)^{+}$ & 2816.6$\pm$0.9 & • & • \\

• & • & & & $\Xi_{c}(2815)^{0}$ & 2819.6$\pm$1.2 & • & • \\
%
%
\noalign{\smallskip}\hline
\end{tabular}
\end{table*}
\par This paper is organized as follows: The hypercentral Constituent Quark Model (hCQM), which has been used for a systematic calculation of various singly charm baryon mass spectroscopy is given in section 2. The details of semi electronic weak decays is presented in section 3. We analyze and discuss different charmed baryons and their obtained results are tabulated in section 4. We also discuss Regge trajectories and their outcomes in same. Lastly, we make our own conclusions in section 5.
\section{The Model}
\label{sec:2}
To deal with three body system (baryons) is always interesting as well as challenging in nuclear and particle physics. The relevant degrees of freedom for the relative motion of the three constituent quarks are provided by the relative Jacobi coordinates ($\vec{\rho}$ and $\vec{\lambda}$) which is given by \cite{Bijker} as
\begin{subequations}
\begin{equation}
\vec{\rho} = \dfrac{1}{\sqrt{2}}(\vec{r_{1}} - \vec{r_{2}})
\end{equation}
\begin{equation}
\vec{\lambda} =\dfrac{m_1\vec{r_1}+m_2\vec{r_2}-(m_1+m_2)\vec{r_3}}{\sqrt{m_1^2+m_2^2+(m_1+m_2)^2}}
\end{equation}
\end{subequations}
Here $m_i$ and $\vec{r_i}$ (i = 1, 2, 3) denote the mass and coordinate of the i-th constituent quark.
The respective reduced masses are given by
\begin{subequations}
\begin{equation}
m_{\rho}=\dfrac{2 m_{1} m_{2}}{m_{1}+ m_{2}}
\end{equation}
\begin{equation}
 m_{\lambda}=\dfrac{2 m_{3} (m_{1}^2 + m_{2}^2+m_1m_2)}{(m_1+m_2)(m_{1}+ m_{2}+ m_{3})}
\end{equation}
\end{subequations}
The constituent quark mass parameters used in our calculations are listed in table 3. The angle of the Hyperspherical coordinates are given by $\Omega_{\rho}= (\theta_{\rho}, \phi_{\rho})$ and $\Omega_{\lambda}= (\theta_{\lambda}, \phi_{\lambda})$. We define hyper radius $x$ and hyper angle $\xi$ in terms of the absolute values $\rho$ and $\lambda$ of the Jacobi coordinates \cite{130,131,132},
\begin{equation}
x= \sqrt{\rho^{2} + \lambda^{2}}\,\,\,and\,\,\, \xi= arctan \left(\dfrac{\rho}{\lambda} \right)
\end{equation}

\begin{table}
\caption{Masses of radial excited states of $\Lambda_{c}^{+}$(in GeV).}
\label{tab:4}
\centering
\begin{adjustbox}{max width=\textwidth}
  \begin{tabular}{*{7}{l}}
\hline\noalign{\smallskip}
State&($\nu$) & A & B & Others  \\
\hline
&0.5	&	2.286	&	2.287 & 2.286 \cite{ebert2011}  \\
&0.7	&	2.286	&	2.286 & 2.285 \cite{yoshida} \\
1S&0.9	&	2.286	&	2.286 &2.286\cite{chen2015} \\
&1.0	&	2.286	&	2.286 &2.268\cite{Roberts2008} \\
	&2.0&	2.286	&	2.286 &2.272\cite{Migura}\\
	&&&&2.286 $\pm$ 0.00014\cite{olive}& \\
\hline
&0.5	&	2.615	&	2.624 &2.769\cite{ebert2011}& \\
&0.7	&	2.641	&	2.656 &2.766 \cite{chen2015}\\
2S&0.9	&	2.667	&	2.680 &2.791 \cite{Roberts2008}\\
&1.0	&	2.678	&	2.699 &2.775 \cite{Capstick}\\
&2.0	&	2.780	&	2.804 & 2.857 \cite{yoshida}\\
&&&&2.766 $\pm$ 0.0024\cite{olive}& \\
\hline
&0.5	&	2.861	&	2.872	&3.130 \cite{ebert2011}\\
&0.7	&	2.924	&	2.948	&3.112 \cite{chen2015}\\
3S&0.9	&	2.987	&	3.009	&3.123 \cite{yoshida}\\
&1.0	&	3.016	&	3.053	&\\
&2.0	&	3.286	&	3.332	&\\
\hline
&0.5	&	3.085	&	3.099	&3.43\cite{ebert2011}\\
&0.7	&	3.191	&	3.223	&3.397\cite{chen2015}\\
4S&0.9	&	3.296	&	3.327	&\\
&1.0	&	3.347	&	3.398	&\\
&2.0	&	3.823	&	3.892	&\\
\hline
&0.5	&	3.295	&	3.311	&\\
&0.7	&	3.446	&	3.486	&\\
5S&0.9	&	3.599	&	3.639	&3.715\cite{ebert2011}\\
&1.0	&	3.673	&	3.739	&\\
&2.0	&	4.385	&	4.480	&\\
\noalign{\smallskip}\hline
\end{tabular}
\end{adjustbox}
\end{table}

\begin{table*}
\caption{ Masses of orbital excited states of $\Lambda_{c}^{+}$ Baryon(in GeV).}
\label{tab:5}
\begin{adjustbox}{max width=\textwidth}
  \begin{tabular}{*{17}{l}}
\hline\noalign{\smallskip}
& \multicolumn{2}{c}{ $\nu$=0.5} & \multicolumn{2}{c}{ $\nu$=0.7}& \multicolumn{2}{c}{ $\nu$=0.9} &\multicolumn{2}{c}{ $\nu$=1.0}& \multicolumn{2}{c}{ $\nu$=2.0} & & \\
\textbf{$n^{2S+1}L_J$}&&&&&&&&&&&&\\
& A & B & A & B & A & B & A & B & A & B &\cite{ebert2011} &\cite{chen2015} &\cite{yoshida}  &Exp.\cite{1} &\cite{98}\\
\hline
$(1^2P_{1/2})$ &2.573	&	2.576	&	2.585	&	2.598	&	2.600	&	2.612	&	2.607	&	2.629	&	2.730	&	2.760	&	2.589 &2.591&2.628&2.592 $\pm$0.0028&2.60 $\pm$ 0.14\\
$(1^2P_{3/2})$ &2.570	&	2.573	&	2.579	&	2.592	&	2.588	&	2.599	&	2.592	&	2.612	&	2.638	&	2.658	&2.627 &2.629&2.630&2.628 $\pm$ 0.11&2.65 $\pm$ 0.14\\
\hline
$(2^2P_{1/2})$ & 2.803	&	2.811	&	2.851	&	2.874	&	2.900	&	2.922	&	2.927	&	2.962	&	3.221	&	3.275	&	2.983  & 2.989 &2.890& $2.939^{+1.4}_{-1.5}$\\
$(2^2P_{3/2})$ & 2.800	&	2.808	&	2.845	&	2.866	&	2.888	&	2.909	&	2.910	&	2.944	&	3.109	&	3.151	&	3.005 & 3.000& 2.917&\\
\hline
$(3^2P_{1/2})$ &3.020	&	3.031	&	3.108	&	3.140	&	3.200	&	3.231	&	3.245	&	3.295	&	3.750	&	3.829	&	3.303& 3.296& 2.933\\
$(3^2P_{3/2})$ &3.017	&	3.028	&	3.102	&	3.132	&	3.187	&	3.216	&	3.228	&	3.276	&	3.620	&	3.686	&	3.322 &3.301&2.956\\
\hline
$(4^2P_{1/2})$ & 3.226	&	3.240	&	3.359	&	3.398	&	3.497	&	3.536	&	3.566	&	3.630	&	4.309	&	4.416	&	3.588\\
$(4^2P_{3/2})$ &3.223	&	3.237	&	3.352	&	3.391	&	3.483	&	3.521	&	3.547	&	3.610	&	4.164	&	4.256	&	3.606\\
\hline
$(5^2P_{1/2})$ & 3.424	&	3.441	&	3.605	&	3.652	&	3.792	&	3.840	&	3.885	&	3.964	&	4.895	&	5.030	&	3.852\\
$(5^2P_{3/2})$ & 3.421	&	3.438	&	3.598	&	3.644	&	3.777	&	3.825	&	3.866	&	3.943	&	4.735	&	4.855	&	3.869\\
\hline
$(1^2D_{3/2})$&	2.764	&	2.771	&	2.794	&	2.816	&	2.825	&	2.845	&	2.842	&	2.873	&	3.048	&	3.098	&2.874	&2.857&&&\\
$(1^2D_{5/2})$&	2.759	&	2.767	&	2.784	&	2.805	&	2.807	&	2.826	&	2.817	&	2.849	&	2.898	&	2.933	&2.880	&2.879&&2.881$\pm$0.035&2.882 $\pm$0.0035\\
\hline
$(2^2D_{3/2})$&	2.980	&	2.992	&	3.050	&	3.081	&	3.122	&	3.152	&	3.158	&	3.207	&	3.565	&	3.641	&3.189	&3.188&&\\
$(2^2D_{5/2})$&	2.976	&	2.987	&	3.040	&	3.070	&	3.103	&	3.132	&	3.133	&	3.179	&	3.392	&	3.450	&3.209	&3.198&&\\
\hline
$(1^2F_{5/2})$&	2.940	&	2.952	&	2.992	&	3.022	&	3.045	&	3.074	&	3.065	&	3.116	&	3.379	&	3.450	&3.097	&3.075\\
$(1^2F_{7/2})$&	2.934	&	2.945	&	2.978	&	3.006	&	3.018	&	3.045	&	3.036	&	3.079	&	3.163	&	3.211	&3.078	&3.092\\
\noalign{\smallskip}\hline
\end{tabular}
\end{adjustbox}
\end{table*}

\begin{table*}
\caption{Masses of radial excited states of $\Sigma_{c}^{0}$(in GeV).}
\label{tab:6}
\centering
\begin{adjustbox}{max width=\textwidth}
  \begin{tabular}{*{7}{l}}
\hline\noalign{\smallskip}
($\nu$) & A & B & Others &  A & B & Others\\
\hline
&\multicolumn{2}{c}{$1^{2}S_{\frac{1}{2}}$ } & & \multicolumn{2}{c}{$1^{4}S_{\frac{3}{2}}$ } & \\
\hline
0.5	&	2.444	&	2.453 	&	2.453\cite{olive}	&	2.506	&	2.523	&	2.518 \cite{olive}	\\
0.7	&	2.444	&	2.453	&	2.443\cite{ebert2011}	&	2.505	&	2.531	&2.519	\cite{ebert2011}	\\
0.9	&	2.444	&	2.453	&	2.459\cite{Migura} 	&	2.508	&	2.541	& 2.539	\cite{Migura}	\\
1.0	&	2.444	&	2.453	&	2.481\cite{lattice}	&	2.506	&	2.529	&	2.559\cite{lattice}\\
2.0	&	2.444	&	2.453	&	2.448 \cite{13}	&	2.508	&	2.564	&	2.505\cite{13}	\\
&&&2.460\cite{yoshida}&&&2.523\cite{yoshida}\\
\hline
&\multicolumn{2}{c}{$2^{2}S_{\frac{1}{2}}$ } & & \multicolumn{2}{c}{$2^{4}S_{\frac{3}{2}}$ } & \\
\hline
0.5	&	2.862	&	2.894	&	2.901\cite{ebert2011}	&	2.909	&	2.944	&2.936\cite{ebert2011}		\\
0.7	&	2.900	&	2.939	& 3.029	\cite{yoshida}	&	2.944	&	2.995	& 3.065	\cite{yoshida}		\\
0.9	&	2.928	&	2.985	&		&	2.979	&	3.046	&		\\
1.0	&	2.947	&	2.992	&		&	2.993	&	3.046	&		\\
2.0	&	3.079	&	3.174	&		&	3.131	&	3.254	&		\\
\hline
 &\multicolumn{2}{c}{$3^{2}S_{\frac{1}{2}}$ } & & \multicolumn{2}{c}{$3^{4}S_{\frac{3}{2}}$ } & \\
\hline
0.5	&	3.164	&	3.212	&	3.271\cite{ebert2011}	&	3.194	&	3.242	&	3.293\cite{ebert2011}		\\
0.7	&	3.253	&	3.318	& 3.103\cite{yoshida}		&	3.280	&	3.351	& 3.094\cite{yoshida}	\\
0.9	&	3.332	&	3.424	&		&	3.364	&	3.461	&		\\
1.0	&	3.373	&	3.449	&		&	3.402	&	3.482	&		\\
2.0	&	3.721	&	3.899	&		&	3.756	&	3.953	&		\\
\hline
 &\multicolumn{2}{c}{$4^{2}S_{\frac{1}{2}}$ } & & \multicolumn{2}{c}{$4^{4}S_{\frac{3}{2}}$ } & \\
\hline
0.5	&	3.445	&	3.507	&		&	3.464	&	3.526	&		\\
0.7	&	3.590	&	3.678	&3.581\cite{ebert2011}		&	3.608	&	3.700	&3.598\cite{ebert2011}		\\
0.9	&	3.726	&	3.851	&		&	3.748	&	3.876	&		\\
1.0	&	3.794	&	3.900	&		&	3.814	&	3.922	&		\\
2.0	&	4.405	&	4.670	&		&	4.430	&	4.708	&		\\
\hline
 &\multicolumn{2}{c}{$5^{2}S_{\frac{1}{2}}$ } & & \multicolumn{2}{c}{$5^{4}S_{\frac{3}{2}}$ } & \\
\hline
0.5	&	3.710	&	3.786	&		&	3.724	&	3.799	&		\\
0.7	&	3.916	&	4.026	&	3.861\cite{ebert2011}	&	3.929	&	4.041	&3.873\cite{ebert2011}		\\
0.9	&	4.114	&	4.272	&		&	4.130	&	4.290	&		\\
1.0	&	4.213	&	4.348	&		&	4.227	&	4.364	&		\\
2.0	&	5.125	&	5.481	&		&	5.145	&	5.509	&		\\
\noalign{\smallskip}\hline
\end{tabular}
\end{adjustbox}
\end{table*}

\begin{table*}
\caption{Masses of radial excited states of $\Sigma_{c}^{+}$(in GeV).}
\label{tab:7}
\centering
\begin{tabular}{lllllll}
\hline\noalign{\smallskip}
($\nu$) & A & B & Others &  A & B & Others\\
\hline
 &\multicolumn{2}{c}{$1^{2}S_{\frac{1}{2}}$ } & & \multicolumn{2}{c}{$1^{4}S_{\frac{3}{2}}$ } & \\
\hline
0.5	&	2.444 & 2.453	&2.453 \cite{olive} &	2.504&2.501	&2.517 \cite{olive}	\\
0.7	&	2.444&2.452 &2.459 \cite{zahra} &	2.501&2.501	&2.497 \cite{zahra}	\\
0.9	&	2.444	&2.452&		&2.501&2.501	\\
1.0	&	2.444 & 2.452	& &	2.506	&2.501 &	\\
2.0	&	2.444&	2.452 & &	2.509	& 2.501&	\\
\hline
 &\multicolumn{2}{c}{$2^{2}S_{\frac{1}{2}}$ } & & \multicolumn{2}{c}{$2^{4}S_{\frac{3}{2}}$ } & \\
\hline
0.5	&	2.871&2.887 & &	2.916 & 2.921	&	\\
0.7	&	2.912 &2.925	&&	2.950&2.960	&	\\
0.9	&	2.942 &2.961 &&	2.984&2.996 &	\\
1.0	&	2.960& 2.978	&&	3.000 & 3.012	&	\\
2.0	&	3.094 & 3.123&&	3.140& 3.160&	\\
\hline
 &\multicolumn{2}{c}{$3^{2}S_{\frac{1}{2}}$ } & & \multicolumn{2}{c}{$3^{4}S_{\frac{3}{2}}$ } & \\
\hline
0.5	&	3.179 &3.199&&	3.206&3.220&	\\
0.7	&	3.270&3.293&&	3.293&3.314&	\\
0.9	&	3.351&3.381	&&	3.378&3.403	&	\\
1.0	&	3.393&3.424&&	3.419&	3.445&	\\
2.0	&	3.746&	3.801 &&	3.777&3.826 &	\\
\hline
 &\multicolumn{2}{c}{$4^{2}S_{\frac{1}{2}}$ } & & \multicolumn{2}{c}{$4^{4}S_{\frac{3}{2}}$ } & \\
\hline
0.5	&	3.464&3.490&&	3.483&	3.503&	\\
0.7	&	3.612&3.645&&	3.628&3.658&	\\
0.9	&	3.751&3.793	&&	3.770&3.808	&	\\
1.0	&	3.822&3.866&&	3.839&3.880&	\\
2.0	&	4.443&4.526&&	4.465&4.543&	\\
\hline
 &\multicolumn{2}{c}{$5^{2}S_{\frac{1}{2}}$ } & & \multicolumn{2}{c}{$5^{4}S_{\frac{3}{2}}$ } & \\
\hline
0.5	&	3.735	&3.766&&	3.748&	3.775&	\\
0.7	&	3.944&3.986	&&	3.956&3.995	&	\\
0.9	&	4.146&	4.200&&	4.160	&4.210&	\\
1.0	&	4.249&	4.306&&	4.261	&4.316&	\\
2.0	&	5.176&5.291	&&	5.193	&5.304&	\\
\noalign{\smallskip}\hline
\end{tabular}
\end{table*}

\begin{table*}
\caption{Masses of radial excited states of $\Sigma_{c}^{++}$(in GeV).}
\label{tab:8}
\centering
\begin{tabular}{lllllll}
\hline\noalign{\smallskip}
($\nu$) & A & B & Others &  A & B & Others\\
\hline
 &\multicolumn{2}{c}{$1^{2}S_{\frac{1}{2}}$ } & & \multicolumn{2}{c}{$1^{4}S_{\frac{3}{2}}$ } & \\
\hline
0.5	&	2.448	&	2.454	&	2.454 \cite{olive}	&	2.505	&	2.530	&	2.518 \cite{olive}		\\
0.7	&	2.446	&	2.454	&	2.454 \cite{zahra}	&	2.506	&	2.531	&	2.492 \cite{zahra}		\\
0.9	&	2.441	&	2.454	&		&	2.509	&	2.561	&			\\
1.0	&	2.449	&	2.454	&		&	2.505	&	2.530	&			\\
2.0	&	2.445	&	2.454	&		&	2.507	&	2.538	&			\\
\hline
 &\multicolumn{2}{c}{$2^{2}S_{\frac{1}{2}}$ } & & \multicolumn{2}{c}{$2^{4}S_{\frac{3}{2}}$ } & \\
\hline
0.5	&	2.883	&	2.917	&		&	2.928	&	2.972	&	\\
0.7	&	2.919	&	2.960	&		&	2.965	&	3.014	&	\\
0.9	&	2.950	&	3.023	&		&	3.001	&	3.097	&	\\
1.0	&	2.973	&	3.016	&		&	3.014	&	3.069	&	\\
2.0	&	3.109	&	3.177	&		&	3.157	&	3.239	&	\\
\hline
 &\multicolumn{2}{c}{$3^{2}S_{\frac{1}{2}}$ } & & \multicolumn{2}{c}{$3^{4}S_{\frac{3}{2}}$ } & \\
\hline
0.5	&	3.198	&	3.253	&		&	3.225	&	3.285	&	\\
0.7	&	3.288	&	3.354	&		&	3.316	&	3.387	&	\\
0.9	&	3.371	&	3.493	&		&	3.403	&	3.538	&	\\
1.0	&	3.417	&	3.492	&		&	3.443	&	3.525	&	\\
2.0	&	3.778	&	3.905	&		&	3.810	&	3.946	&	\\
\hline
 &\multicolumn{2}{c}{$4^{2}S_{\frac{1}{2}}$ } & & \multicolumn{2}{c}{$4^{4}S_{\frac{3}{2}}$ } & \\
\hline
0.5	&	3.490	&	3.563	&		&	3.509	&	3.584	&	\\
0.7	&	3.640	&	3.729	&		&	3.659	&	3.750	&	\\
0.9	&	3.783	&	3.949	&		&	3.805	&	3.980	&	\\
1.0	&	3.856	&	3.962	&		&	3.874	&	3.984	&	\\
2.0	&	4.491	&	4.682	&		&	4.515	&	4.711	&	\\
\hline
 &\multicolumn{2}{c}{$5^{2}S_{\frac{1}{2}}$ } & & \multicolumn{2}{c}{$5^{4}S_{\frac{3}{2}}$ } & \\
\hline
0.5	&	3.757	&	3.856	&		&	3.772	&	3.871	&	\\
0.7	&	3.980	&	4.091	&		&	3.994	&	4.106	&	\\
0.9	&	4.188	&	4.398	&		&	4.204	&	4.419	&	\\
1.0	&	4.293	&	4.429	&		&	4.306	&	4.445	&	\\
2.0	&	5.243	&	5.500	&		&	5.260	&	5.521	&	\\
\noalign{\smallskip}\hline
\end{tabular}
\end{table*}
\noindent In the center of mass frame ($R_{c.m.}= 0$), the kinetic energy operator can be written as
\begin{equation}
-\frac{\hbar^2}{2m}(\bigtriangleup_{\rho} + \bigtriangleup_{\lambda})= -\frac{\hbar^2}{2m}\left(\frac{\partial^2}{\partial x^2}+\frac{5}{x}\frac{\partial}{\partial x}+\frac{L^2(\Omega)}{x^2}\right)
\end{equation}
where $L^2(\Omega)$=$L^2(\Omega_{\rho},\Omega_{\lambda},\xi)$ is the quadratic Casimir operator of the six-dimensional rotational group O(6) and its eigenfunctions are the hyperspherical harmonics, $Y_{[\gamma]l_{\rho}l_{\lambda}}$ ($\Omega_{\rho}$,$\Omega_{\lambda}$,$\xi$) satisfying the eigenvalue relation, \\
 $L^2Y_{[\gamma]l_{\rho}l_{\lambda}}$
($\Omega_{\rho}$,$\Omega_{\lambda},\xi)$=-$\gamma (\gamma +4) Y_{[\gamma]l_{\rho}l_{\lambda}}(\Omega_{\rho},\Omega_{\lambda},\xi)$. Here, $\gamma$ is the grand angular momentum quantum number.\\

\par In present paper, the confining three-body potential is chosen within a string-like picture, where the quarks are connected by gluonic strings and the potential increases linearly with a collective radius $r_{3q}$ as mentioned in \cite{ginnani2015}. Accordingly the effective two body interactions can be written as
\begin{equation}
\sum_{i<j}V(r_{ij})=V(x)+. . . .
\end{equation}
In the hypercentral approximation, the potential is only depends on hyper radius(x). More details can be seen in references \cite{ginnani2015,M. Ferraris}. On the other hand, the hyper radius $x$ is a collective coordinate and therefore the hypercentral potential contains also the three-body effects. The  Hamiltonian of three body baryonic system in the hCQM is then expressed as
\begin{equation}
H=\dfrac{P_{x}^{2}}{2m} +V(x)
\end{equation}where, $m=\frac{2 m_{\rho} m_{\lambda}}{m_{\rho} + m_{\lambda}}$, is the reduced mass and $x$ is the six dimensional radial hyper central coordinate of the three body system. The hyperradial Schrodinger equation corresponds to the above Hamiltonian can be written as,

\begin{equation}
\left[\dfrac{d^{2}}{d x^{2}} + \dfrac{5}{x} \dfrac{d}{dx} - \dfrac{\gamma(\gamma +4)}{x^{2}} \right] \Psi_{ \gamma}(x) = -2m[E- V(x)]\Psi_{ \gamma}(x)
\end{equation}
where $\Psi_{\gamma}$(x) is the hypercentral wave function. We consider a reduced hypercentral radial function, $\phi_{\gamma}(x)$ = $x^{\frac{5}{2}}\Psi_{ \gamma}(x)$. Thus, six dimensional hyperradial Schrodinger equation reduces to,
\begin{equation}\label{eq:6}
\left[\dfrac{-1}{2m}\dfrac{d^{2}}{d x^{2}} + \dfrac{\frac{15}{4}+ \gamma(\gamma+4)}{2mx^{2}}+ V(x)\right]\phi_{ \gamma}(x)= E\phi_{\gamma}(x)
\end{equation}
For the present study, we consider the hypercentral potential V(x) as the color coulomb plus power potential with first order correction \cite{koma,11,20},
\begin{equation}\label{eq:7}
V(x) =  V^{0}(x) + \left(\dfrac{1}{m_{\rho}}+ \dfrac{1}{m_{\lambda}}\right) V^{(1)}(x)+V_{SD}(x)
\end{equation}
where $V^{0}(x)$ is given by
\begin{subequations}
\begin{equation}
V^{(0)}(x)= \dfrac{\tau}{x}+ \beta x^{\nu}
\end{equation}
and first order correction as similar to the one given by \cite{koma},
\begin{equation}
V^{(1)}(x)= - C_{F}C_{A} \dfrac{\alpha_{s}^{2}}{4 x^{2}}
\end{equation}
\end{subequations}
Here, the hyper-Coulomb strength $\tau = -\frac{2}{3} \alpha_{s}$ where $\alpha_{s}$ corresponds to the strong running coupling constant; $\frac{2}{3}$ is the color factor for baryon, $\beta$ corresponds to the string tension of the confinement. $\nu$ is a potential index, varying from 0.5 to 2.0. $C_{F}$ and $C_{A}$ are the Casimir charges of the fundamental and adjoint representation. The strong running coupling constant $\alpha_{s}$ is given by,
\begin{equation}
\alpha_{s}= \dfrac{\alpha_{s}(\mu_{0})}{1+\dfrac{33-2n_{f}}{12 \pi} \alpha_{s}(\mu_{0}) ln \left(\dfrac{m_{1}+ m_{2}+ m_{3}}{\mu_{0}}\right)}
\end{equation}

\noindent If we compare Eqn.(\ref{eq:6}) with the usual three dimensional radial Schrodinger equation, the resemblance between angular momentum and hyper angular momentum is given by \cite{9}, ${l(l+1)\rightarrow \frac{15}{4}+ \gamma(\gamma+4)}$. The spin-dependent part of Eqn. (\ref{eq:7}), $V_{SD}(x)$ contains three types of the interaction terms, such as the spin-spin term $V_{SS} (x)$, the spin-orbit term $V_{\gamma S}(x)$ and tensor term $V_{T}(x)$ given by \cite{12},
\begin{eqnarray}
V_{SD}(x)= V_{SS}(x)(\vec{S_{\rho}}.\vec{S_\lambda})
+ V_{\gamma S}(x) (\vec{\gamma} \cdot \vec{S})&&  \nonumber \\ + V_{T} (x)
\left[ S^2-\dfrac{3(\vec{S }\cdot \vec{x})(\vec{S} \cdot \vec{x})}{x^{2}} \right]
\end{eqnarray}
The Spin-orbit and the tensor term describe the fine structure of the states, while the spin- spin term gives the spin singlet triplet splittings. The coefficient of these spin dependent terms of Eqn.(10) can be written in terms of the vector, $V_{V}(x)=\frac{\tau}{x}$ and scalar, $V_{S}(x)=\beta x^{\nu}$ parts of the static potential as
\begin{equation}
V_{\gamma S} (x) = \dfrac{1}{2 m_{\rho} m_{\lambda}x}  \left(3\dfrac{dV_{V}}{dx} -\dfrac{dV_{S}}{dx} \right)
\end{equation}
\begin{equation}
V_{T}(x)=\dfrac{1}{6 m_{\rho} m_{\lambda}} \left(3\dfrac{d^{2}V_{V}}{dx^{2}} -\dfrac{1}{x}\dfrac{dV_{V}}{dx} \right)
\end{equation}

\begin{equation}
V_{SS}(x)= \dfrac{1}{3 m_{\rho} m_{\lambda}} \bigtriangledown^{2} V_{V}
\end{equation}

\begin{table*}
\caption{Masses of orbital excited states of $\Sigma_{c}^{0}$(in GeV).}
\label{tab:9}
\begin{adjustbox}{max width=\textwidth}
  \begin{tabular}{*{15}{l}}
\hline\noalign{\smallskip}
&  \multicolumn{2}{c}{$\nu$=0.5} & \multicolumn{2}{c}{$\nu$=0.7}& \multicolumn{2}{c}{$\nu$=0.9}&\multicolumn{2}{c}{$\nu$=1.0}& \multicolumn{2}{c}{$\nu$=2.0}&  \\
\textbf{$n^{2S+1}L_J$}&&&&&&&&&&&\\
& A & B & A & B & A & B & A & B & A & B &\cite{ebert2011} &\cite{Roberts2008}& \cite{yoshida} & Others\\
 \hline
$(1^2P_{1/2})$&	2.768	&	2.799	&	2.789	&	2.834	&	2.812	&	2.873	&	2.824	&	2.873	&	3.011	&	3.150	&2.799& 2.748 &2.802& 2.806 $\pm$ 0.018 \cite{olive}	\\
$(1^2P_{3/2})$&	2.763	&	2.794	&	2.779	&	2.822	&	2.793	&	2.849	&	2.799	&	2.844	&	2.860	&	2.959	&2.798	&2.763 &2.807& 2.740(48)(51) \cite{4}\\
$(1^4P_{1/2})$&	2.770	&	2.802	&	2.794	&	2.840	&	2.822	&	2.884	&	2.836	&	2.887	&	3.086	&	3.246	&2.713	&2.768 & &2.73$\pm$0.18 \cite{98}\\
$(1^4P_{3/2})$&	2.765	&	2.797	&	2.784	&	2.828	&	2.802	&	2.861	&	2.811	&	2.858	&	2.935	&	3.054	&2.773	&2.776&&2.80$\pm$0.15 \cite{98}\\
$(1^4P_{5/2})$&	2.759	&	2.790	&	2.771	&	2.812	&	2.777	&	2.830	&	2.778	&	2.821	&	2.734	&	2.799	&2.789	 &2.790&2.839&2.89$\pm$0.15 \cite{98}\\
\hline
$(2^2P_{1/2})$ &	3.064	&	3.111	&	3.134	&	3.201	&	3.203	&	3.294	&	3.237	&	3.316	&	3.653	&	3.877	&3.172&&2.826	\\
$(2^2P_{3/2})$ &	3.059	&	3.105	&	3.123	&	3.188	&	3.182	&	3.270	&	3.210	&	3.284	&	3.468	&	3.643		&3.172&&2.837\\
$(2^4P_{1/2})$ &	3.067	&	3.114	&	3.139	&	3.207	&	3.213	&	3.306	&	3.251	&	3.332	&	3.746	&	3.994		&3.125\\
$(2^4P_{3/2})$ &	3.062	&	3.108	&	3.128	&	3.194	&	3.192	&	3.282	&	3.224	&	3.300	&	3.561	&	3.760		&3.151\\
$(2^4P_{5/2})$ &	3.055	&	3.101	&	3.115	&	3.177	&	3.165	&	3.250	&	3.187	&	3.257	&	3.314	&	3.447		&3.161&&3.316\\
 \hline
$(3^2P_{1/2})$ &	3.343	&	3.403	&	3.468	&	3.556	&	3.591	&	3.714	&	3.652	&	3.758	&	4.343	&	4.656	&3.488&&2.909	\\
$(3^2P_{3/2})$ &	3.338	&	3.398	&	3.457	&	3.542	&	3.569	&	3.688	&	3.623	&	3.724	&	4.129	&	4.385	&3.486	&&2.910\\
$(3^4P_{1/2})$ &	3.345	&	3.406	&	3.474	&	3.562	&	3.602	&	3.726	&	3.667	&	3.774	&	4.450	&	4.791		&3.455\\
$(3^4P_{3/2})$ &	3.340	&	3.400	&	3.462	&	3.549	&	3.580	&	3.701	&	3.637	&	3.741	&	4.236	&	4.521		&3.469\\
$(3^4P_{5/2})$ &	3.334	&	3.393	&	3.447	&	3.532	&	3.550	&	3.667	&	3.598	&	3.697	&	3.951	&	4.160	&	3.475&&3.521\\
\hline
$(4^2P_{1/2})$ &	3.608	&	3.681	&	3.793	&	3.901	&	3.976	&	4.130	&	4.067	&	4.202	&	5.071	&	5.478	&3.770	\\
$(4^2P_{3/2})$ &	3.603	&	3.676	&	3.781	&	3.888	&	3.953	&	4.104	&	4.036	&	4.166	&	4.832	&	5.176		&3.768\\
$(4^4P_{1/2})$ &	3.611	&	3.684	&	3.799	&	3.908	&	3.987	&	4.143	&	4.083	&	4.220	&	5.190	&	5.629		&3.743\\
$(4^4P_{3/2})$ &	3.606	&	3.679	&	3.787	&	3.895	&	3.964	&	4.117	&	4.052	&	4.184	&	4.951	&	5.327		&3.753\\
$(4^4P_{5/2})$ &	3.599	&	3.672	&	3.771	&	3.877	&	3.934	&	4.081	&	4.010	&	4.137	&	4.633	&	4.923		&3.757\\
\hline
$(5^2P_{1/2})$ &	3.864	&	3.950	&	4.111	&	4.240	&	4.359	&	4.544	&	4.482	&	4.643	&	5.832	&	6.337		\\
$(5^2P_{3/2})$ &	3.859	&	3.944	&	4.099	&	4.226	&	4.335	&	4.516	&	4.450	&	4.608	&	5.570	&	6.006		\\
$(5^4P_{1/2})$ &	3.866	&	3.952	&	4.117	&	4.246	&	4.371	&	4.558	&	4.497	&	4.661	&	5.962	&	6.503		\\
$(5^4P_{3/2})$ &	3.861	&	3.947	&	4.105	&	4.233	&	4.347	&	4.530	&	4.466	&	4.626	&	5.701	&	6.171		\\
$(5^4P_{5/2})$ &	3.855	&	3.940	&	4.089	&	4.215	&	4.315	&	4.493	&	4.423	&	4.578	&	5.352	&	5.729		\\
\hline
$(1^4D_{1/2})$ &	3.022	&	3.067	&	3.080	&	3.146	&	3.140	&	3.229	&	3.174	&	3.246	&	3.714	&	3.994	&3.041		\\
$(1^2D_{3/2})$&	3.013	&	3.058	&	3.061	&	3.124	&	3.106	&	3.191	&	3.129	&	3.198	&	3.436	&	3.642	&3.043		\\
$(1^4D_{3/2})$&	3.016	&	3.061	&	3.067	&	3.131	&	3.117	&	3.204	&	3.144	&	3.214	&	3.529	&	3.759	&3.040		\\
$(1^2D_{5/2})$&	3.006	&	3.050	&	3.044	&	3.105	&	3.075	&	3.157	&	3.089	&	3.156	&	3.189	&	3.330	&3.038		\\
$(1^4D_{5/2})$&	3.009	&	3.053	&	3.050	&	3.112	&	3.087	&	3.169	&	3.104	&	3.172	&	3.282	&	3.447	&3.023		\\
$(1^4D_{7/2})$&	2.999	&	3.043	&	3.029	&	3.088	&	3.048	&	3.127	&	3.054	&	3.119	&	2.974	&	3.056	&3.013		\\
\hline
$(2^4D_{1/2})$&	3.300	&	3.361	&	3.412	&	3.499	&	3.526	&	3.652	&	3.586	&	3.694	&	4.432	&	4.812	&3.370		\\
$(2^2D_{3/2})$&	3.292	&	3.351	&	3.393	&	3.478	&	3.490	&	3.610	&	3.539	&	3.641	&	4.112	&	4.407	&3.366		\\
$(2^4D_{3/2})$&	3.295	&	3.355	&	3.400	&	3.485	&	3.502	&	3.624	&	3.555	&	3.659	&	4.218	&	4.542	&3.364		\\
$(2^2D_{5/2})$&	3.285	&	3.343	&	3.376	&	3.459	&	3.459	&	3.572	&	3.498	&	3.593	&	3.827	&	4.046	&3.365		\\
$(2^4D_{5/2})$&	3.287	&	3.346	&	3.383	&	3.466	&	3.471	&	3.586	&	3.513	&	3.611	&	3.933	&	4.181	&3.349		\\
$(2^4D_{7/2})$&	3.278	&	3.336	&	3.361	&	3.442	&	3.432	&	3.539	&	3.461	&	3.552	&	3.577	&	3.730	&3.342		\\
\hline
$(1^4F_{3/2})$&	3.254	&	3.312	&	3.345	&	3.428	&	3.437	&	3.557	&	3.490	&	3.593	&	4.271	&	4.650	&3.288		\\
$(1^2F_{5/2})$&	3.242	&	3.309	&	3.319	&	3.399	&	3.390	&	3.503	&	3.428	&	3.523	&	3.879	&	4.154	&3.283		\\
$(1^4F_{5/2})$&	3.245	&	3.303	&	3.326	&	3.407	&	3.403	&	3.518	&	3.445	&	3.542	&	3.986	&	4.289	&3.254		\\
$(1^4F_{7/2})$&	3.234	&	3.291	&	3.302	&	3.381	&	3.361	&	3.468	&	3.388	&	3.478	&	3.630	&	3.838	&3.253		\\
$(1^2F_{7/2})$&	3.231	&	3.288	&	3.295	&	3.373	&	3.348	&	3.453	&	3.371	&	3.459	&	3.523	&	3.703	&3.227		\\
$(1^4F_{9/2})$&	3.221	&	3.277	&	3.274	&	3.350	&	3.310	&	3.409	&	3.320	&	3.402	&	3.203	&	3.297	&3.209		\\
\hline
\end{tabular}
\end{adjustbox}
\end{table*}

\begin{table*}
\caption{Masses of orbital excited states of $\Sigma_{c}^{++}$(in GeV).}
\centering
\label{tab:10}
\begin{tabular}{c|ccccccccccc}
\hline
&  \multicolumn{2}{c}{$\nu$=0.5} & \multicolumn{2}{c}{$\nu$=0.7}& \multicolumn{2}{c}{$\nu$=0.9}&\multicolumn{2}{c}{$\nu$=1.0}& \multicolumn{2}{c}{$\nu$=2.0} & Exp.\cite{olive} \\
\textbf{$n^{2S+1}L_J$}&&&&&&&&\\
& A & B & A & B & A & B & A & B & A & B \\
 \hline
$(1^2P_{1/2})$&	2.781	&	2.820	&	2.805	&	2.849	&	2.829	&	2.911	&	2.842	&	2.890	&	3.048	&	3.148	& 2.801$^{+0.004}_{-0.006}$ 	 \\
$(1^2P_{3/2})$&	2.776	&	2.814	&	2.793	&	2.835	&	2.808	&	2.885	&	2.814	&	2.860	&	2.879	&	2.948	& 	\\
$(1^4P_{1/2})$&	2.784	&	2.823	&	2.811	&	2.855	&	2.840	&	2.924	&	2.856	&	2.906	&	3.132	&	3.249		\\
$(1^4P_{3/2})$&	2.779	&	2.817	&	2.799	&	2.842	&	2.818	&	2.898	&	2.828	&	2.875	&	2.963	&	3.048		\\
$(1^4P_{5/2})$&	2.772	&	2.809	&	2.783	&	2.824	&	2.789	&	2.863	&	2.791	&	2.835	&	2.738	&	2.780		\\
\hline
$(2^2P_{1/2})$&	3.091	&	3.147	&	3.165	&	3.230	&	3.238	&	3.357	&	3.273	&	3.352	&	3.721	&	3.885		\\
$(2^2P_{3/2})$&	3.086	&	3.141	&	3.152	&	3.217	&	3.214	&	3.333	&	3.247	&	3.318	&	3.514	&	3.639	\\
$(2^4P_{1/2})$&	3.094	&	3.151	&	3.171	&	3.237	&	3.250	&	3.370	&	3.288	&	3.369	&	3.825	&	4.008		\\
$(2^4P_{3/2})$&	3.089	&	3.144	&	3.159	&	3.224	&	3.226	&	3.345	&	3.259	&	3.335	&	3.618	&	3.762		\\
$(2^4P_{5/2})$&	3.082	&	3.136	&	3.142	&	3.206	&	3.195	&	3.312	&	3.219	&	3.290	&	3.342	&	3.434		\\
\hline
$(3^2P_{1/2})$&	3.383	&	3.455	&	3.513	&	3.602	&	3.641	&	3.809	&	3.708	&	3.817	&	4.443	&	4.674	\\
$(3^2P_{3/2})$&	3.377	&	3.449	&	3.500	&	3.587	&	3.618	&	3.779	&	3.675	&	3.779	&	4.204	&	4.390		\\
$(3^4P_{1/2})$&	3.385	&	3.458	&	3.519	&	3.609	&	3.653	&	3.824	&	3.724	&	3.836	&	4.563	&	4.816		\\
$(3^4P_{3/2})$&	3.380	&	3.452	&	3.507	&	3.594	&	3.630	&	3.794	&	3.691	&	3.798	&	4.324	&	4.532		\\
$(3^4P_{5/2})$&	3.372	&	3.443	&	3.490	&	3.574	&	3.598	&	3.754	&	3.648	&	3.747	&	4.005	&	4.154		\\
\hline
$(4^2P_{1/2})$&	3.660	&	3.748	&	3.853	&	3.962	&	4.045	&	4.252	&	4.142	&	4.279	&	5.205	&	5.507	\\
$(4^2P_{3/2})$&	3.655	&	3.741	&	3.840	&	3.947	&	4.019	&	4.221	&	4.107	&	4.240	&	4.938	&	5.189		\\
$(4^4P_{1/2})$&	3.663	&	3.751	&	3.859	&	3.969	&	4.058	&	4.268	&	4.160	&	4.299	&	5.339	&	5.665		\\
$(4^4P_{3/2})$&	3.658	&	3.745	&	3.846	&	3.954	&	4.032	&	4.237	&	4.125	&	4.260	&	5.072	&	5.348		\\
$(4^4P_{5/2})$&	3.650	&	3.736	&	3.828	&	3.934	&	3.998	&	4.195	&	4.078	&	4.207	&	4.715	&	4.925		\\
\hline
$(5^2P_{1/2})$&	3.928	&	4.029	&	4.184	&	4.314	&	4.445	&	4.690	&	4.575	&	4.740	&	6.001	&	6.376		\\
$(5^2P_{3/2})$&	3.922	&	4.023	&	4.171	&	4.299	&	4.418	&	4.660	&	4.540	&	4.701	&	5.709	&	6.028		\\
$(5^4P_{1/2})$&	3.930	&	4.032	&	4.190	&	4.322	&	4.458	&	4.704	&	4.593	&	4.760	&	6.148	&	6.550	\\
$(5^4P_{3/2})$&	3.925	&	4.026	&	4.178	&	4.307	&	4.431	&	4.675	&	4.557	&	4.720	&	5.855	&	6.202		\\
$(5^4P_{5/2})$&	3.918	&	4.018	&	4.161	&	4.287	&	4.396	&	4.635	&	4.510	&	4.668	&	5.465	&	5.738		\\
 \hline
$(1^4D_{1/2})$&	3.038	&	3.092	&	3.088	&	3.151	&	3.136	&	3.254	&	3.161	&	3.237	&	3.497	&	3.649			\\
$(1^2D_{3/2})$&	3.041	&	3.095	&	3.095	&	3.159	&	3.149	&	3.269	&	3.178	&	3.257	&	3.600	&	3.772			\\
$(1^4D_{3/2})$&	3.047	&	3.101	&	3.109	&	3.174	&	3.175	&	3.301	&	3.211	&	3.296	&	3.808	&	4.018			\\
$(1^2D_{5/2})$&	3.033	&	3.086	&	3.076	&	3.138	&	3.115	&	3.228	&	3.133	&	3.205	&	3.324	&	3.444			\\
$(1^4D_{5/2})$&	3.030	&	3.083	&	3.069	&	3.131	&	3.102	&	3.212	&	3.117	&	3.185	&	3.221	&	3.322			\\
$(1^4D_{7/2})$&	3.023	&	3.076	&	3.053	&	3.113	&	3.072	&	3.175	&	3.078	&	3.140	&	2.979	&	3.035			\\
\hline
$(2^4D_{1/2})$&	3.330	&	3.400	&	3.435	&	3.521	&	3.539	&	3.697	&	3.589	&	3.695	&	4.205	&	4.424			\\
$(2^2D_{3/2})$&	3.333	&	3.404	&	3.443	&	3.529	&	3.552	&	3.713	&	3.606	&	3.715	&	4.324	&	4.566			\\
$(2^4D_{3/2})$&	3.339	&	3.411	&	3.457	&	3.545	&	3.579	&	3.746	&	3.640	&	3.756	&	4.563	&	4.850			\\
$(2^2D_{5/2})$&	3.325	&	3.395	&	3.423	&	3.507	&	3.517	&	3.670	&	3.561	&	3.662	&	4.005	&	4.187			\\
$(2^4D_{5/2})$&	3.322	&	3.391	&	3.416	&	3.499	&	3.503	&	3.654	&	3.544	&	3.641	&	3.886	&	4.045			\\
$(2^4D_{7/2})$&	3.315	&	3.383	&	3.400	&	3.481	&	3.472	&	3.617	&	3.505	&	3.594	&	3.607	&	3.714			\\
\hline
$(1^4F_{3/2})$&	3.290	&	3.329	&	3.386	&	3.471	&	3.491	&	3.648	&	3.547	&	3.654	&	4.403	&	4.690			\\
$(1^2F_{5/2})$&	3.277	&	3.314	&	3.357	&	3.439	&	3.436	&	3.585	&	3.528	&	3.574	&	3.965	&	4.170			\\
$(1^4F_{5/2})$&	3.281	&	3.318	&	3.365	&	3.448	&	3.451	&	3.602	&	3.495	&	3.596	&	4.084	&	4.312			\\
$(1^4F_{7/2})$&	3.269	&	3.304	&	3.339	&	3.419	&	3.402	&	3.544	&	3.431	&	3.523	&	3.686	&	3.839			\\
$(1^2F_{7/2})$&	3.265	&	3.300	&	3.331	&	3.410	&	3.387	&	3.527	&	3.411	&	3.501	&	3.566	&	3.697			\\
$(1^4F_{9/2})$&	3.255	&	3.287	&	3.308	&	3.384	&	3.342	&	3.475	&	3.353	&	3.435	&	3.208	&	3.271			\\
\hline
\end{tabular}
\end{table*}

\begin{table*}
\caption{Masses of orbital excited states of $\Sigma_{c}^{+}$(in GeV).}
\label{tab:table11}
\centering
\begin{tabular}{c|ccccccccccc}
\hline
&  \multicolumn{2}{c}{$\nu$=0.5} & \multicolumn{2}{c}{$\nu$=0.7}& \multicolumn{2}{c}{$\nu$=0.9}&\multicolumn{2}{c}{$\nu$=1.0}& \multicolumn{2}{c}{$\nu$=2.0} & Exp\cite{olive} \\
\textbf{$n^{2S+1}L_J$}&&&&&&&&&&\\
& A & B & A & B & A & B & A & B & A & B \\
 \hline
$(1^2P_{1/2})$ &2.772& 2.783& 2.795&2.809& 2.818	&	2.832	&2.831&2.849&3.026& 3.066 & 2.792$^{+0.014}_{-0.005}$\\
$(1^2P_{3/2})$ &2.768&2.777	&	2.784&2.796	&2.798	&	2.812	&	2.805&2.820	&	2.866&2.892 & \\
 $(1^4P_{1/2})$ &2.775	& 2.786 & 2.801 &2.815 &2.828	&	2.841	&	2.844& 2.863 &	3.105& 3.153 \\
$(1^4P_{3/2})$ &2.770&	2.780&2.790&2.802&2.808	&	2.822	&2.818&2.834&	2.946&2.979\\
$(1^4P_{5/2})$ & 2.764&2.772&	2.775&2.786&	2.781	&	2.796	&2.783&2.796&2.734&2.747\\
\hline
$(2^2P_{1/2})$ &  3.076&3.092&	3.146&3.170&3.217	&	3.245	&	3.254&3.284&	3.682&3.752 \\
 $(2^2P_{3/2})$ &  3.070	&	3.086	&	3.135	&	3.506	&3.157	&	3.223	&3.224	&	3.253	&	3.487	&	3.539	\\
$(2^4P_{1/2})$ &3.078	&	3.094	&	3.151	&	3.176	&3.228	&	3.256	&	3.268	&	3.300	&	3.779	&	3.859
 \\
 $(2^4P_{3/2})$ & 3.073	&	3.089	&	3.140	&	3.163	&3.206	&	3.233	&3.239	&	3.269	&	3.584	&3.646\\
 $(2^4P_{5/2})$ & 3.066	&	3.082	&	3.125	&	3.147	&3.176	&	3.203	&	3.200	&	3.228	&	3.325	&3.362 \\
 \hline
 $(3^2P_{1/2})$ & 3.360	&	3.383	&	3.487	&	3.519	&3.611	&	3.652	&	3.677	&	3.721	&	4.386 &	4.488	 \\
 $(3^2P_{3/2})$ &  3.354	&	3.377	&	3.474	&	3.506	&3.588	&	3.628	&	3.645	&	3.688	&	4.161	&	4.242\\
$(3^4P_{1/2})$ &3.362	&	3.385	&	3.493	&	3.525	&3.622	&	3.665	&	3.692	&	3.737	&	4.498	&	4.611\\
 $(3^4P_{5/2})$ & 3.357	&	3.380	&	3.481	&	3.512	&3.600	&	3.640	&	3.661	&	3.704	&	4.273	&	4.365\\
 $(3^4P_{{1/2}})$ & 3.350	&	3.372	&	3.464	&	3.495	&3.570	&	3.608	&	3.619	&	3.661	&	3.973	&	4.037\\
    \hline
$(4^2P_{1/2})$ &3.631	&	3.659	&	3.818	&	3.859	&	4.100	&	4.158	&	4.100	&	4.158	& 5.128& 5.265 \\
$(4^2P_{3/2})$ &3.625	&	3.654	&	3.805	&	3.846	&3.980	&	4.032	&	4.056	&	4.123	&	4.877	&	4.990\\
$(4^4P_{1/2})$ &  3.633	&	3.662	&	3.824	&	3.865	&	4.016	&	4.071	&4.116	&	4.175	&	5.254	&	5.403
\\
 $(4^4P_{3/2})$ &3.628	&	3.657	&	3.812	&	3.853	&	3.992	&	4.045	&4.084	&	4.141	&	5.003	&	5.128
\\
 $(4^4P_{5/2})$ & 3.621	&	3.650	&	3.795	&	3.835	&	3.960	& 3.835	&4.040		&	4.095	&	4.667	&	4.761\\
 \hline
$(5^2P_{1/2})$ &3.891	&	3.926	&	4.142	&	4.192	&	4.392	&	4.460	&	4.524	&	4.596	&	5.905	&	6.077\\
 $(5^2P_{3/2})$ &3.886	&	3.920	&	4.130&	4.179	&	4.369	&	4.433	&	4.489	&	4.559	&	5.629	&	5.776\\
 $(5^4P_{1/2})$ &3.894	&	3.928	&	4.148	&	4.199	&	4.404	&	4.473	&	4.541	&	4.614	&	6.042	&	6.228\\
 $(5^4P_{3/2})$ &3.889	&	3.923	&	4.136	&	4.186	&	4.380	&	4.446	&	4.507	&	4.578	&	4.136	&	5.767\\
 $(5^4P_{5/2})$ &3.882	&	3.916	&	4.119	&	4.168	&	4.349	&	4.411	&	4.461	&	4.529	&	5.400	&	5.525\\
  \hline
$(1^4D_{1/2})$&	3.024	&	3.041	&	3.071	&	3.094	&	3.118	&	3.146	&	3.142	&	3.173	&	3.461	&	3.528			\\
$(1^2D_{3/2})$&	3.027	&	3.044	&	3.077	&	3.102	&	3.130	&	3.160	&	3.158	&	3.190	&	3.559	&	3.635			\\
$(1^4D_{3/2})$&	3.033	&	3.050	&	3.090	&	3.116	&	3.154	&	3.186	&	3.190	&	3.224	&	3.754	&	3.848			\\
$(1^2D_{5/2})$&	3.018	&	3.035	&	3.060	&	3.083	&	3.097	&	3.125	&	3.116	&	3.145	&	3.299	&	3.351			\\
$(1^4D_{5/2})$&	3.015	&	3.032	&	3.054	&	3.075	&	3.085	&	3.112	&	3.100	&	3.128	&	3.202	&	3.244			\\
$(1^4D_{7/2})$&	3.008	&	3.024	&	3.038	&		3.046&	3.057	&	3.081	&	3.063	&	3.088	&	2.974	&	2.996			\\
\hline
$(2^4D_{1/2})$&	3.308	&	3.331	&	3.410	&	3.442	&	3.509	&	3.552	&	3.562	&	3.606	&	4.151	&	4.250			\\
$(2^2D_{3/2})$&	3.311	&	3.334	&	3.417	&	3.449	&	3.521	&	3.565	&	3.579	&	3.624	&	4.263	&	4.373			\\
$(2^4D_{3/2})$&	3.316	&	3.341	&	3.430	&	3.463	&	3.546	&	3.592	&	3.613	&	3.660	&	4.488	&	4.619			\\
$(2^2D_{5/2})$&	3.303	&	3.326	&	3.399	&	3.431	&	3.489	&	3.529	&	3.534	&	3.576	&	3.963	&	4.045			\\
$(2^4D_{5/2})$&	3.300	&	3.323	&	3.392	&	3.424	&	3.477	&	3.515	&	3.518	&	3.559	&	3.851	&	3.922			\\
$(2^4D_{7/2})$&	3.282	&	3.316	&	3.377	&	3.407	&	3.448	&	3.484	&	3.478	&	3.517	&	3.589	&	3.635			\\
\hline
$(1^4F_{3/2})$&	3.269	&	3.292	&	3.360	&	3.393	&	3.455	&	3.500	&	3.515	&	3.562	&	4.215	&	4.460			\\
$(1^2F_{5/2})$&	3.257	&	3.279	&	3.334	&	3.365	&	3.407	&	3.448	&	3.449	&	3.491	&	3.915	&	4.009			\\
$(1^4F_{5/2})$&	3.260	&	3.283	&	3.341	&	3.373	&	3.420	&	3.462	&	3.467	&	3.510	&	4.028	&	4.132			\\
$(1^4F_{7/2})$&	3.249	&	3.271	&	3.317	&	3.347	&	3.377	&	3.414	&	3.406	&	3.446	&	3.653	&	3.722			\\
$(1^2F_{7/2})$&	3.245	&	3.268	&	3.310	&	3.340	&	3.363	&	3.400	&	3.388	&	3.426	&	3.541	&	3.599			\\
$(1^4F_{9/2})$&	3.235	&	3.257	&	3.288	&	3.317	&	3.324	&	3.357	&	3.334	&	3.369	&	3.203	&	3.230			\\
\hline
\end{tabular}
\end{table*}
\noindent Instead of the six dimensional delta function which appear into spin-spin interaction term of $Eq^n.$(13), we use smear function similar to the one given by \cite{15,95}
\begin{equation}
V_{SS}(x)= \dfrac{-A}{6 m_{\rho} m_{\lambda}}  \frac{e^{-x/x_0}}{x x_0^2}
\end{equation}

where $x_{0}$ is the hyperfine parameter of the model. We take $A ={A_{0}}/{(n + \gamma +\frac{3}{2})^{2}} $ , where $A_{0}$ is arbitary constant. For the calculation of ground state masses ($J^{P}=\frac{1}{2}^{+}$ and $J^{P}=\frac{3}{2}^{+}$),  we fix $\beta$ in each case of $\nu$ and then calculated the excited states of the heavy baryons. $\beta$ is state dependent parameter \cite{93}. The baryon spin average mass in this hypercentral model is $M_B = \sum_{i=1} m_{i} + BE$. We have numerically solved the six dimensional Schrodinger equation using Mathematica notebook \cite{lucha}.\\

\par We have calculated the masses of orbital and radial excited heavy baryons and followed the $^{(2S+1)} {\gamma}_{J}$ notations for the spectra [See Table 4-17]. Masses are obtained in the present study for each case of the model potential, $\nu$= 0.5 to $\nu$=2.0. Note that, A are masses without first order correction and B are masses with first order correction in Table [4-18].  We plot graphs for  calculated excited state (S and P state) with different potential index (M $\rightarrow$ $\nu$) (See Fig. 1-5). The Regge trajectories are also drawn in $M^{2} \rightarrow$ n and $M^{2} \rightarrow J$ plane (See Fig. 6-14).

\begin{table*}
\caption{\label{tab:table4}Masses of radial excited states of $\Xi_{c}^{0}$(in GeV).}
\label{tab:12}
\centering
\begin{tabular}{lllllll}
\hline
$n^{2S+1}L_{J}$ & A & B & Others &  A & B & Others\\
\hline
 &\multicolumn{2}{c}{$1^{2}S_{\frac{1}{2}}$ } & & \multicolumn{2}{c}{$1^{4}S_{\frac{3}{2}}$ } & \\
\hline
0.5	&	2.470	&	2.471	&	2.471 \cite{olive}	&	2.585	&	2.601	&	2.646 \cite{olive}\\
0.7	&	2.470	&	2.471	&2.476\cite{ebert2011}		&	2.616	&	2.613	&	2.654\cite{ebert2011}\\
0.9	&	2.471	&	2.470	&2.467\cite{chen2015}		&	2.602	&	2.615	&	2.608(13\cite{4}	\\
1.0	&	2.470	&	2.470	&2.466\cite{Roberts2008}		&	2.584	&	2.610	&	2.655\cite{lattice}	\\
2.0	&	2.470	&	2.473	&2.473\cite{29}		&	2.611	&	2.601	&		\\
&&&2.442(11)(20)\cite{4}&&&\\
\hline
 &\multicolumn{2}{c}{$2^{2}S_{\frac{1}{2}}$ } & & \multicolumn{2}{c}{$2^{4}S_{\frac{3}{2}}$ } & \\
\hline
0.5	&	2.831	&	2.854	&	2.968 $\pm$2.6\cite{olive}	&	2.919	&	2.946	&		\\
0.7	&	2.885	&	2.897	&2.959\cite{ebert2011}		&	2.993	&	2.996	&		\\
0.9	&	2.905	&	2.929	&2.959\cite{chen2015}		&	3.002	&	3.030	&		\\
1.0	&	2.903	&	2.940	&	2.924\cite{Roberts2008}	&	2.859	&	3.038	&		\\
2.0	&	3.040	&	3.057	&	3.137\cite{13}	&	3.148	&	3.151	&		\\
\hline
 &\multicolumn{2}{c}{$3^{2}S_{\frac{1}{2}}$ } & & \multicolumn{2}{c}{$3^{4}S_{\frac{3}{2}}$ } & \\
\hline
0.5	&	3.097	&	3.134	&		&	3.151	&	3.189	&		\\
0.7	&	3.213	&	3.230	&3.323\cite{ebert2011}		&	3.279	&	3.290			\\
0.9	&	3.267	&	3.310	&	3.325\cite{chen2015}	&	3.327	&	3.371	&		\\
1.0	&	3.272	&	3.339	&3.183\cite{Roberts2008}		&	3.325	&	3.399	&		\\
2.0	&	3.614&	3.645	&		&	3.687	&	3.708	&		\\
\hline
 &\multicolumn{2}{c}{$4^{2}S_{\frac{1}{2}}$ } & & \multicolumn{2}{c}{$4^{4}S_{\frac{3}{2}}$ } & \\
\hline
0.5	&	3.339	&	3.387	&		&	3.375	&		&		\\
0.7	&	3.518	&	3.541	&		&	3.563	&	3.580	&		\\
0.9	&	3.614	&	3.673	&3.632\cite{ebert2011}		&	3.655	&	3.714	&		\\
1.0	&	3.631	&	3.726	&3.629\cite{chen2015}		&	3.668	&	3.766	&		\\
2.0	&	4.220	&	4.267	&		&	4.272	&	4.312	&		\\
\hline
 &\multicolumn{2}{c}{$5^{2}S_{\frac{1}{2}}$ } & & \multicolumn{2}{c}{$5^{4}S_{\frac{3}{2}}$ } & \\
\hline
0.5	&	3.566	&	3.624	&		&	3.591	&	3.650	&		\\
0.7	&	3.810	&	3.839	&		&	3.842	&	3.866	&		\\
0.9	&	3.953	&	4.029	&		&	3.982	&	4.058	&		\\
1.0	&	3.986	&	4.107	&	3.909\cite{ebert2011}	&	4.012	&	4.136	&		\\
2.0	&	4.854	&	4.920	&		&	4.894&	4.953	&		\\
\hline
\end{tabular}
\end{table*}

\begin{table*}
\caption{Masses of radial excited states of $\Xi_{c}^{+}$(in GeV).}
\label{tab:13}
\centering
\begin{tabular}{lllllll}
\hline
$n^{2S+1}L_{J}$ & A & B & Others &  A & B & Others\\
\hline
 &\multicolumn{2}{c}{$1^{2}S_{\frac{1}{2}}$ } & & \multicolumn{2}{c}{$1^{4}S_{\frac{3}{2}}$ } & \\
\hline
0.5	&	2.467	&	2.467 	&	2.467 \cite{olive}	&	2.635	&	2.625	&	2.646 $\pm$0.0005	\cite{olive}\\
0.7	&	2.467	&	2.467	&	2.576 \cite{zahra}	&	2.605	&	2.603	&2.634 \cite{zahra}\\
0.9	&	2.467	&	2.467	&		&	2.635	&	2.624	&		\\
1.0	&	2.467	&	2.467	&		&	2.625	&	2.619	&		\\
2.0	&	2.467	&	2.467	&		&	2.628	&	2.635	&		\\

\hline
 &\multicolumn{2}{c}{$2^{2}S_{\frac{1}{2}}$ } & & \multicolumn{2}{c}{$2^{4}S_{\frac{3}{2}}$ } & \\
\hline
0.5	&	2.844	&	2.879	&		&	2.949	&	2.993	&	2.971 $\pm$ 0.0033 \cite{olive}	\\
0.7	&	2.883	&	2.895	&		&	2.986	&	2.990	&		\\
0.9	&	2.937	&	2.944	&		&	3.059	&	3.052	&		\\
1.0	&	2.944	&	2.956	&		&	3.059	&	3.061	&		\\
2.0	&	3.067	&	3.098	&		&	3.189	&	3.219	&		\\

\hline
 &\multicolumn{2}{c}{$3^{2}S_{\frac{1}{2}}$ } & & \multicolumn{2}{c}{$3^{4}S_{\frac{3}{2}}$ } & \\
\hline
0.5	&	3.172	&	3.182	&		&	3.248	&	3.249	&		\\
0.7	&	3.212	&	3.230	&		&	3.275	&	3.287	&		\\
0.9	&	3.328	&	3.339	&		&	3.404	&	3.404	&		\\
1.0	&	3.350	&	3.370	&		&	3.423	&	3.435	&		\\
2.0	&	3.671	&	3.731	&		&	3.753	&	3.811	&		\\

\hline
&\multicolumn{2}{c}{$4^{2}S_{\frac{1}{2}}$ } & & \multicolumn{2}{c}{$4^{4}S_{\frac{3}{2}}$ } & \\
\hline
0.5	&	3.441	&	3.453	&		&	3.492	&	3.497	&		\\
0.7	&	3.518	&	3.543	&		&	3.561	&	3.580	&		\\
0.9	&	3.702	&	3.716	&		&	3.753	&	3.760	&		\\
1.0	&	3.744	&	3.772	&		&	3.793	&	3.815	&		\\
2.0	&	4.306	&	4.398	&		&	4.366	&	4.455	&		\\
\hline
 &\multicolumn{2}{c}{$5^{2}S_{\frac{1}{2}}$ } & & \multicolumn{2}{c}{$5^{4}S_{\frac{3}{2}}$ } & \\
\hline
0.5	&	3.692	&	3.707	&		&	3.729	&	3.738	&		\\
0.7	&	3.812	&	3.842	&		&	3.842	&	3.869	&		\\
0.9	&	4.066	&	4.115	&		&	4.103	&	0.031	&		\\
1.0	&	4.131	&	4.167	&		&	4.166	&	4.198	&		\\
2.0	&	4.971	&	5.097	&		&	5.016	&	5.139	&		\\
\hline
\end{tabular}
\end{table*}

\begin{table*}
\caption{Masses of orbital excited states of $\Xi_{c}^{0}$(in GeV).}
\label{tab:14}
\begin{adjustbox}{max width=\textwidth}
  \begin{tabular}{*{17}{l}}
\hline
$n^{2S+1}L_{J}$&  \multicolumn{2}{c}{0.5} & \multicolumn{2}{c}{0.7}& \multicolumn{2}{c}{0.9}&\multicolumn{2}{c}{1.0}& \multicolumn{2}{c}{2.0} & & &\\
& A & B & A & B & A & B & A & B & A & B & \cite{ebert2011}& \cite{chen2015}&\cite{98}&\cite{4}&  Exp.\cite{olive}\\
 \hline
$(1^2P_{1/2})$&	2.709	&	2.731	&	2.758	&	2.765	&	2.765	&	2.787	&	2.756	&	2.796	&	2.915	&	2.931&2.792 &2.779&2.790&2.761(77)(79)&2.792$\pm$0.0033		\\
$(1^2P_{3/2})$&	2.707	&	2.728	&	2.752	&	2.759	&	2.755	&	2.777	&	2.744	&	2.781	&	2.828	&	2.839 &2.819 &2.814&2.830	 &2.891(32)(36)&2.820$\pm$0.0012	\\
$(1^4P_{1/2})$&	2.710	&	2.732	&	2.761	&	2.768	&	2.770	&	2.792	&	2.763	&	2.803	&	2.958	&	2.977	&&&&&2.931$\pm$0.0008	\\
$(1^4P_{3/2})$&	2.708	&	2.730	&	2.755	&	2.762	&	2.760	&	2.782	&	2.750	&	2.788	&	2.872	&	2.885	\\
$(1^5P_{5/2})$&	2.705	&	2.726	&	2.748	&	2.754	&	2.747	&	2.769	&	2.733	&	2.769	&	2.756	&	2.762		\\
\hline
$(2^2P_{1/2})$&	2.967	&	3.000	&	3.067	&	3.081	&	3.107	&	3.145	&	3.107	&	3.170	&	3.467	&	3.500	&3.179 &3.195	& \\
$(2^2P_{3/2})$&	2.964	&	2.997	&	3.061	&	3.073	&	3.095	&	3.133	&	3.092	&	3.154	&	3.355	&	3.381	&3.201 &3.204	\\
$(2^4P_{1/2})$&	2.968	&	3.002	&	3.070	&	3.084	&	3.113	&	3.151	&	3.114	&	3.178	&	3.522	&	3.559		\\
$(2^4P_{3/2})$&	2.966	&	2.999	&	3.064	&	3.077	&	3.101	&	3.139	&	3.099	&	3.162	&	3.411	&	3.440		\\
$(2^4P_{5/2})$&	2.962	&	2.995	&	3.056	&	3.068	&	3.085	&	3.123	&	3.079	&	3.140	&	3.262	&	3.281		\\
\hline
$(3^2P_{1/2})$&	3.205	&	3.250	&	3.364	&	3.384	&	3.443	&	3.499	&	3.456	&	3.546	&	4.060	&	4.112	&3.500&3.521	\\
$(3^2P_{3/2})$&	3.203	&	3.247	&	3.357	&	3.376	&	3.430	&	3.485	&	3.439	&	3.527	&	3.927	&	3.971	&3.519 &3.525	\\
$(3^4P_{1/2})$&	3.207	&	3.251	&	3.367	&	3.387	&	3.449	&	3.507	&	3.464	&	3.555	&	4.126	&	4.182		\\
$(3^4P_{3/2})$&	3.204	&	3.248	&	3.360	&	3.380	&	3.436	&	3.492	&	3.447	&	3.536	&	3.994	&	4.041		\\
$(3^4P_{5/2})$&	3.200	&	3.244	&	3.351	&	3.370	&	3.419	&	3.472	&	3.425	&	3.511	&	3.817	&	3.853		\\
\hline
$(4^2P_{1/2})$&	3.431	&	3.486	&	3.651	&	3.677	&	3.776	&	3.848	&	3.804	&	3.919	&	4.687	&	4.760	&3.785	\\
$(4^2P_{3/2})$&	3.429	&	3.483	&	3.644	&	3.670	&	3.762	&	3.833	&	3.786	&	3.899	&	4.537	&	4.600	&3.804	\\
$(4^4P_{1/2})$&	3.433	&	3.488	&	3.655	&	3.681	&	3.783	&	3.855	&	3.812	&	3.929	&	4.762	&	4.840		\\
$(4^4P_{3/2})$&	3.430	&	3.485	&	3.647	&	3.673	&	3.769	&	3.840	&	3.795	&	3.909	&	4.612	&	4.680		\\
$(4^4P_{5/2})$&	3.426	&	3.480	&	3.638	&	3.663	&	3.750	&	3.820	&	3.772	&	3.882	&	4.412	&	4.466		\\
\hline
$(5^2P_{1/2})$&	3.648	&	3.713	&	3.931	&	3.964	&	4.106	&	4.194	&	4.151	&	4.293	&	5.344	&	5.438	&4.048	\\
$(5^2P_{3/2})$&	3.645	&	3.709	&	3.924	&	3.956	&	4.092	&	4.178	&	4.133 &	4.272	&	5.178	&	5.261	&4.066	\\
$(5^4P_{1/2})$&	3.649	&	3.714	&	3.935	&	3.968	&	4.114	&	4.202	&	4.159	&	4.304	&	5.427	&	5.526		\\
$(5^4P_{3/2})$&	3.647	&	3.711	&	3.928	&	3.960	&	4.099	&	4.186	&	4.142	&	4.283	&	5.261	&	5.349		\\
$(5^4P_{5/2})$&	3.643	&	3.707	&	3.918	&	3.950	&	4.080	&	4.165	&	4.119	&	4.254	&	5.040	&	5.114		\\
\hline
$(1^4D_{1/2})$&	2.927	&	2.963	&	3.017	&	3.032	&	3.041	&	3.087	&	3.047	&	3.108	&	3.446	&	3.490	&		&		&		\\
$(1^2D_{3/2})$&	2.927	&	2.959	&	3.006	&	3.020	&	3.018	&	3.065	&	3.027	&	3.080	&	3.279	&	3.311	&	3.059	&	3.055	&	&&	\\
$(1^4D_{3/2})$&	2.928	&	2.960	&	3.009	&	3.024	&	3.034	&	3.073	&	3.162	&	3.089	&	3.335	&	3.370	&		&		&		\\
$(1^2D_{5/2})$&	2.923	&	2.955	&	2.996	&	3.009	&	3.009	&	3.045	&	3.026	&	3.054	&	3.130	&	3.152	&	3.076	&	3.076	&&&	3.079$\pm$0.014	\\
$(1^4D_{5/2})$&	2.924	&	2.956	&	3.000	&	3.013	&	3.016	&	3.053	&	2.997	&	3.064	&	3.186	&	3.211	&		&		&		\\
$(1^4D_{7/2})$&	2.919	&	2.952	&	2.988	&	3.000	&	2.993	&	3.028	&	2.979	&	3.032	&	2.999	&	3.013	&		&		&		\\
\hline
$(2^4D_{1/2})$&	3.169	&	3.213	&	3.312	&	3.333	&	3.381	&	3.437	&	3.390	&	3.480	&	4.055	&	4.119	&		&		&		\\
$(2^2D_{3/2})$&	3.165	&	3.208	&	3.300	&	3.320	&	3.360	&	3.413	&	3.364	&	3.450	&	3.857	&	3.907	&	3.388	&	3.407	&		\\
$(2^4D_{3/2})$&	3.166	&	3.210	&	3.304	&	3.324	&	3.367	&	3.421	&	3.372	&	3.460	&	3.923	&	3.978	&		&	3.416	&		\\
$(2^2D_{5/2})$&	3.161	&	3.204	&	3.290	&	3.310	&	3.341	&	3.393	&	3.340	&	3.422	&	3.680	&	3.719	&	3.407	&		&		\\
$(2^4D_{5/2})$&	3.162	&	3.205	&	3.294	&	3.314	&	3.348	&	3.401	&	3.349	&	3.432	&	3.746	&	3.790	&		&		&		\\
$(2^4D_{7/2})$&	3.157	&	3.200	&	3.281	&	3.300	&	3.324	&	3.375	&	3.319	&	3.398	&	3.526	&	3.555	&		&		&		\\
\hline
$(1^4F_{3/2})$&	3.131	&	3.174	&	3.253	&	3.246	&	3.305	&	3.360	&	3.307	&	3.392	&	3.895	&	3.959	&		&		&		\\
$(1^2F_{5/2})$&	3.124	&	3.167	&	3.237	&	3.257	&	3.277	&	3.328	&	3.272	&	3.352	&	3.653	&	3.700	&	3.278	&	3.286	&		\\
$(1^4F_{5/2})$&	3.126	&	3.169	&	3.241	&	3.261	&	3.284	&	3.337	&	3.282	&	3.363	&	3.719	&	3.771	&		&		&		\\
$(1^4F_{7/2})$&	3.120	&	3.162	&	3.227	&	3.246	&	3.258	&	3.308	&	3.250	&	3.327	&	3.499	&	3.536	&		&		&		\\
$(1^2F_{7/2})$&	3.118	&	3.160	&	3.223	&	3.242	&	3.250	&	3.299	&	3.240	&	3.316	&	3.432	&	3.465	&	3.292	&	3.301	&		\\
$(1^4F_{9/2})$&	3.113	&	3.154	&	3.210	&	3.228	&	3.227	&	3.274	&	3.211	&	3.283	&	3.234	&	3.254	&		&		&		\\
\hline
\end{tabular}
\end{adjustbox}
\end{table*}

\begin{table*}
\caption{Masses of orbital excited states of $\Xi_{c}^{+}$(in GeV).}
\label{tab:15}
\begin{tabular}{c|ccccccccccc}
\hline
$n^{2S+1}L_{J}$&  \multicolumn{2}{c}{0.5} & \multicolumn{2}{c}{0.7}& \multicolumn{2}{c}{0.9}&\multicolumn{2}{c}{1.0}& \multicolumn{2}{c}{2.0}& Exp.\cite{olive}  \\
& A & B & A & B & A & B & A & B & A & B \\
\hline
$(1^2P_{1/2})$&	2.729	&	2.729	&	2.751	&	2.759	&	2.801	&	2.802	&	2.802	&	2.810	&	2.949	&	2.986	&2789.1 $\pm$ 0.032	\\
$(1^2P_{3/2})$&	2.725	&	2.726	&	2.745	&	2.752	&	2.789	&	2.790	&	2.787	&	2.794	&	2.853	&	2.879	&2819.6$\pm$ 0.0012		\\
$(1^4P_{1/2})$&	2.730	&	2.730	&	2.754	&	2.762	&	2.807	&	2.808	&	2.809	&	2.818	&	2.997	&	3.039	&2931.6$\pm$ 0.0008	\\
$(1^4P_{3/2})$&	2.727	&	2.727	&	2.748	&	2.756	&	2.795	&	2.796	&	2.794	&	2.802	&	2.901	&	2.933	&		\\
$(1^4P_{5/2})$&	2.723	&	2.723	&	2.741	&	2.747	&	2.779	&	2.780	&	2.775	&	2.780	&	2.772	&	2.790	&		\\
\hline
$(2^2P_{1/2})$&	3.013	&	3.017	&	3.056	&	3.077	&	3.166	&	3.173	&	3.183	&	3.198	&	3.528	&	3.597	&		\\
$(2^2P_{3/2})$&	3.010	&	3.014	&	3.049	&	3.070	&	3.152	&	3.158	&	3.165	&	3.180	&	3.404	&	3.459	&		\\
$(2^4P_{1/2})$&	3.015	&	3.019	&	3.059	&	3.081	&	3.173	&	3.180	&	3.192	&	3.208	&	3.590	&	3.665	&		\\
$(2^4P_{3/2})$&	3.011	&	3.015	&	3.053	&	3.074	&	3.159	&	3.165	&	3.174	&	3.189	&	3.466	&	3.528	&		\\
$(2^4P_{5/2})$&	3.007	&	3.011	&	3.043	&	3.064	&	3.140	&	3.146	&	3.150	&	3.165	&	3.301	&	3.344	&		\\
\hline
$(3^2P_{1/2})$&	3.276	&	3.284	&	3.355	&	3.382	&	3.526	&	3.537	&	3.561	&	3.586	&	4.151	&	4.253	&		\\
$(3^2P_{3/2})$&	3.273	&	3.284	&	3.347	&	3.375	&	3.511	&	3.522	&	3.542	&	3.566	&	4.004	&	4.090	&		\\
$(3^4P_{1/2})$&	3.278	&	3.286	&	3.358	&	3.386	&	3.533	&	3.545	&	3.571	&	3.596	&	4.225	&	4.334	&		\\
$(3^4P_{3/2})$&	3.275	&	3.282	&	3.351	&	3.378	&	3.518	&	3.529	&	3.551	&	3.576	&	4.078	&	4.171	&		\\
$(3^4P_{5/2})$&	3.270	&	3.278	&	3.341	&	3.369	&	3.498	&	3.509	&	3.525	&	3.549	&	3.882	&	3.954	&		\\
\hline
$(4^2P_{1/2})$&	3.526	&	3.537	&	3.644	&	3.678	&	3.882	&	3.898	&	3.939	&	3.974	&	4.810	&	4.947	&		\\
$(4^2P_{3/2})$&	3.523	&	3.533	&	3.637	&	3.671	&	3.866	&	3.882	&	3.919	&	3.952	&	4.643	&	4.762	&		\\
$(4^4P_{1/2})$&	3.528	&	3.539	&	3.648	&	3.682	&	3.890	&	3.907	&	3.949	&	3.985	&	4.893	&	5.039	&		\\
$(4^4P_{3/2})$&	3.524	&	3.535	&	3.641	&	3.674	&	3.874	&	3.890	&	3.929	&	3.963	&	4.726	&	4.854	&		\\
$(4^4P_{5/2})$&	3.520	&	3.530	&	3.630	&	3.664	&	3.852	&	3.868	&	3.902	&	3.933	&	4.505	&	4.608	&		\\
\hline
$(5^2P_{1/2})$&	3.765	&	3.779	&	3.927	&	3.968	&	4.235	&	4.257	&	4.318	&	4.360	&	5.499	&	5.673	&		\\
$(5^2P_{3/2})$&	3.762	&	3.775	&	3.920	&	3.960	&	4.218	&	4.240	&	4.296	&	4.338	&	5.315	&	5.469	&		\\
$(5^4P_{1/2})$&	3.767	&	3.781	&	3.931	&	3.972	&	4.243	&	4.265	&	4.329	&	4.372	&	5.591	&	5.775	&		\\
$(5^4P_{3/2})$&	3.763	&	3.777	&	3.923	&	3.964	&	4.227	&	4.248	&	4.307	&	4.349	&	5.407	&	5.571	&		\\
$(5^4P_{5/2})$&	3.759	&	3.772	&	3.913	&	3.953	&	4.204	&	4.226	&	4.277	&	4.319	&	5.162	&	5.299	&		\\
\hline
$(1^4D_{1/2})$&	2.962	&2.965		&	3.013	&	3.027	&	3.104	&	3.110	&	3.117	&	3.134	&	3.519	&	3.603	&	\\
$(1^2D_{3/2})$&	2.957	&2.960		&	3.001	&	3.015	&	3.081	&	3.087	&	3.088	&	3.104	&	3.333	&	3.396	&3.054$\pm$0.017	\\
$(1^4D_{3/2})$&	2.959	&2.962		&	3.005	&	3.019	&	3.089	&	3.095	&	3.097	&	3.114	&	3.395	&	3.465	&	\\
$(1^2D_{5/2})$&	2.953	&2.956		&	2.991	&	3.005	&	3.060	&	3.067	&	3.062	&	3.077	&	3.167	&	3.212	&	3.077$\pm$0.0004\\
$(1^4D_{5/2})$&		2.954	&2.958		&	2.995	&	3.009	&	3.068	&	3.074	&	3.071	&	3.087	&	3.229	&	3.281	&	\\
$(1^4D_{7/2})$&	2.949	&2.952		&	2.982	&	2.996	&	3.042	&	3.049	&	3.039	&	3.053	&	3.022	&	3.052	&	3.123$\pm$0.016\\
\hline
$(2^4D_{1/2})$&3.088	&3.219		&	3.308	&	3.330	&	3.461	&	3.474	&	3.490	&	3.516	&	4.160	&	4.281	&	\\
$(2^2D_{3/2})$&	3.082	&	3.214	&	3.297	&	3.318	&	3.436	&	3.449	&	3.460	&	3.485	&	3.940	&	4.036	&	\\
$(2^4D_{3/2})$&	3.084	&3.215		&	3.301	&	3.322	&	3.444	&	3.457	&	3.470	&	3.495	&	4.013	&	4.118	&	\\
$(2^2D_{5/2})$&	3.076	&	3.209	&	3.287	&	3.308	&	3.415	&	3.427	&	3.434	&	3.458	&	3.744	&	3.819	&	\\
$(2^4D_{5/2})$&	3.078	&	3.211	&	3.291	&	3.312	&	3.423	&	3.435	&	3.444	&	3.468	&	3.817	&	3.901	&	\\
$(2^4D_{7/2})$&	3.071	&	3.205	&	3.279	&	3.298	&	3.396	&	3.407	&	3.411	&	3.434	&	3.573	&	3.629	&	\\
\hline
$(1^4F_{3/2})$&	3.172	&	3.179	&	3.250	&	3.271	&	3.339	&	3.387	&	3.465	&	3.432	&	3.997	&	4.116	&	\\
$(1^2F_{5/2})$&	3.165	&3.171		&	3.234	&	3.255	&	3.306	&	3.357	&	3.421	&	3.394	&	3.728	&	3.817	&	\\
$(1^4F_{5/2})$&	3.167	&	3.173	&	3.238	&	3.259	&	3.315	&	3.365	&	3.433	&	3.406	&	3.801	&	3.898	&	\\
$(1^4F_{7/2})$&3.160	&3.166		&	3.224	&	3.244	&	3.284	&	3.338	&	3.393	&	3.366	&	3.556	&	3.627	&	\\
$(1^2F_{7/2})$&	3.158	&3.164		&	3.219	&	3.239	&	3.275	&	3.330	&	3.381	&	3.354	&	3.483	&	3.545	&	\\
$(1^4F_{9/2})$&	3.152	&3.158		&	3.206	&	3.226	&	3.248	&	3.305	&	3.345	&	3.318	&	3.263	&	3.301	&	\\
\hline
\end{tabular}
\end{table*}

\begin{table*}
\caption{Masses of radial excited states of $\Omega_{c}^{0}$(in GeV).}
\centering
\label{tab:16}
\begin{tabular}{llllllllllll}
\hline
($\nu$) & A & B & Others &  A & B & Others\\
\hline
 &\multicolumn{2}{c}{$1^{2}S_{\frac{1}{2}}$ } & & \multicolumn{2}{c}{$1^{4}S_{\frac{3}{2}}$ } & \\
\hline
0.5	&	2.695	&	2.695	&	2.695$\pm$0.0017\cite{olive}	&	2.748	&	2.754	&2.766$\pm$0.0002\cite{olive}		\\
0.7	&	2.695	&	2.695	&	2.698\cite{ebert2011}	&	2.742	&	2.752	& 2.768\cite{ebert2011}		\\
0.9	&	2.695	&	2.695	&2.718\cite{Roberts2008}&	2.748	&	2.750	& 2.776\cite{Roberts2008}		\\
1.0	&	2.695	&	2.695	&2.699\cite{13}	&	2.740	&	2.745	&2.768\cite{13}	\\
2.0	&	2.695	&	2.695 &2.731\cite{yoshida}&	2.751	&	2.757	&	2.779\cite{yoshida}\\
\hline
&\multicolumn{2}{c}{$2^{2}S_{\frac{1}{2}}$ } & & \multicolumn{2}{c}{$2^{4}S_{\frac{3}{2}}$ } & \\
\hline
0.5	&	3.075	&	3.089	&3.088\cite{ebert2011}		&	3.113	&	3.128	&3.123\cite{ebert2011}		\\
0.7	&	3.104	&	3.123	&	3.227\cite{yoshida}	&	3.137	&	3.160	&	3.257\cite{yoshida}	\\
0.9	&	3.140	&	3.154	&		&	3.176	&	3.190	&		\\
1.0	&	3.147	&	3.164	&		&	3.178	&	3.197	&		\\
2.0	&	3.279	&	3.306	&		&	3.320	&	3.349	&		\\
\hline
 &\multicolumn{2}{c}{$3^{2}S_{\frac{1}{2}}$ } & & \multicolumn{2}{c}{$3^{4}S_{\frac{3}{2}}$ } & \\
\hline
0.5	&	3.349	&	3.371	&	3.489\cite{ebert2011}	&	3.371	&	3.394	&	3.510\cite{ebert2011}	\\
0.7	&	3.421	&	3.454	&3.292\cite{yoshida}		&	3.441	&	3.476	&	3.285\cite{yoshida}	\\
0.9	&	3.505	&	3.531	&		&	3.527	&	3.552	&		\\
1.0	&	3.529	&	3.561	&		&	3.548	&	3.580	&		\\
2.0	&	3.866	&	3.921	&		&	3.892	&	3.948	&		\\
\hline
 &\multicolumn{2}{c}{$4^{2}S_{\frac{1}{2}}$ } & & \multicolumn{2}{c}{$4^{4}S_{\frac{3}{2}}$ } & \\
\hline
0.5	&	3.602	&	3.633	&3.814\cite{ebert2011}		&	3.617	&	3.647	&	3.830\cite{ebert2011}	\\
0.7	&	3.724	&	3.769	&		&	3.737	&	3.783	&		\\
0.9	&	3.863	&	3.900	&		&	3.877	&	3.914	&		\\
1.0	&	3.907	&	3.953	&		&	3.920	&	3.966	&		\\
2.0	&	4.491	&	4.576	&		&	4.510	&	4.595	&		\\
\hline
 &\multicolumn{2}{c}{$5^{2}S_{\frac{1}{2}}$ } & & \multicolumn{2}{c}{$5^{4}S_{\frac{3}{2}}$ } & \\
\hline
0.5	&	3.843	&	3.880	&4.102\cite{ebert2011}		&	3.853	&	3.890	&4.114\cite{ebert2011}		\\
0.7	&	4.017	&	4.074	&		&	4.026	&	4.084	&		\\
0.9	&	4.215	&	4.263	&		&	4.225	&	4.273	&		\\
1.0	&	4.283	&	4.343	&		&	4.292	&	4.352	&		\\
2.0	&	5.150	&	5.267	&		&	5.164	&	5.281	&		\\
\hline
\end{tabular}
\end{table*}
 
 \begin{table*}
\caption{Masses of orbital excited states of $\Omega_{c}^{0}$(in GeV).}
\label{tab:17}
\begin{adjustbox}{max width=\textwidth}
  \begin{tabular}{*{17}{l}}
\hline
&  \multicolumn{2}{c}{$\nu$=0.5} & \multicolumn{2}{c}{$\nu$=0.7}& \multicolumn{2}{c}{$\nu$=0.9} &\multicolumn{2}{c}{$\nu$=1.0}& \multicolumn{2}{c}{$\nu$=2.0} & & \\
$n^{2S+1}L_{J}$&&&&&&&&\\
& A & B & A & B & A & B & A & B & A & B &\cite{ebert2011} & \cite{Roberts2008}& \cite{yoshida}& \cite{98} &\cite{87}\\
\hline
$(1^2P_{1/2})$&	2.985	&	3.000	&	2.997	&	3.019	&	3.022	&	3.037	&	3.022	&	3.041	&	3.164	&	3.221	&	3.055 &2.977&3.030& 3.250	& 2.98$\pm$0.16\\
$(1^2P_{3/2})$&	2.982	&	2.996	&	2.991	&	3.012	&	3.010	&	3.024	&	3.007	&	3.024	&	3.067	&	3.110	&	3.054 &2.986 & 3.033& 3.260& 2.98$\pm$0.16\\
$(1^4P_{1/2})$&	2.987	&	3.001	&	3.000	&	3.023	&	3.028	&	3.043	&	3.030	&	3.050	&	3.212	&	3.277	&	2.966&2.977	\\
$(1^4P_{3/2})$&	2.984	&	2.998	&	2.994	&	3.015	&	3.016	&	3.030	&	3.014	&	3.033	&	3.115	&	3.165	&	3.029 & 2.959& 	\\
$(1^4P_{5/2})$&	2.979	&	2.993	&	2.986	&	3.006	&	3.000	&	3.013	&	2.994	&	3.010	&	2.986	&	3.017	&	3.051& 3.014&3.057& 3.320	\\
\hline
$(2^2P_{1/2})$&	3.254	&	3.277	&	3.308	&	3.342	&	3.377	&	3.402	&	3.394	&	3.427	&	3.740	&	3.839	&	3.435	&&3.048\\
$(2^2P_{3/2})$&	3.251	&	3.273	&	3.301	&	3.334	&	3.364	&	3.389	&	3.377	&	3.408	&	3.622	&	3.703	&	3.433&&3.056	\\
$(2^4P_{1/2})$&	3.256	&	3.278	&	3.312	&	3.346	&	3.383	&	3.408	&	3.403	&	3.436	&	3.800	&	3.907	&	3.384&	\\
$(2^4P_{3/2})$&	3.253	&	3.275	&	3.305	&	3.338	&	3.370	&	3.396	&	3.386	&	3.417	&	3.681	&	3.771	&	3.415	\\
$(2^4P_{5/2})$&	3.248	&	3.270	&	3.295	&	3.328	&	3.353	&	3.378	&	3.363	&	3.393	&	3.523	&	3.589	&	3.427&&3.477	\\
\hline
$(3^2P_{1/2})$&	3.507	&	3.537	&	3.608	&	3.653	&	3.729	&	3.767	&	3.767	&	3.813	&	4.362	&	4.504	&	3.754	&&3.048\\
$(3^2P_{3/2})$&	3.504	&	3.533	&	3.601	&	3.646	&	3.716	&	3.752	&	3.748	&	3.793	&	4.224	&	4.347	&	3.752	&&3.056\\
$(3^4P_{1/2})$&	3.508	&	3.538	&	3.612	&	3.657	&	3.736	&	3.775	&	3.776	&	3.823	&	4.430	&	4.583	&	3.717	\\
$(3^4P_{3/2})$&	3.505	&	3.535	&	3.605	&	3.650	&	3.722	&	3.760	&	3.757	&	3.803	&	4.293	&	4.426	&	3.737	\\
$(3^4P_{5/2})$&	3.501	&	3.531	&	3.595	&	3.640	&	3.704	&	3.739	&	3.733	&	3.777	&	0.000	&	4.216	&	3.744&&3.620	\\
\hline
$(4^2P_{1/2})$&	3.748	&	3.784	&	3.901	&	3.958	&	4.079	&	4.127	&	4.140	&	4.199	&	5.019	&	5.209	&	4.037	\\
$(4^2P_{3/2})$&	3.744	&	3.781	&	3.893	&	3.950	&	4.065	&	4.112	&	4.120	&	4.179	&	4.866	&	5.033	&	4.036	\\
$(4^4P_{1/2})$&	3.749	&	3.786	&	3.904	&	3.962	&	4.086	&	4.135	&	4.149	&	4.209	&	5.096	&	5.296	&	4.009	\\
$(4^4P_{3/2})$&	3.746	&	3.783	&	3.897	&	3.954	&	4.072	&	4.120	&	4.130	&	4.189	&	4.942	&	5.121	&	4.023	\\
$(4^4P_{5/2})$&	3.742	&	3.778	&	3.887	&	3.943	&	4.053	&	4.099	&	4.104	&	4.162	&	4.738	&	4.887	&	4.028	\\
\hline
$(5^2P_{1/2})$&	3.979	&	4.022	&
	4.186	&	4.255	&	4.427	&	4.486	&	4.512	&	4.587	&	5.707	&	5.946	&	\\
$(5^2P_{3/2})$&	3.976	&	4.019	&
	4.179	&	4.247	&	4.412	&	4.470	&	4.492	&	4.565	&	5.540	&	5.754	&	\\
$(5^4P_{1/2})$ &	3.980	&	4.024	&
	4.190	&	4.259	&	4.434	&	4.494	&	4.522	&	4.598	&	5.791	&	6.042	&	\\
$(5^4P_{3/2})$&	3.977	&	4.021	&
	4.183	&	4.251	&	4.419	&	4.478	&	4.502	&	4.576	&	5.624	&	5.850	&	\\
$(5^4P_{5/2})$&	3.973	&	4.017	&
	4.174	&	4.241	&	4.399	&	4.457	&	4.476	&	4.547	&	5.400	&	5.593	&	\\
	 \hline
$(1^4D_{3/2})$&	3.213	&	3.236	&	3.250	&	3.287	&	3.310	&	3.337	&	3.322	&	3.354	&	3.712	&	3.792	&	3.287\\ 
$(1^2D_{3/2})$&	3.208	&	3.230	&	3.240	&	3.273	&	3.288	&	3.313	&	3.294	&	3.325	&	3.534	&	3.595	&	3.298\\
$(1^4D_{3/2})$&	3.210	&	3.232	&	3.243	&	3.278	&	3.295	&	3.321	&	3.303	&	3.335	&	3.593	&	3.661	&	3.282\\
$(1^2D_{5/2})$&	3.203	&	3.225	&	3.230	&	3.262	&	3.268	&	3.292	&	3.269	&	3.299	&	3.376	&	3.420	&	3.297\\
$(1^4D_{5/2})$&	3.205	&	3.227	&	3.234	&	3.266	&	3.275	&	3.300	&	3.278	&	3.308	&	3.435	&	3.486	&	3.286\\
$(1^4D_{7/2})$&	3.198	&	3.220	&	3.222	&	3.252	&	3.250	&	3.274	&	3.248	&	3.276	&	3.237	&	3.267	&	3.283\\
\hline
$(2^4D_{3/2})$&	3.465	&	3.496	&	3.551	&	3.597	&	3.658	&	3.695	&	3.693	&	3.741	&	4.346	&	4.460	&3.623	\\
$(2^2D_{3/2})$&	3.460	&	3.490	&	3.540	&	3.584	&	3.637	&	3.673	&	3.664	&	3.709	&	4.141	&	4.233	&	3.627\\
$(2^4D_{3/2})$&	3.462	&	3.492	&	3.544	&	3.589	&	3.644	&	3.680	&	3.674	&	3.719	&	4.209	&	4.309	&3.613	\\
$(2^2D_{5/2})$&	3.456	&	3.485	&	3.530	&	3.573	&	3.617	&	3.652	&	3.637	&	3.680	&	3.958	&	4.032	&	3.626\\
$(2^4D_{5/2})$&	3.457	&	3.487	&	3.533	&	3.577	&	3.624	&	3.660	&	3.647	&	3.691	&	4.027	&	4.107	&3.614	\\
$(2^4D_{7/2})$&	3.452	&	3.481	&	3.521	&	3.563	&	3.600	&	3.635	&	3.614	&	3.656	&	3.798	&	3.855	&	3.611\\
\hline
$(1^4F_{3/2})$&	3.422	&	3.451	&	3.489	&	3.532	&	3.575	&	3.609	&	3.593	&	3.643	&	4.170	&	4.282	&	3.533\\
$(1^2F_{5/2})$&	3.414	&	3.443	&	3.472	&	3.514	&	3.545	&	3.579	&	3.558	&	3.602	&	3.919	&	4.004	&	3.522\\
$(1^4F_{5/2})$&	3.416	&	3.446	&	3.477	&	3.519	&	3.553	&	3.587	&	3.568	&	3.613	&	3.988	&	4.080	&	3.515\\
$(1^4F_{7/2})$&	3.410	&	3.438	&	3.462	&	3.504	&	3.526	&	3.559	&	3.535	&	3.577	&	3.759	&	3.828	&	3.514\\
$(1^2F_{7/2})$&	3.408	&	3.436	&	3.457	&	3.499	&	3.518	&	3.551	&	3.526	&	3.565	&	3.691	&	3.752	&	3.498\\
$(1^4F_{9/2})$&	3.401	&	3.430	&	3.444	&	3.485	&	3.493	&	3.526	&	3.496	&	3.532	&	3.485	&	3.525	&	3.485\\
\hline
\end{tabular}
\end{adjustbox}
\end{table*}

\begin{table*}
\caption{Semi-electronic decays in $s \rightarrow u$ transition for charm baryons are listed.}
\label{tab:18}
\centering
\begin{tabular}{ccccccccc}
\hline
Mode & $J^{P} \rightarrow J'^{P{'}}$ & $s^{l} \rightarrow s'^{l'}$& \multicolumn{2}{c}{$\bigtriangleup m$ (GeV)}& \multicolumn{3}{c}{Decay Rates
 (GeV)} & \\
\cline{4-5} \cline{6-8}\\
&&&A&B&A&B&Ref.  \cite{sven}\\
\hline
$\Xi_c^0 \rightarrow \Lambda_c^{+} e^{-} \bar{\nu}$ & $\frac{1}{2}^{+} \rightarrow \frac{1}{2}^{+}$& $0 \rightarrow 0$ &0.184&0.184& $7.839\times 10^{-19}$& $7.839\times 10^{-19}$ & $7.91\times 10^{-19} $\\
$\Xi_c^0 \rightarrow \Sigma_c^{+} e^{-} \bar{\nu}$ & $\frac{1}{2}^{+} \rightarrow \frac{1}{2}^{+}$ & $0 \rightarrow 0$ &0.026&0.018&$4.416\times 10^{-23}$ & $7.023\times 10^{-24}$ &$6.97\times 10^{-24} $ \\
~~$\Xi_c^+ \rightarrow \Sigma_c^{*++} e^{-} \bar{\nu}$ & $\frac{1}{2}^{+} \rightarrow \frac{3}{2}^{+}$ & $0 \rightarrow 1$ &0.019&0.013& $1.840 \times 10^{-23}$  & $2.760\times 10^{-24}$ &$3.74\times 10^{-24} $ \\
$\Omega_c^0 \rightarrow \Xi_c^{+} e^{-} \bar{\nu}$ & $\frac{1}{2}^{+} \rightarrow \frac{1}{2}^{+}$ & $1 \rightarrow 0$ &0.228&0.228&  $2.143\times 10^{-18} $ &$2.290\times 10^{-18} $& $2.26\times 10^{-18} $   \\
$\Omega_c^0 \rightarrow \Xi_c^{*+} e^{-} \bar{\nu}$ & $\frac{1}{2}^{+} \rightarrow \frac{3}{2}^{+}$ & $1 \rightarrow 1$ &0.07&0.071& $2.057\times 10^{-28} $ & $2.436\times 10^{-28} $&$1.49\times 10^{-29} $\\
\hline
\end{tabular}
\end{table*}

\section{Semi-electronic decays of $\Xi_{c}$ and $\Omega_{c}$ bayons }
In this section, we discuss, weak decays of heavy baryons. In which, the heavy quark(c or b) acts as spectator and the strange quark inside heavy hadron decays in weak interaction \cite{sven,Hai1,Hai}. These kind of small phase space transition could be possible in semi-electronic, semi-muonic and non leptonic decays of the heavy baryons and mesons. Such calculation has already been reported in heavy meson sector. For instance, Ref. \cite{sven} reported branching ratios and decays rates while  Ref. \cite{Hai} reported S-wave amplitudes and branching ratios of heavy-flavor conserving charmed and bottom baryons. In reference to that, we calculate here, the semi-electronic decays for strange-charm heavy flavour baryons $\Omega_{c}$ and $\Xi_{c}$ using our spectral parameters.\\
 The strange baryon with heavy flavor charm quark undergo weak transitions. The quark transitions of singly charmed baryon $\Xi_c^{0}$, $\Xi_c^{+}$ and $\Omega_c^{0}$  are  manifest in $s \rightarrow u$ in Table 18. In second column (of Table 18) we list initial and final total angular momentum (J) and parity (P) and in third column we give initial and final total spin $s_{l}$ of the light degree of freedom. Where, $\bigtriangleup m$ = M - m, is a the mass difference between initial and final state of Baryons. Here, we use our calculated ground state masses ($\nu$=1.0) of without first order correction(A) and with first order correction(B). The differential decay rates for exclusive semi-electronic decays are given by \cite{sven,nakul},
\begin{eqnarray}
\dfrac{d \Gamma}{dw} = \dfrac{G_{F}^{2} M^{5} \vert V_{CKM}\vert^{2}}{192 \pi^{3}} \sqrt{w^{2}-1} P(w)
\end{eqnarray}
where $P(w)$ contains the hadronic and leptonic tensor. After evaluating the integration over $w$=1 in the hadronic form factors we study the following decay given below. Different  $s_l \rightarrow s_{l^{'}}$ are noticeable. \\
For the final state with ``$\Lambda$-like" baryon,
\begin{eqnarray}
\Gamma^{\frac{1}{2}^{+} \rightarrow \frac{1}{2}^{+}}_{0^{+} \rightarrow 0^{+}} = \dfrac{G_{F}^{2}\vert V_{CKM}\vert^{2}}{60 \pi^{3}} (M-m)^{5}
\end{eqnarray}
For the final state with ``$\Sigma$-like" baryon,
\begin{eqnarray}
\Gamma^{\frac{1}{2}^{+} \rightarrow \frac{3}{2}^{+}}_{0^{+} \rightarrow 1^{+}} = \dfrac{G_{F}^{2}\vert V_{CKM}\vert^{2}}{30 \pi^{3}} (M-m)^{5}\vert A(1)\vert^{2}
\end{eqnarray}
For the final state with ``$\Xi$-like" baryon,
\begin{equation}
\Gamma^{\frac{1}{2}^{+} \rightarrow \frac{1}{2}^{+}}_{1^{+} \rightarrow 1^{+}} = \dfrac{G_{F}^{2}\vert V_{CKM}\vert^{2}}{15 \pi^{3}} (M-m)^{5}
\end{equation}
where $G_{F}$ is the Fermi Coupling constant and $\vert A(1)\vert^{2}$=1.  The superscript of $\Gamma$ in $Eq^n$.(16), (17) and (18) indicates spin parity transition ($J^{P} \rightarrow J'^{P{'}}$) of baryon , while the subscripts of $\Gamma$ indicate spin parity transition ($s^l \rightarrow s'^{{l}{'}}$) of light degrees of freedom. The results are tabulated in Table 18.

\section{Results and Discussions}

\begin{figure*}
\centering
\begin{minipage}[b]{0.40\linewidth}
\includegraphics[scale=0.30]{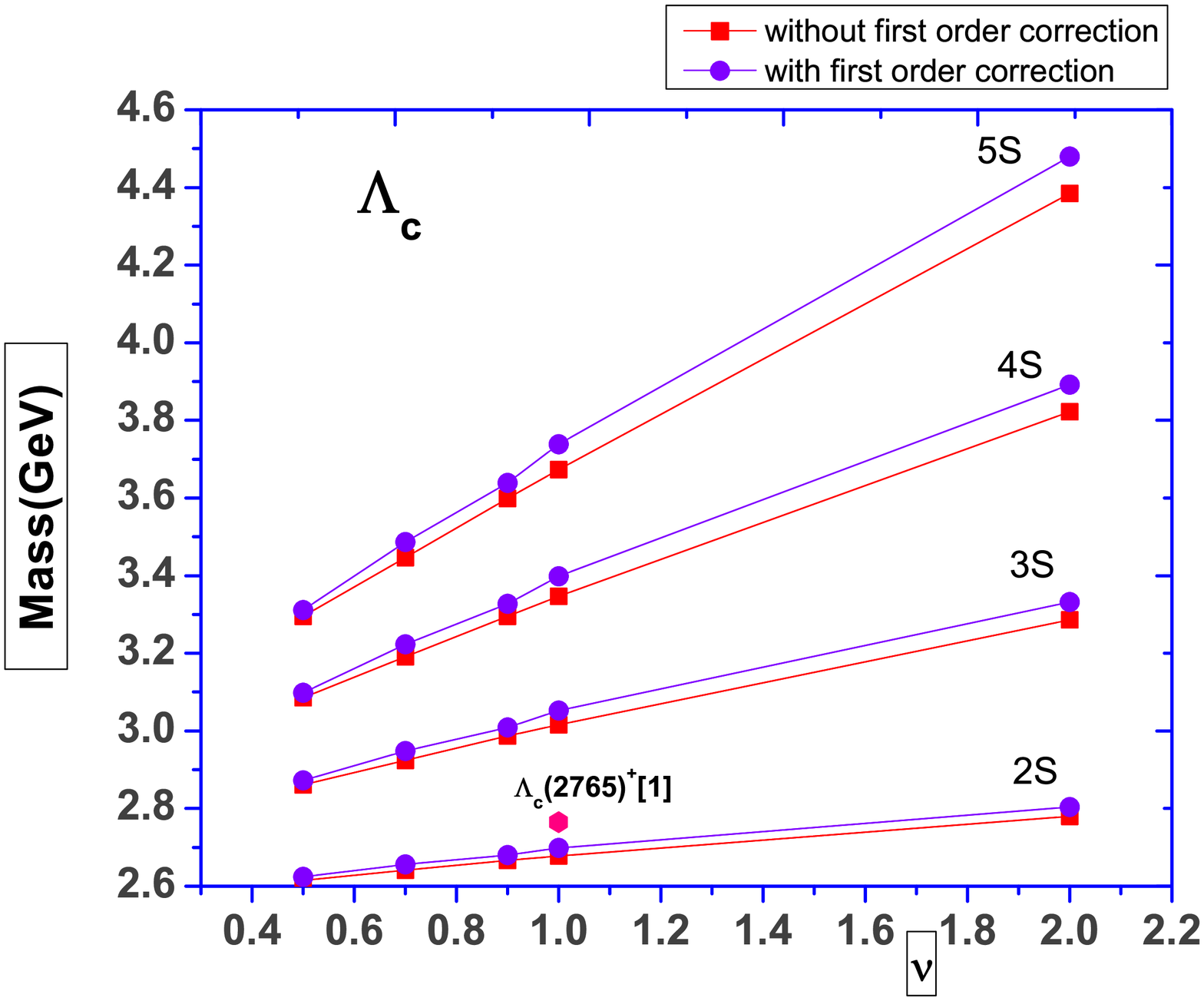}
\label{fig:minipage1}
\end{minipage}
\quad
\begin{minipage}[b]{0.40\linewidth}
\includegraphics[scale=0.30]{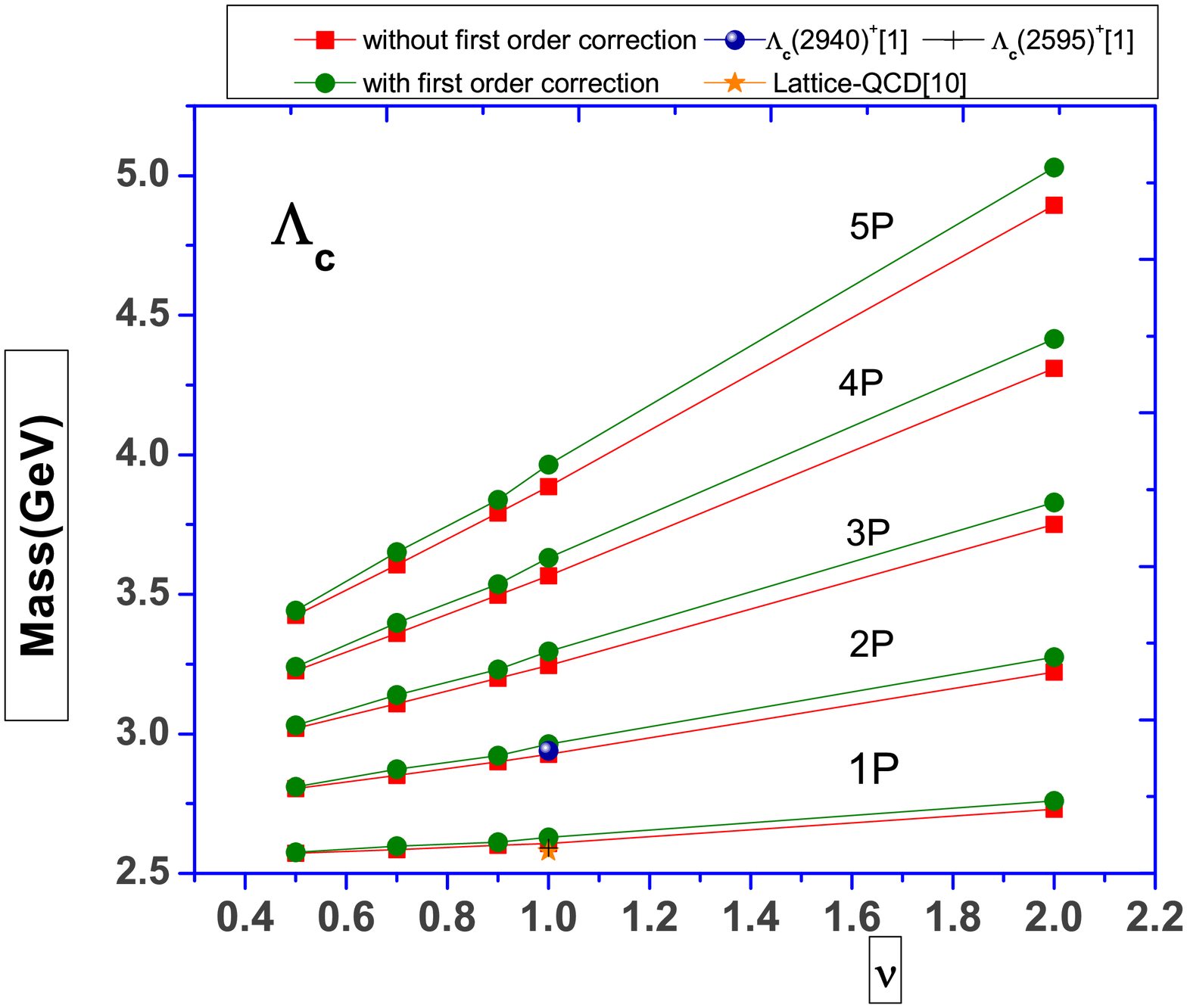}
\label{fig:minipage2}
\end{minipage}
\caption{\label{fig:epsart} Variation of mass with potential index in S and P states.}
\end{figure*}

\begin{figure*}
\centering
\begin{minipage}[b]{0.30\linewidth}
\includegraphics[scale=0.20]{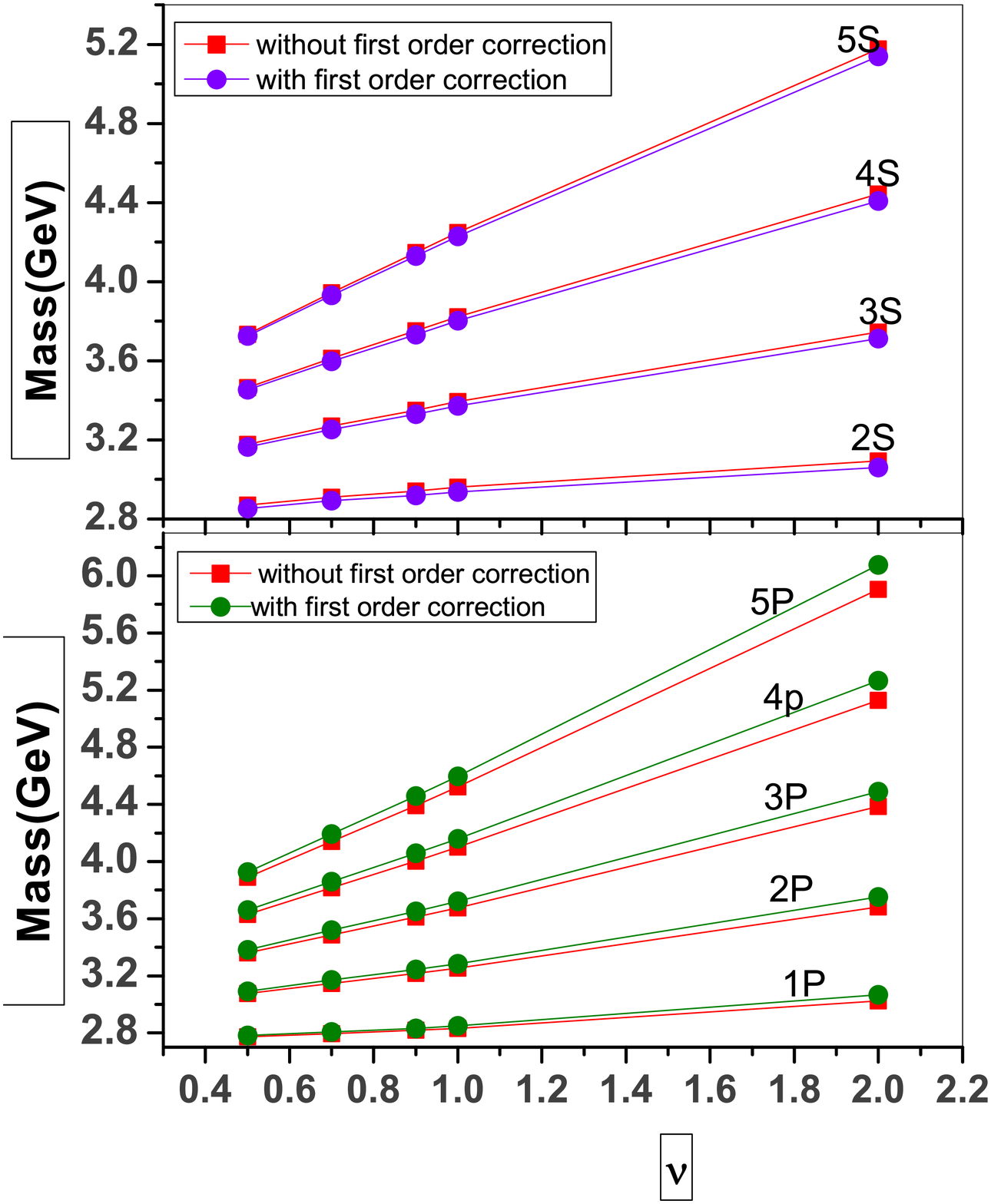}
\label{fig:minipage1}
\end{minipage}
\quad
\begin{minipage}[b]{0.30\linewidth}
\includegraphics[scale=0.20]{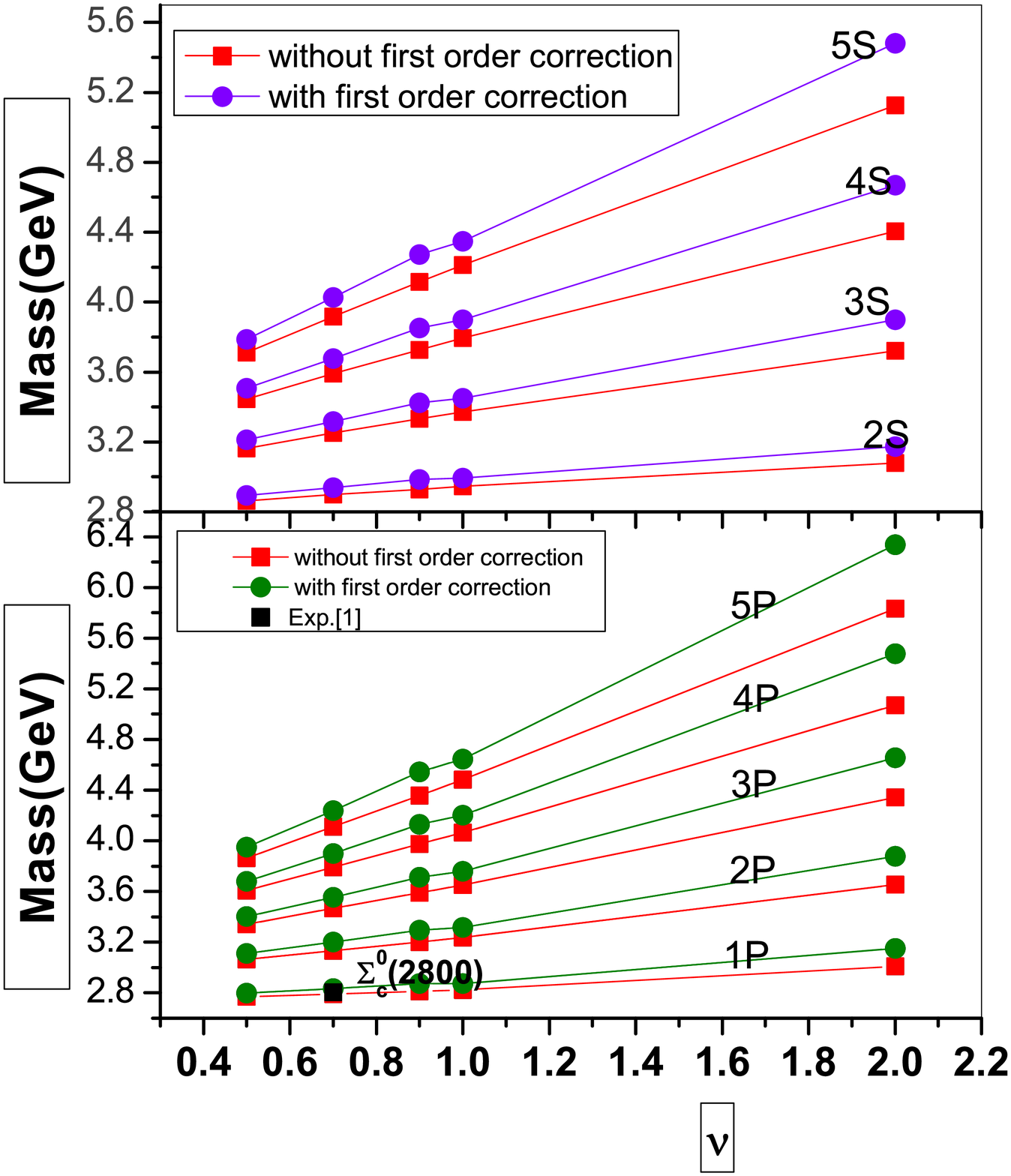}
\label{fig:minipage2}
\end{minipage}
\quad
\begin{minipage}[b]{0.30\linewidth}
\includegraphics[scale=0.20]{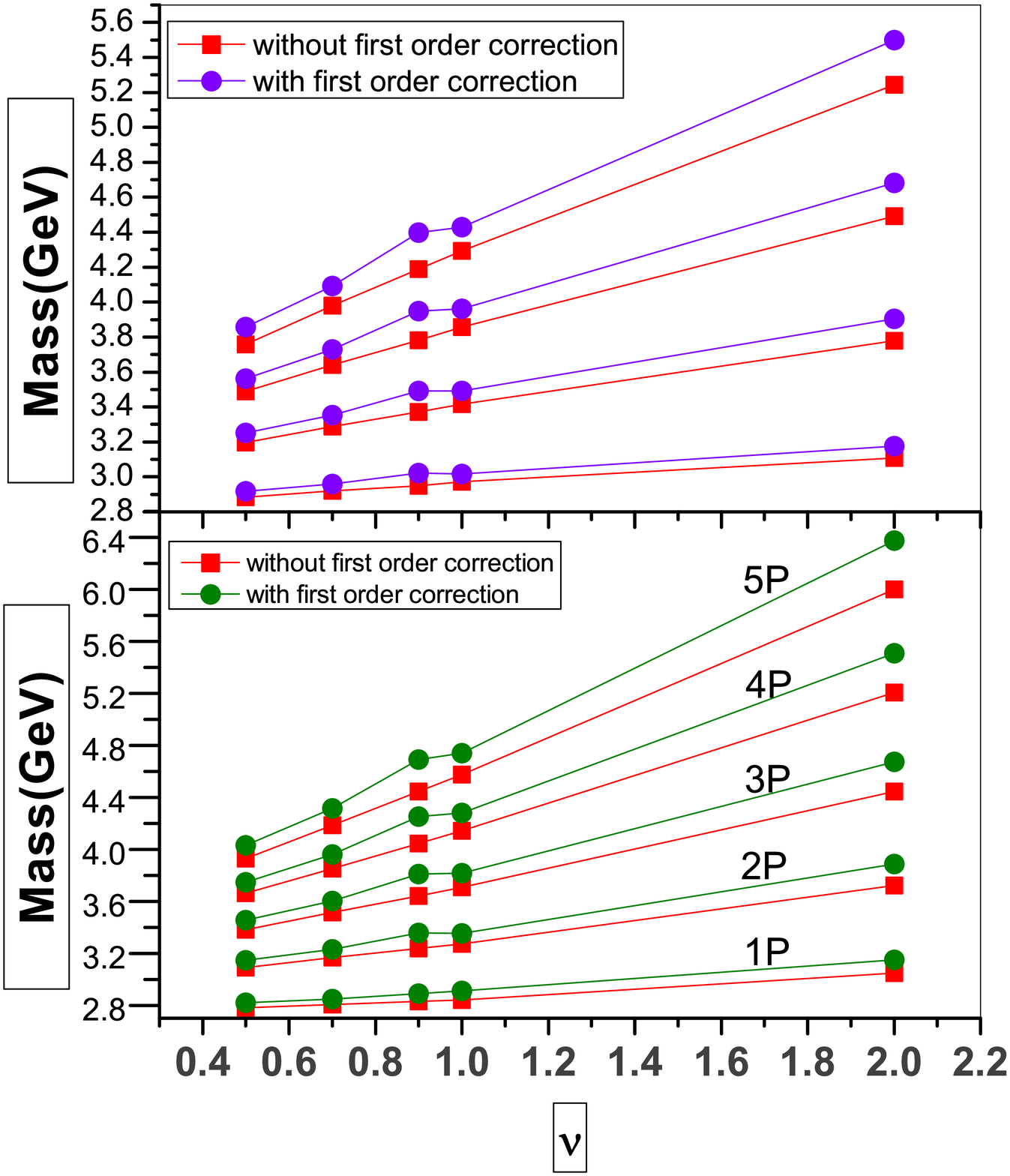}
\label{fig:minipage2}
\end{minipage}
\caption{\label{fig:epsart} Variation of mass with potential index in S and P states for $\Sigma_{c}^{+}$ (first), $\Sigma_{c}^{0}$ (second) and $\Sigma_{c}^{++}$ (third) baryons.}
\end{figure*}

The mass spectroscopy of single charmed baryons have been studied in the framework of Hypercentral Constituent Quark Model (hCQM) with coulomb plus power potential scheme. The ground state as well as the excited states (both radial and orbital) masses are calculated by solving the six dimensional hyper radial Schrodinger equation numerically. We also included the first order correction to the potential (see Eqn. (\ref{eq:7})). The calculations have been done without first order correction(A) as well as by adding $\frac{\alpha_{s}^{2}}{mx^{2}}$ first order correction(B) (See Table 4-17). The calculated values with first order correction are more than that of without first order correction. We also discuss and compare the experimental known states (mentioned in Table 2) with the obtained results. Present results  are in good accordance with experimental and other model results. Our  predicted $J^{P}$ values of known experimental states are listed in Table 19. We have plotted Regge Trajectories of all charmed baryons in (n, $M^2$) plane (See Fig. (6-8)) and also in (J, $M^2$) plane (See Fig. (9-13)). We have done linear fitting in (J, $M^2$) plane to obtain straight lines (See Fig. (9-12)). We observe that the square of the calculated masses fit very well to the linear trajectory and almost parallel for S, P and D states. 

\par A brief description of the singly charmed baryons (mentioned in Table-1) are given in following sub-sections. Mass spectra and the individual graph aspects for each system are also mentioned. M $\rightarrow$ $\nu$ graphs are plotted for S state (with $J^{P}$= $\frac{1}{2}^{+}$) and P state (with $J^{P}$= $\frac{1}{2}^{-}$) only, but mass spectra have been calculated for  S, P, D and F states.

\subsection{$\Lambda_c^{+}$}
The first singly charm baryon was $\Lambda_c^{+}$ and its experimental known ground state is $\Lambda_{c}(2286)^{+}$, assigned as $J^P =\frac{1}{2}^{+}$ by BaBar Collaboration \cite{3}. The first orbital excited states with quantum number $J^P = \frac{1}{2}^{-}$ and $J^P =\frac{3}{2}^{-}$ are $\Lambda_{c}(2595)^{+}$ and $\Lambda_{c}(2625)^{+}$ respectively. Their masses are also measured in CDF experiment \cite{CDF} recently. $\Lambda_{c}(2880)^{+}$ is also known experimentaly with $J^P = \frac{5}{2}^{+}$. Some known Lattice results for ground state $\Lambda_c^+$ are given below,
\begin{center}
$m_{\Lambda_{c}}({1}/{2}^{+})$ = 2280(17)(24) \cite{4}\\
$m_{\Lambda_{c}}({1}/{2}^{+})$ = 2333(112)(10) \cite{pacs}\\
$m_{\Lambda_{c}}({1}/{2}^{+})$ = 2254(48)(31) \cite{brown}\\
$m_{\Lambda_{c}}({1}/{2}^{+})$ = 2272(26) \cite{alex}\\
\end{center}
We have calculated the mass spectra of $\Lambda_{c}^{+}$ baryon for 1S-5S, 1P-5P, 1D-2D and 1F states with and without first order corrections (see Table 4-5). Our results are close to the experimental measurements and other theoretical predictions \cite{ebert2011,chen2015,yoshida} at potential index $\nu$= 1.0. The ground state lattice QCD calculations are also shown in Table (IV). The recent Lattice QCD result for P state is reported by Ref.\cite{4} is 2.578(144)(145) for J$^{P}$ value $\frac{1}{2}^{-}$. S. Migura et al. \cite{Migura} also reported masses 2.530 and 2.580 GeV for $J^{P}$ values, $\frac{1}{2}^{-}$ and $\frac{3}{2}^{-}$, respectively.\\ 

Two more states namely $\Lambda_{c}(2765)^{+}$ and $\Lambda_{c}(2940)^{+}$ are observed by CLEO collaboration \cite{cleo} and Babar collaboration \cite{3}. The spin and parity of these states have not been assigned yet. Therefore, many theoretical models \cite{{chen2015},yoshida,ebert2011,Roberts2008} have compared $\Lambda_{c}(2765)^{+}$ with $J^P = \frac{1}{2}^{+}$, first radial excited state. Our model observation also concluded it as 2S state. $\Lambda_{c}(2940)^{+}$ is compared with the orbital excited state (2P) in our calculation. Similar kind of predictions are given by \cite{chen2015}. Hence, we assigned, $J^P = \frac{1}{2}^{-}$ value. These two states (2S and 2P) are reasonably close to our present results at $\nu$=1.0.\\

We have plotted the graph $M \rightarrow \nu$ (Fig.1) for radial (upto 5S) and orbital (upto 5P) excited states. We have also included experimental \cite{olive} as well as lattice results \cite{4} in Fig.1. It is  observed that as $\nu$ increases the mass of $\Lambda_{c}$ baryon increases. The Regge trajectories of $\Lambda_{c}^{+}$ are plotted in Fig. (6,14). In both figures 1S, 1P and our predicted 2P states are close to experimental results while our predicted 2S and 1D states shown few MeV difference. We can observe that all results are parallel to each other without any fitting the experimental states are also shown for particular states.

\subsection{$\Sigma_c$}

$\Sigma_{c}$ baryons have triplet states, namely $\Sigma_{c}^{++}$, $\Sigma_{c}^{+}$ and $\Sigma_{c}^0$. We have calculated ground state as well as higher excited states with and without first order corrections for $\Sigma_{c}^0$, $\Sigma_{c}^+$ and $\Sigma_{c}^{++}$ (See Table 6-11). The separate calculations have been performed for all $\Sigma$'s due to their unequal light quarks (u,d) masses. The PDG (2014) has listed $\Sigma_{c}(2455)^0$, $\Sigma_{c}(2455)^+$ and $\Sigma_{c}(2455)^{++}$ state with $J^P = \frac{1}{2}^{+}$ and the states $\Sigma_{c}(2520)^{0*}$, $\Sigma_{c}(2520)^{+*}$ and $\Sigma_{c}(2520)^{++*}$  with $J^P = \frac{3}{2}^{+}$. Ref. \cite{zahra} has also done separate calculations of the ground state of three $\Sigma$'s.\\

\begin{figure}
\centering
\includegraphics[scale=0.30]{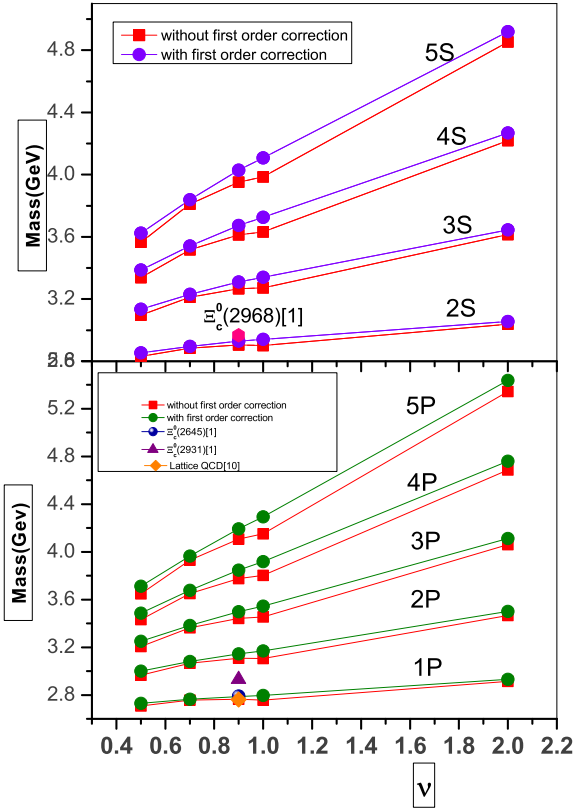}
\label{fig:minipage1}
\caption{\label{fig:epsart} Variation of mass with potential index in S and P states for $\Xi_{c}^{0}$ baryon.}
\end{figure}

\begin{figure}
\centering
\includegraphics[scale=0.30]{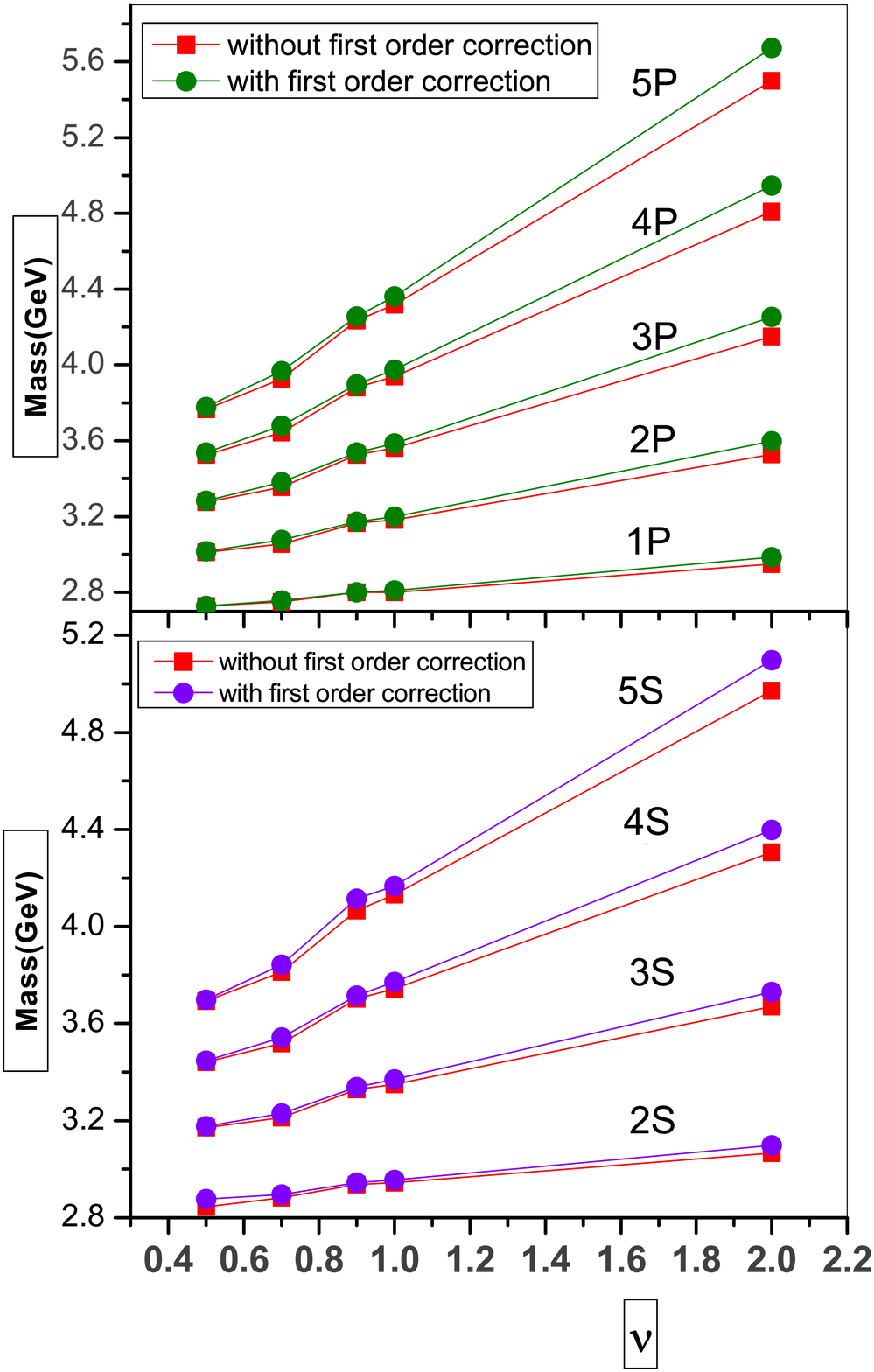}
\label{fig:minipage2}
\caption{\label{fig:epsart} Variation of mass with potential index in S and P states for $\Xi_{c}^{+}$ baryon.}
\end{figure}

\par For higher radial excited states, we have compared other theoretical results only with $\Sigma_c^0$ baryon as we do not have distinguished triplet states. The $J^{P}$ value of $\Sigma_{c}(2800)$ state is unknown experimentally. We compare, 1P state of our triplet $\Sigma_{c}^{++,+,0}$ with $J^P$($\frac{3}{2}^{-}$ and $\frac{5}{2}^{-}$) \cite{152}. Ref. \cite{87} has also calculated for $J^P$($\frac{1}{2}^{-}$ and $\frac{3}{2}^{-}$) and obtained 2.74$\pm$0.20 GeV. In addition, Ref. \cite{92} has also calculated m($\Sigma_{c}$)= 2.84 $\pm$ 0.11 GeV for $J^{P}=\frac{3}{2}^{-}$. Masses predicted by different Lattice QCD calculations for the ground states are listed below.
\begin{center}
$m_{\Sigma_{c}}({1}/{2}^{+})$ = 2434(20)(26) \cite{4}\\
$m_{\Sigma_{c}}({1}/{2}^{+})$ = 2467(39)(11) \cite{pacs}\\
$m_{\Sigma_{c}}({1}/{2}^{+})$ = 2474(48)(31) \cite{brown}\\
$m_{\Sigma_{c}}({1}/{2}^{+})$ = 2445(32) \cite{alex}\\
{\rule{\linewidth}{0.15mm} }
$m_{\Sigma_{c}}({3}/{2}^{+})$ =2506(18)(25) \cite{4}\\
$m_{\Sigma_{c}}({3}/{2}^{+})$ = 2551(43)(25) \cite{brown}\\
$m_{\Sigma_{c}}({3}/{2}^{+})$ = 2538(70)(11) \cite{pacs}\\
$m_{\Sigma_{c}}({3}/{2}^{+})$ = 2513(38) \cite{alex}\\
\end{center}
The present mass predicted  for $\Sigma_c$ baryons is reasonably close to experimental as well as other theoretical predictions for potential index $\nu$ = 0.9 [See Table 6-11]. We have plotted the M $\rightarrow \nu$ graph for $\Sigma_c^{0},\Sigma_c^{++}$ and $\Sigma_c^{+}$ with and without adding first order correction. We have compared the excited states masses of experimental, Lattice-QCD and theoretical prediction for $\Sigma_c^{0}$ and are shown in Fig.2. Once again masses are increasing with increasing $\nu$ values.  Present masses with correction term are greater than without correction in case of $\Sigma_c^{0}, \Sigma_c^{+}$ and $\Sigma_c^{++}$ baryons [See Fig. 2].\\

We have calculated the mass spectra of $\Sigma_{c}^{0}$, $\Sigma_{c}^{+}$, $\Sigma_{c}^{++}$ baryon for 1S-5S, 1P-5P, 1D-2D and 1F states of $\Sigma_c^{0}$, $\Sigma_c^{+}$ and $\Sigma_c^{++}$ baryons. Only Ref. \cite{ebert2011} has calculated the mass of such states. Our results are in accordance with the ref. \cite{ebert2011} results. The Regge trajectory are shown in Fig. [8-10] for all these baryons for S ,P and D states. 1S and our predicted 1P states are matched with available experimental results in both planes with $J^{P}$ ($\frac{1}{2}^{+}$ and $\frac{3}{2}^{+}$) for 1S and ($\frac{3}{2}^{-}$ and $\frac{5}{2}^{-}$) for 1P. Figures show that all lines are almost linear, parallel and at equidistance. 
\begin{figure*}
\centering
\begin{minipage}[]{0.40\linewidth}
\includegraphics[scale=0.30]{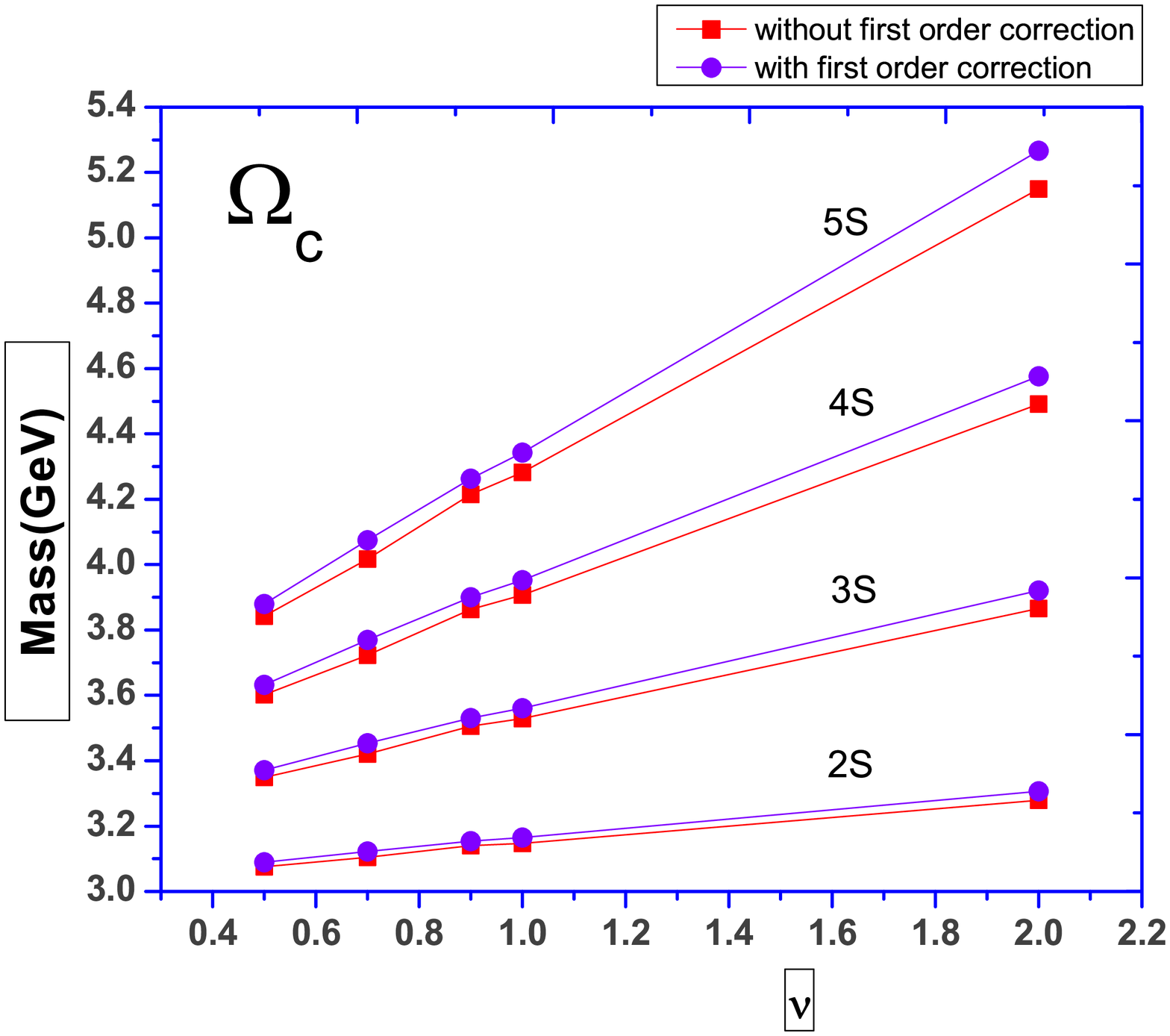}
\label{fig:minipage1}
\end{minipage}
\quad
\begin{minipage}[]{0.40\linewidth}
\includegraphics[scale=0.30]{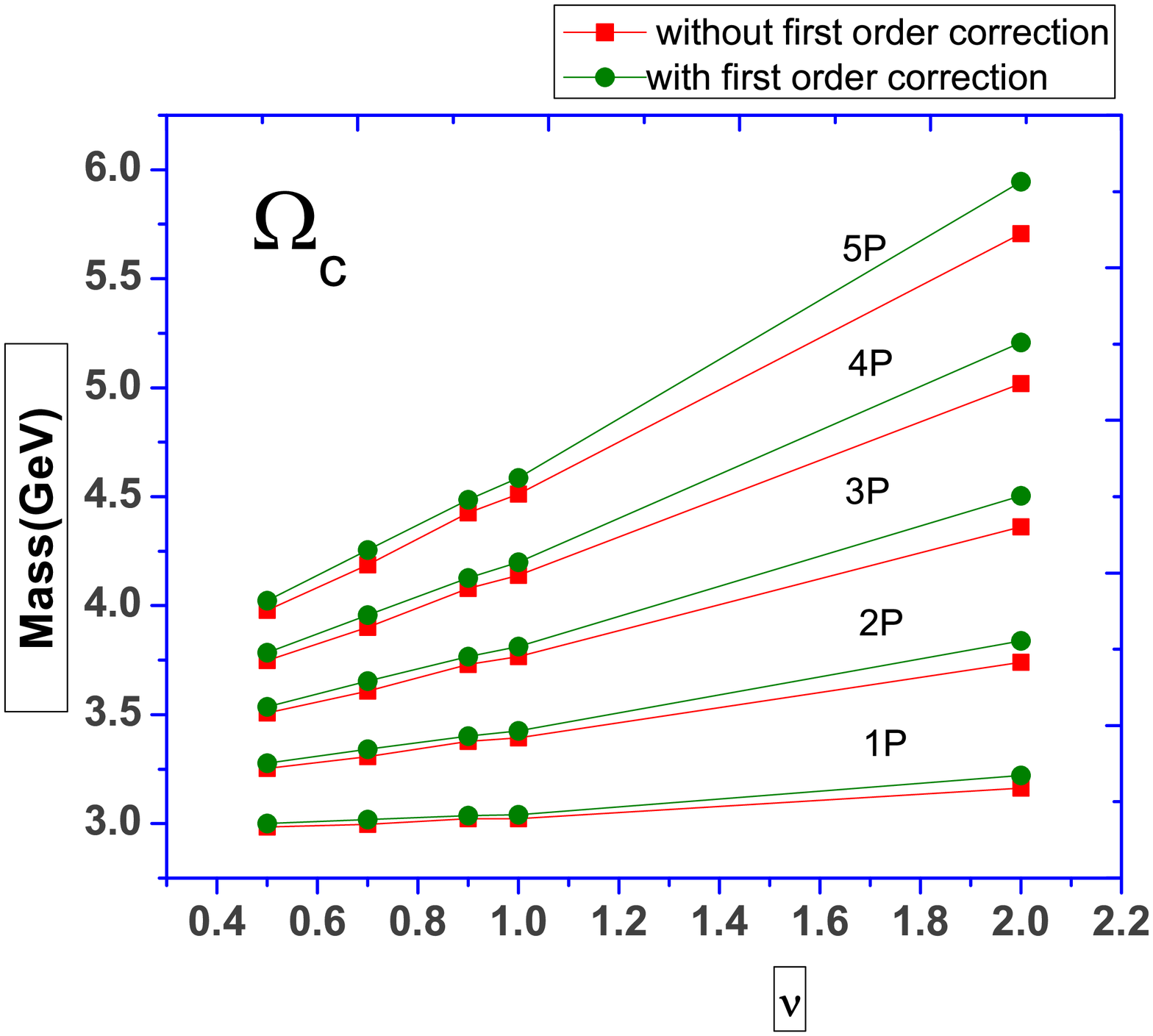}
\label{fig:minipage2}
\end{minipage}
\caption{\label{fig:epsart} Variation of mass with potential index in S and P states.}
\end{figure*}
\subsection{$\Xi_c$}
We discuss  the two lowest states $\Xi_{c}(2470)$ and $\Xi_{c}^{*}(2645)$ listed by PDG (2014). They are assigned as ground state with $J^{P}= \frac{1}{2}^{+}$ and $J^{P}= \frac{3}{2}^{+}$ respectively by Ref. \cite{olive,ebert2011,{chen2015},Roberts2008}. The present calculated excited states are in good agreement with other theoretical predictions (See Table 12-13). Various Lattice QCD predicted masses(in GeV) for $J^{P}= \frac{1}{2}^{+}$ and $J^{P}= \frac{3}{2}^{+}$ are 
\begin{center}
$m_{\Xi_{c}}({1}/{2}^{+})$ = 2.442(11)(20) \cite{4}\\
$m_{\Xi_{c}}({1}/{2}^{+})$ = 2.455(12)(11) \cite{pacs}\\
$m_{\Xi_{c}}({1}/{2}^{+})$ = 2.433(35)(30) \cite{brown}\\
$m_{\Xi_{c}}({1}/{2}^{+})$ = 2.469(28) \cite{alex}\\
{\rule{\linewidth}{0.15mm} }
$m_{\Xi_{c}}({3}/{2}^{+})$ =2.608(13)(22) \cite{4}\\
$m_{\Sigma_{c}}({3}/{2}^{+})$ = 2.674(26)(12) \cite{pacs}\\
$m_{\Xi_{c}}({3}/{2}^{+})$ = 2.648(70)(11) \cite{brown}\\
$m_{\Xi_{c}}({1}/{2}^{+})$ = 2.628(33) \cite{alex}\\
\end{center}

\par The orbital excited state masses are mentioned in Table [14-15]. The first orbital excited state $\Xi_{c}(2790)$ with $J^{P}= \frac{1}{2}^{-}$ and $\Xi_{c}^{*}(2815)$ with $J^{P}= \frac{3}{2}^{-}$ are well established experimentally \cite{olive} and also by other theoretical models (Ref. \cite{ebert2011}, \cite{chen2015}) and lattice results (Ref. \cite{4}) as 1P state. Moreover, Ref. \cite{92} has also calculated m($\Xi_{c}^{*}$)= 2.93 $\pm$ 0.11 GeV. All are in good agreement with our calculations of 1P state.\\

\begin{figure*}
\centering
\begin{minipage}[]{0.40\linewidth}
\includegraphics[scale=0.30]{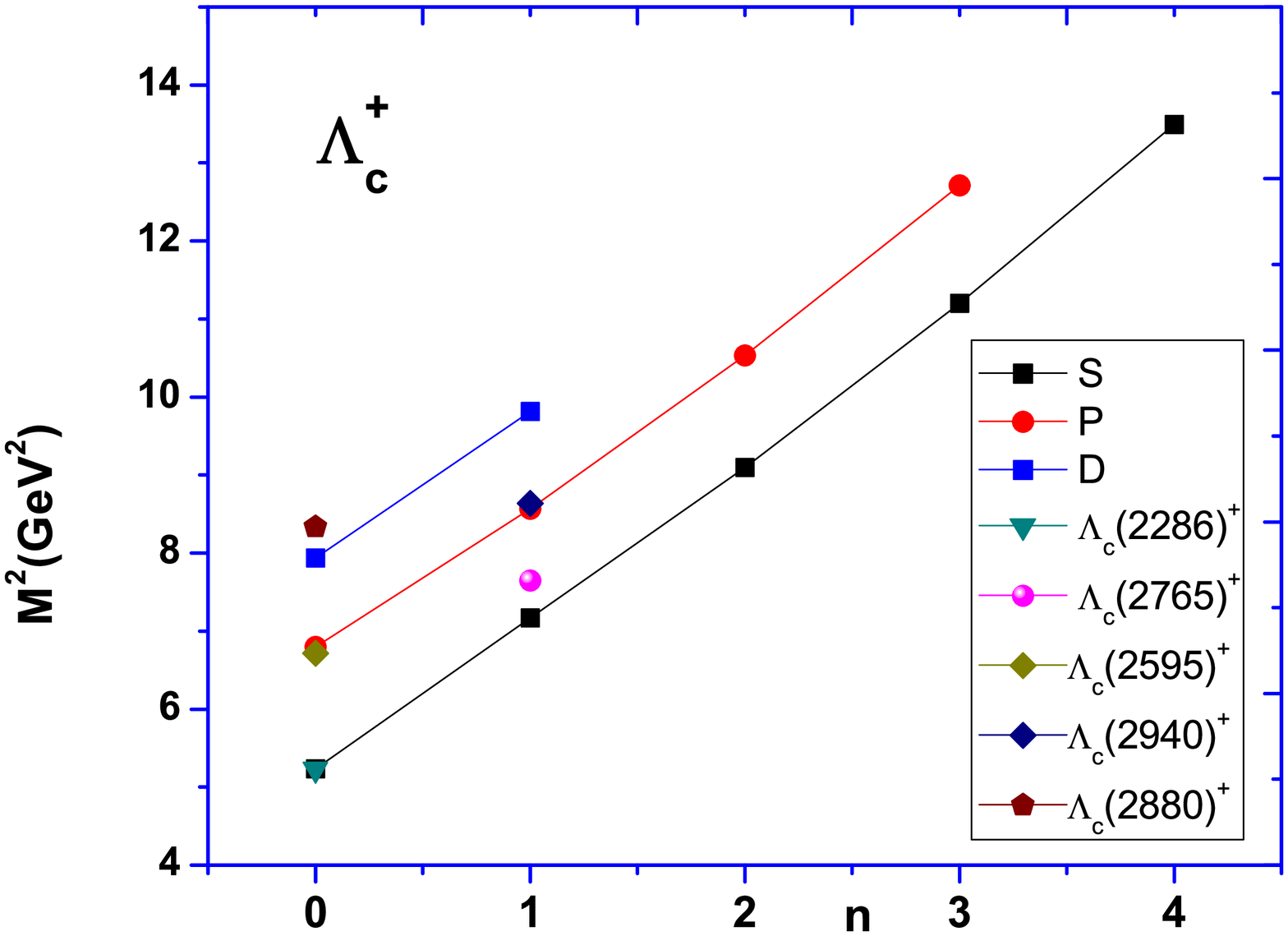}
\label{fig:minipage1}
\end{minipage}
\quad
\begin{minipage}[]{0.40\linewidth}
\includegraphics[scale=0.30]{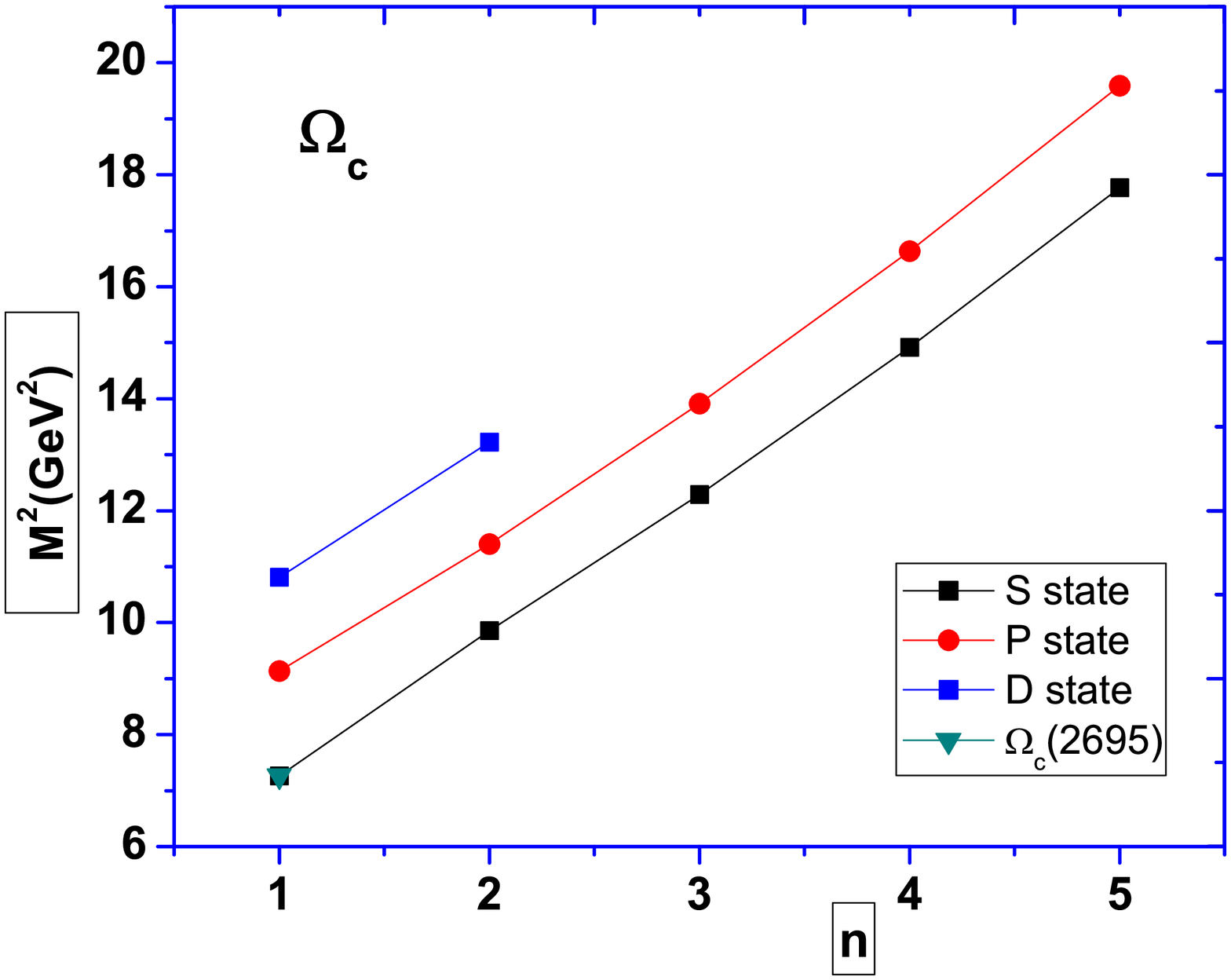}
\label{fig:minipage2}
\end{minipage}
\caption{\label{fig:epsart} Variation of mass for $\Lambda_{c}^{+}$(left) and $\Omega_{c}^{0}$ (right) with different states. The (M$^{2}\rightarrow$ n) Regge trajectories for $J^{P}$ values $\frac{1}{2}^{+}$, $\frac{1}{2}^{-}$ and $\frac{5}{2}^{+}$ are shown from bottom to top. Available Expt. data is also given with particle name.}
\end{figure*}

\begin{figure*}
\centering
\begin{minipage}[]{0.40\linewidth}
\includegraphics[scale=0.30]{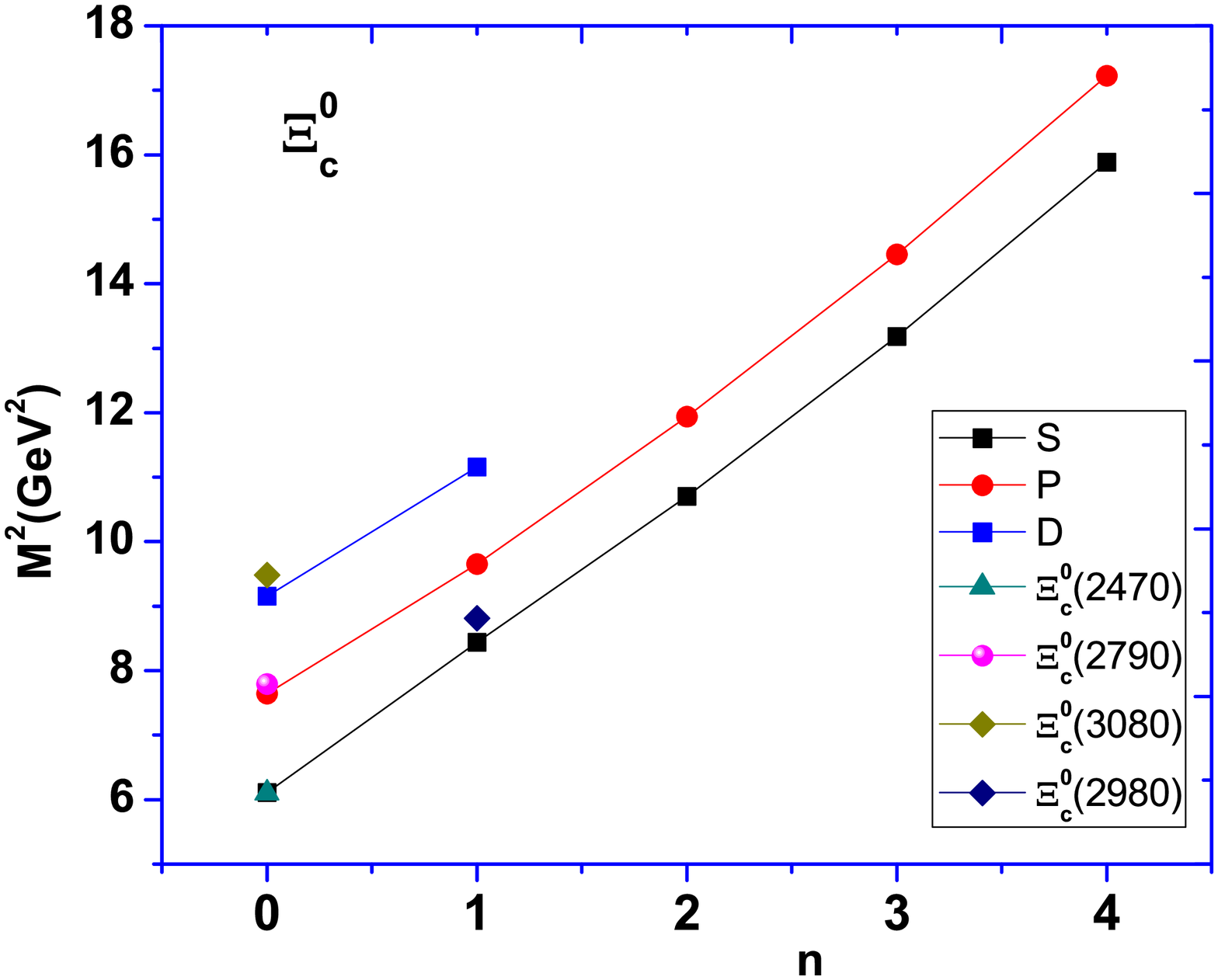}
\label{fig:minipage1}
\end{minipage}
\quad
\begin{minipage}[]{0.40\linewidth}
\includegraphics[scale=0.30]{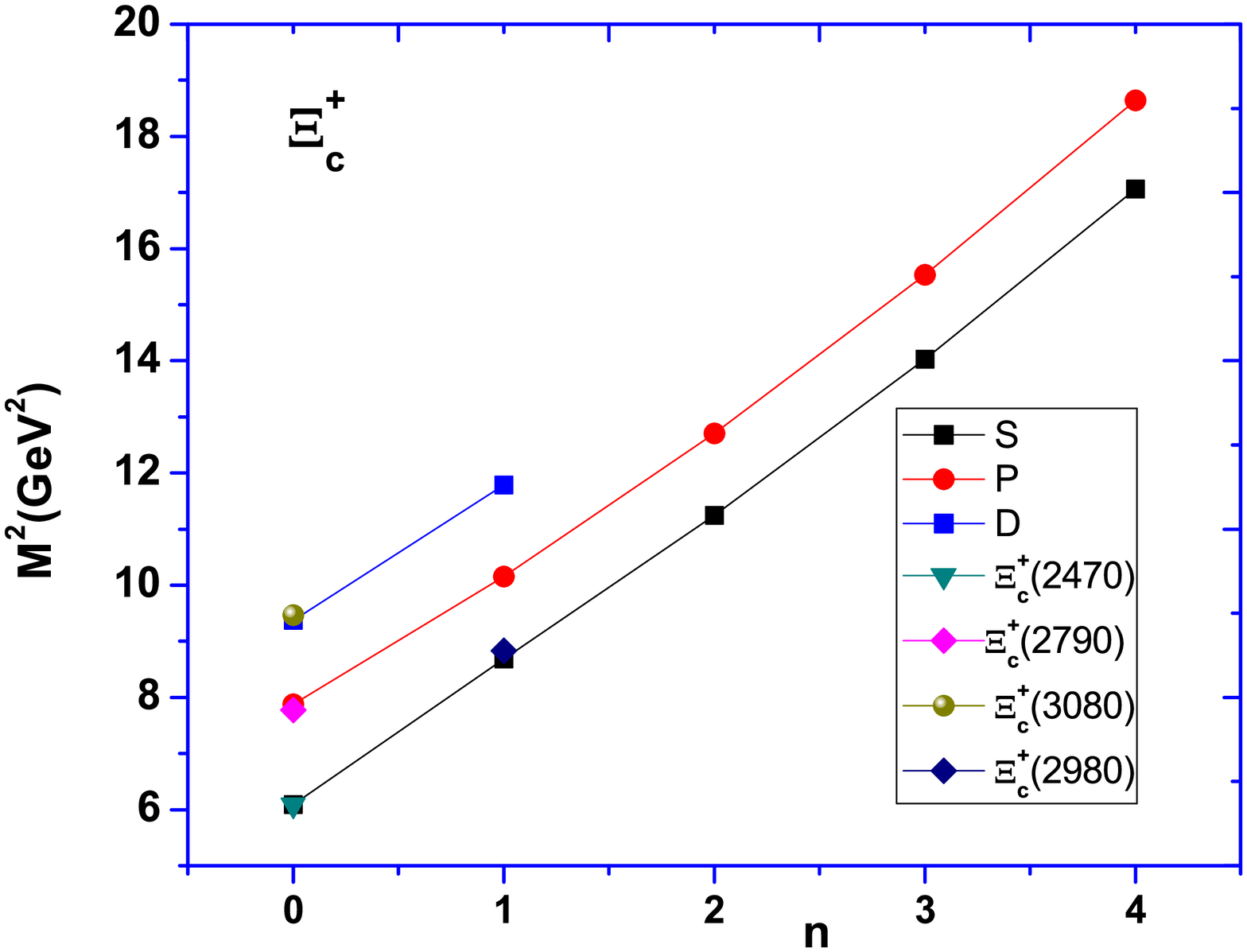}
\label{fig:minipage2}
\end{minipage}
\caption{\label{fig:epsart} Variation of mass for  $\Xi_{c}^{0}$(left) and $\Xi_{+}^{0}$ (right) with different states. The (M$^{2}\rightarrow$ n) Regge trajectories for $J^{P}$ values $\frac{1}{2}^{+}$, $\frac{1}{2}^{-}$ and $\frac{5}{2}^{+}$ are shown from bottom to top. Available Expt. data is also given with particle name.}
\end{figure*}

\begin{figure*}
\centering
\begin{minipage}[]{0.30\linewidth}
\includegraphics[scale=0.25]{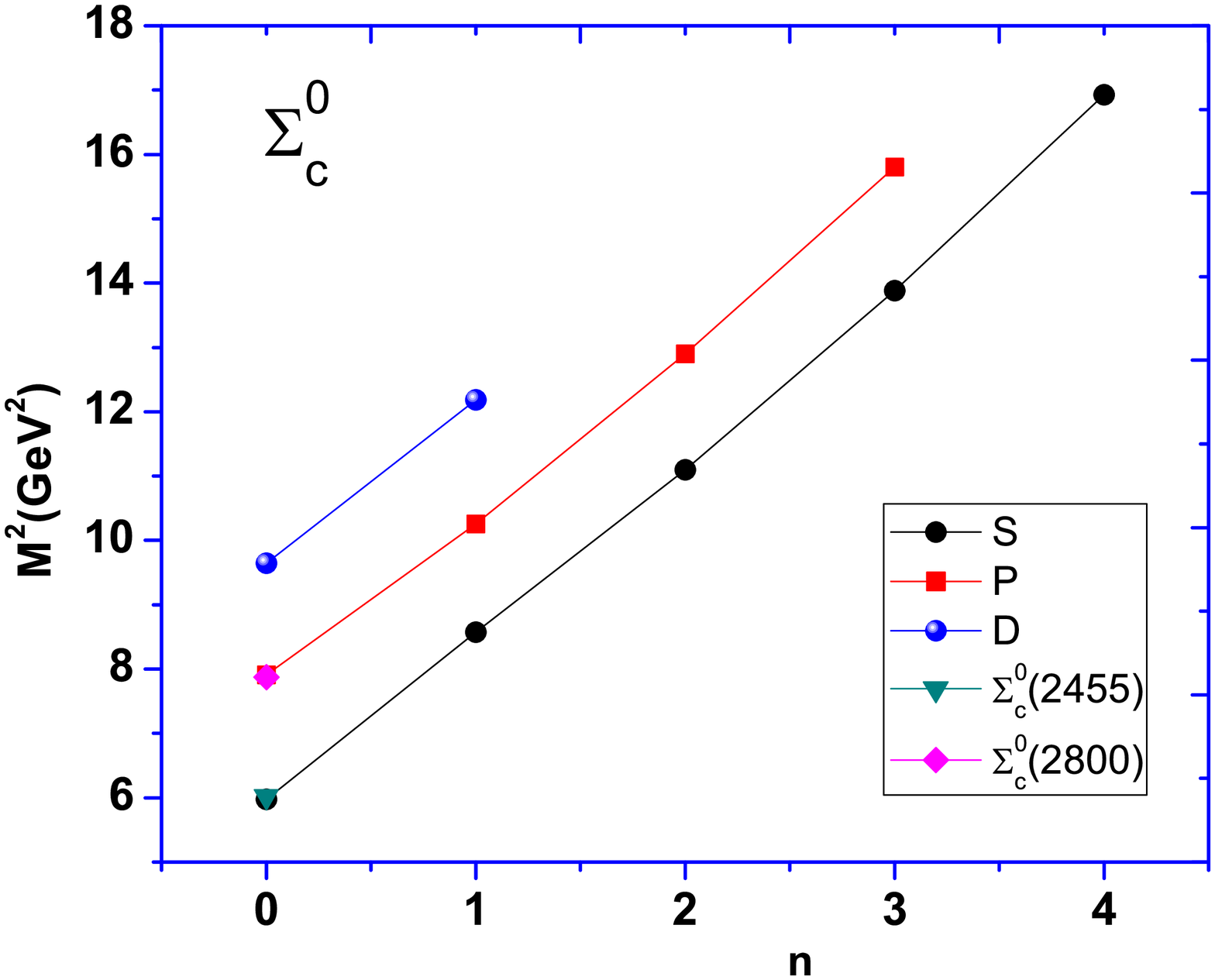}
\label{fig:minipage1}
\end{minipage}
\quad
\begin{minipage}[]{0.30\linewidth}
\includegraphics[scale=0.25]{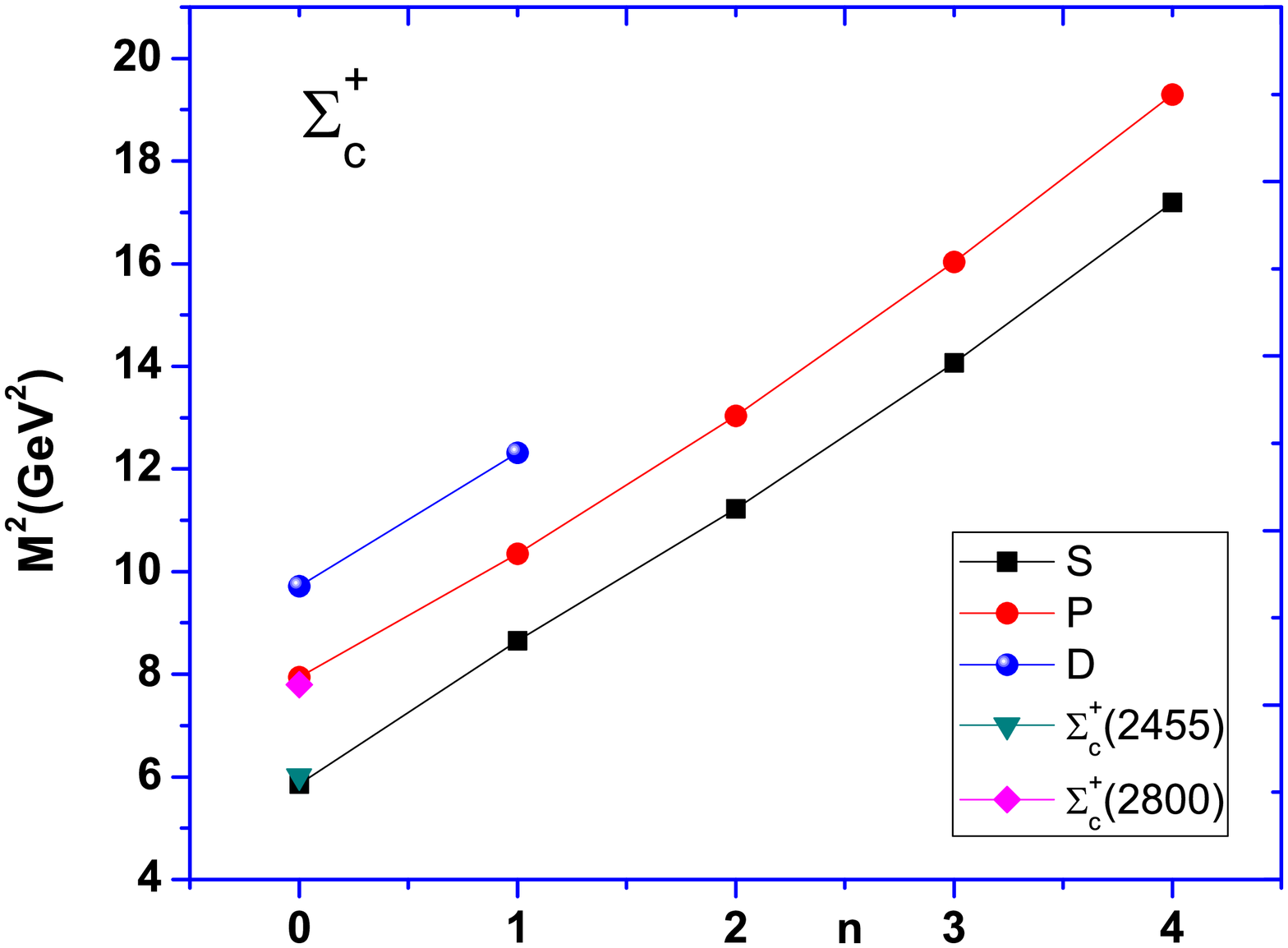}
\label{fig:minipage2}
\end{minipage}
\quad
\begin{minipage}[]{0.30\linewidth}
\includegraphics[scale=0.25]{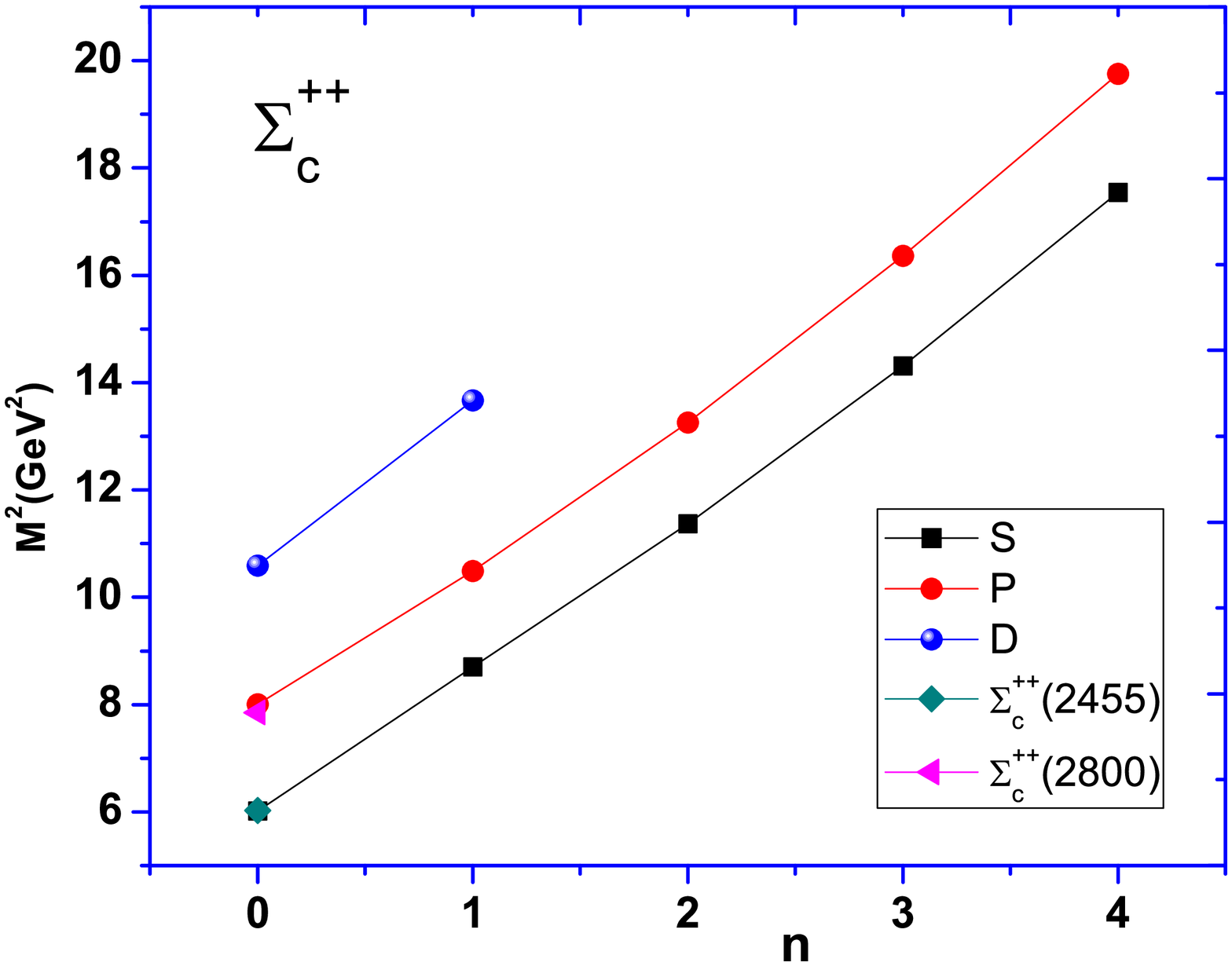}
\label{fig:minipage2}
\end{minipage}
\caption{\label{fig:epsart}  Variation of mass for $\Sigma_{c}^{0}$(first), $\Sigma_{c}^{+}$ (second) and $\Sigma_{c}^{++}$(third)with different states. The (M$^{2}\rightarrow$ n) Regge trajectories for $J^{P}$ values $\frac{1}{2}^{+}$, $\frac{1}{2}^{-}$ and $\frac{5}{2}^{+}$ are shown from bottom to top. Available Expt. data is also given with particle name..}
\end{figure*}
\begin{figure*}
\centering
\begin{minipage}[]{0.30\linewidth}
\includegraphics[scale=0.25]{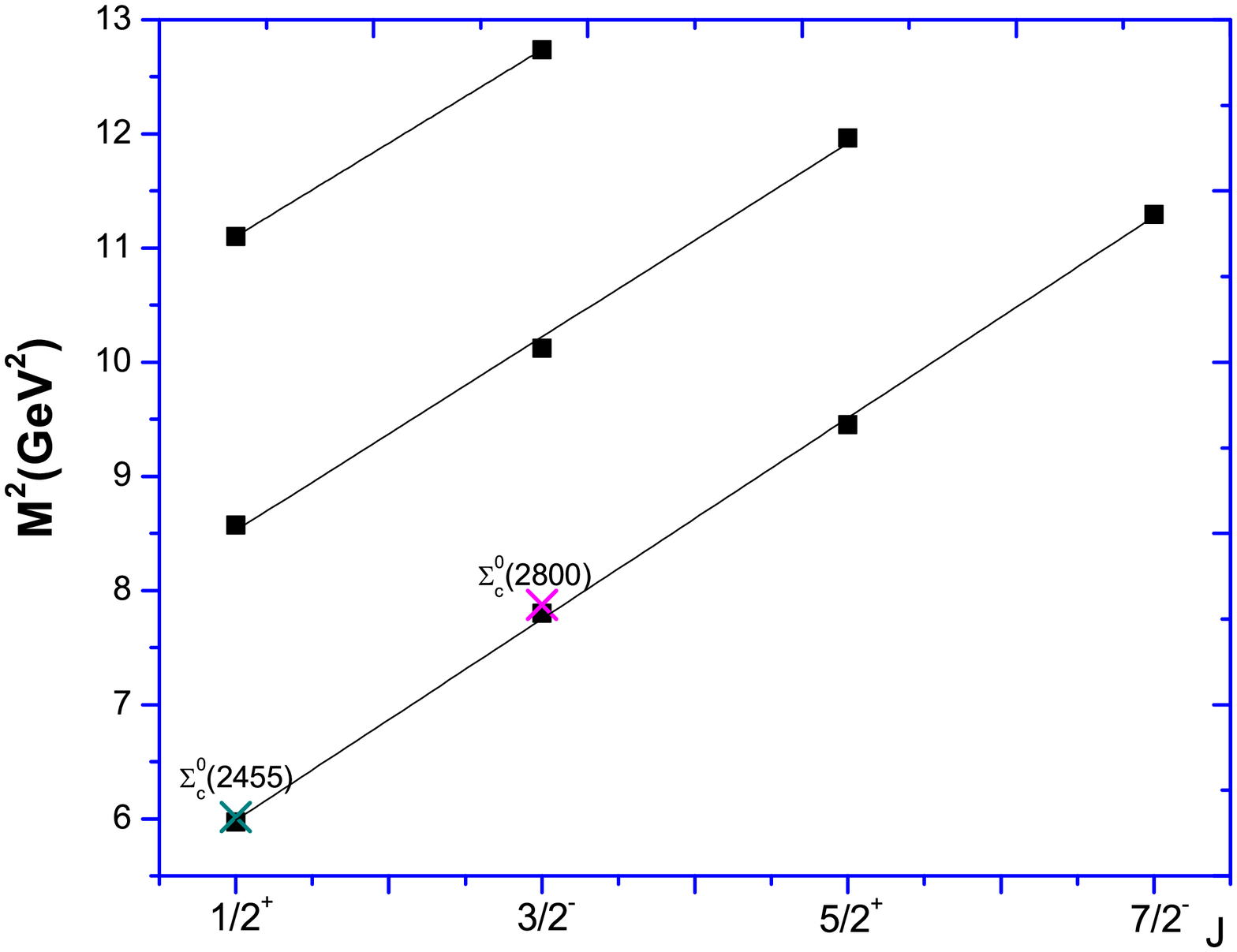}
\end{minipage}
\quad
\begin{minipage}[]{0.30\linewidth}
\includegraphics[scale=0.25]{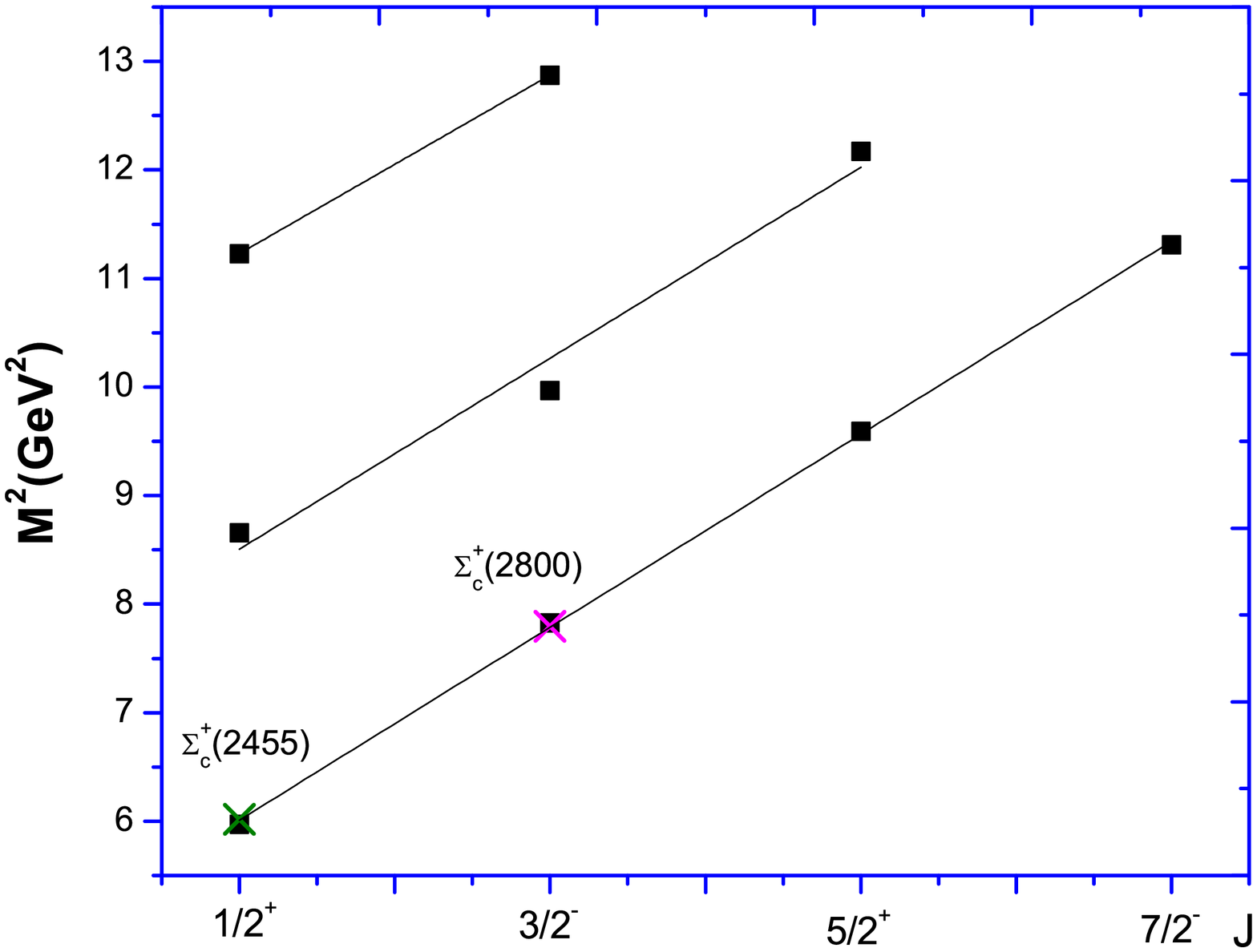}
\end{minipage}
\quad
\begin{minipage}[]{0.30\linewidth}
\includegraphics[scale=0.25]{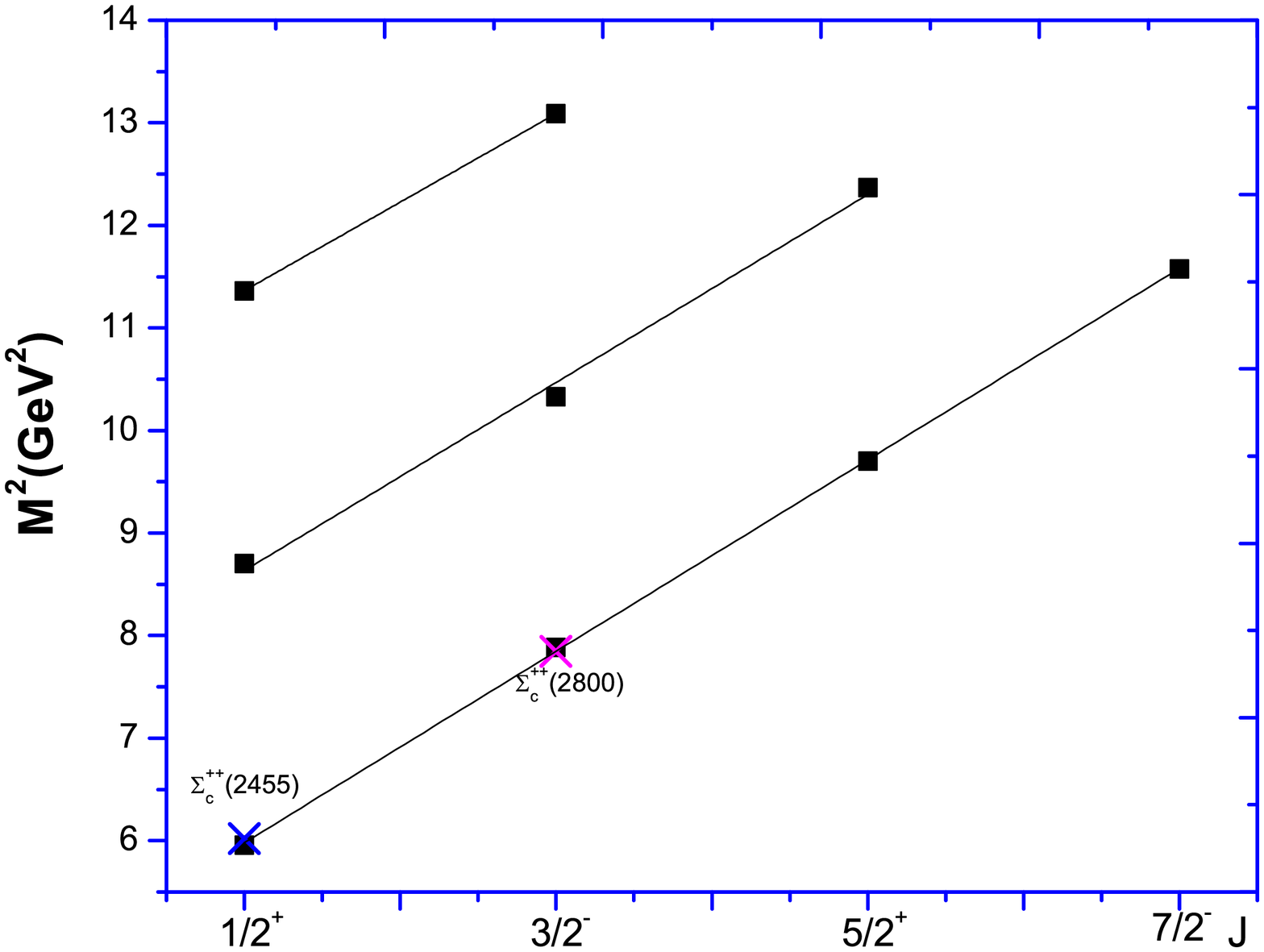}
\end{minipage}
\caption{\label{fig:epsart} Parent and daughter(J, $M^{2}$) Regge trajectories for $\Sigma_{c}^{0}$(first), $\Sigma_{c}^{+}$ (second) and $\Sigma_{c}^{++}$(third) baryons with natural parities. Available experimental data are also given with particle names.}
\end{figure*}

\begin{figure*}
\centering
\begin{minipage}[]{0.30\linewidth}
\includegraphics[scale=0.25]{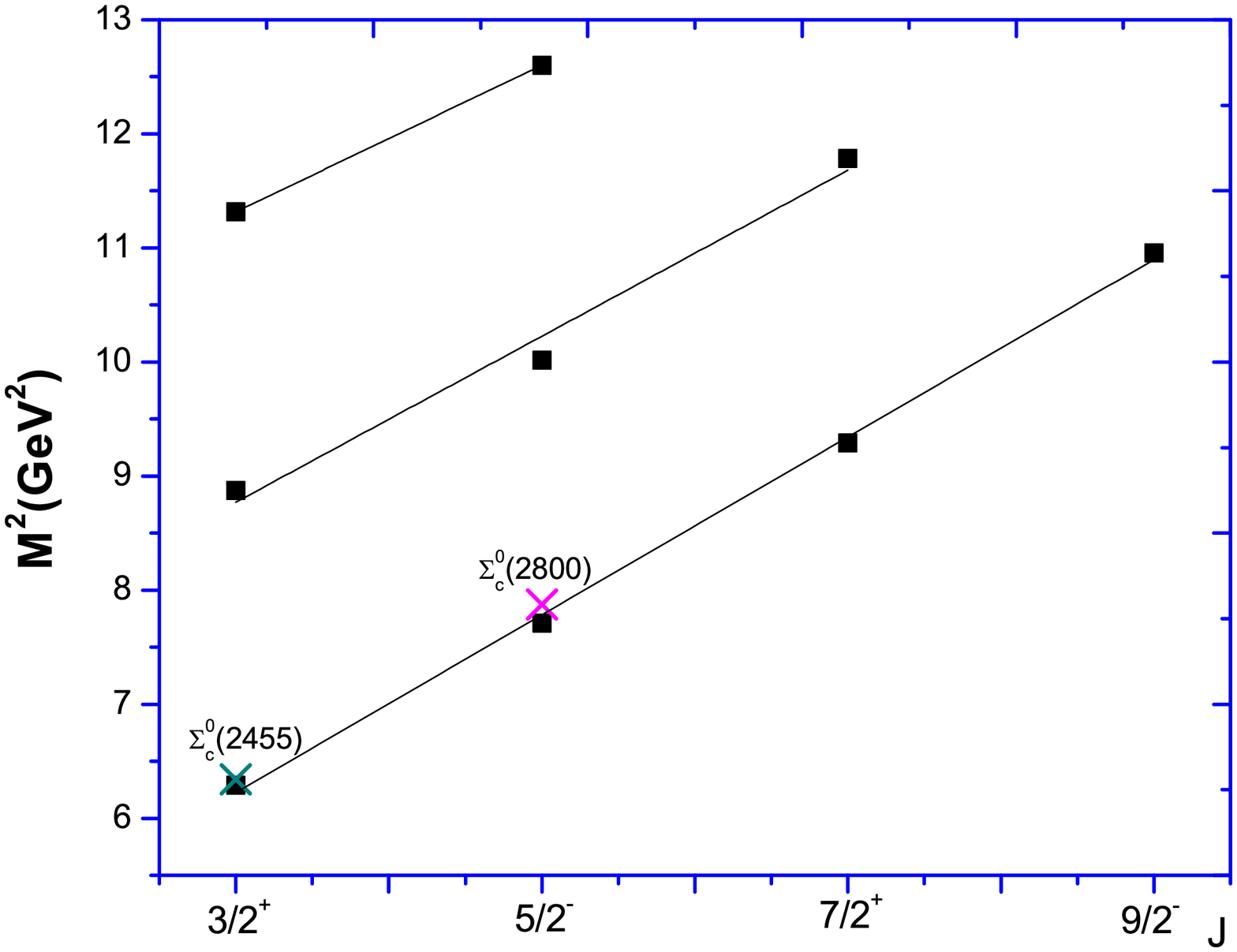}
\end{minipage}
\quad
\begin{minipage}[]{0.30\linewidth}
\includegraphics[scale=0.25]{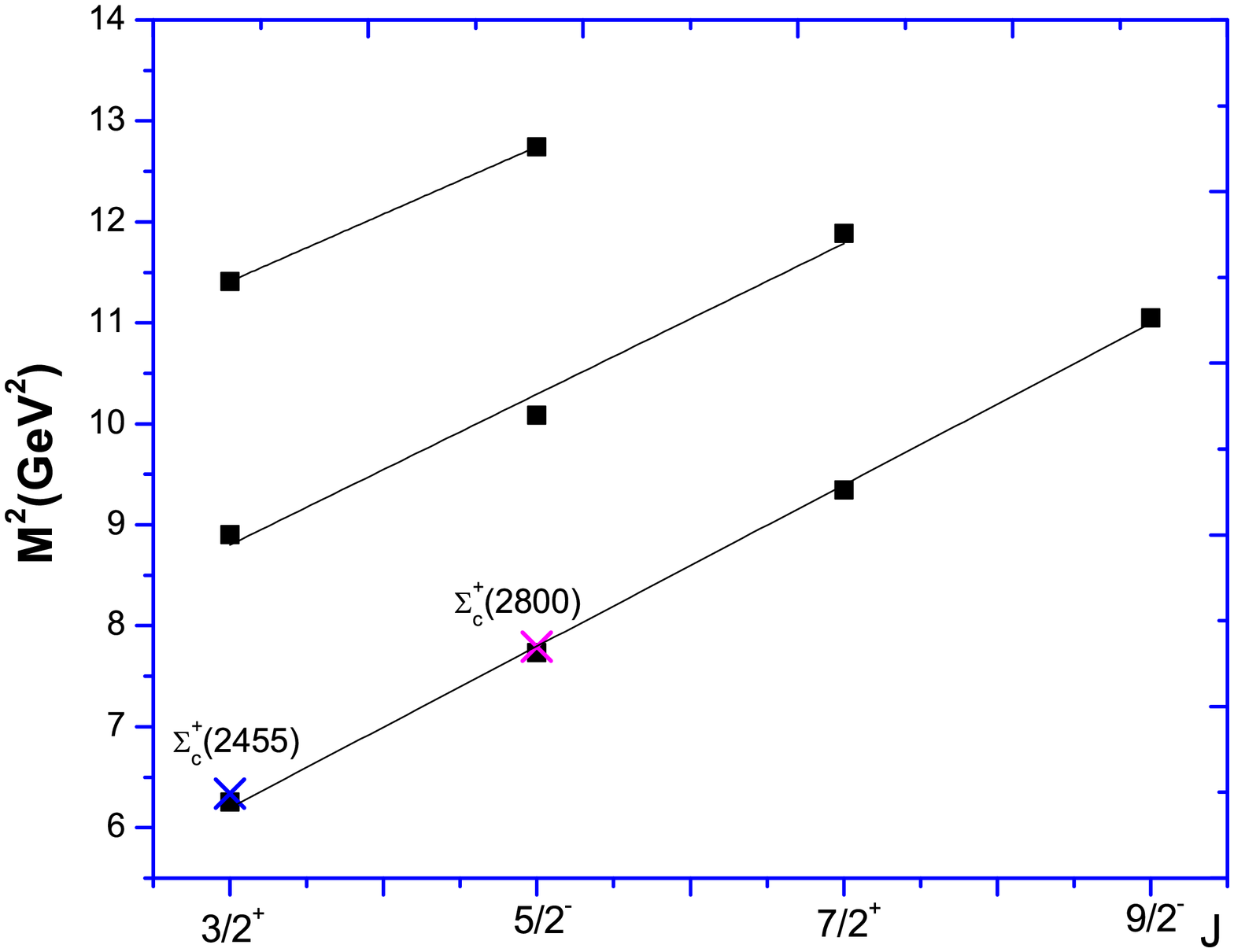}
\end{minipage}
\quad
\begin{minipage}[]{0.30\linewidth}
\includegraphics[scale=0.25]{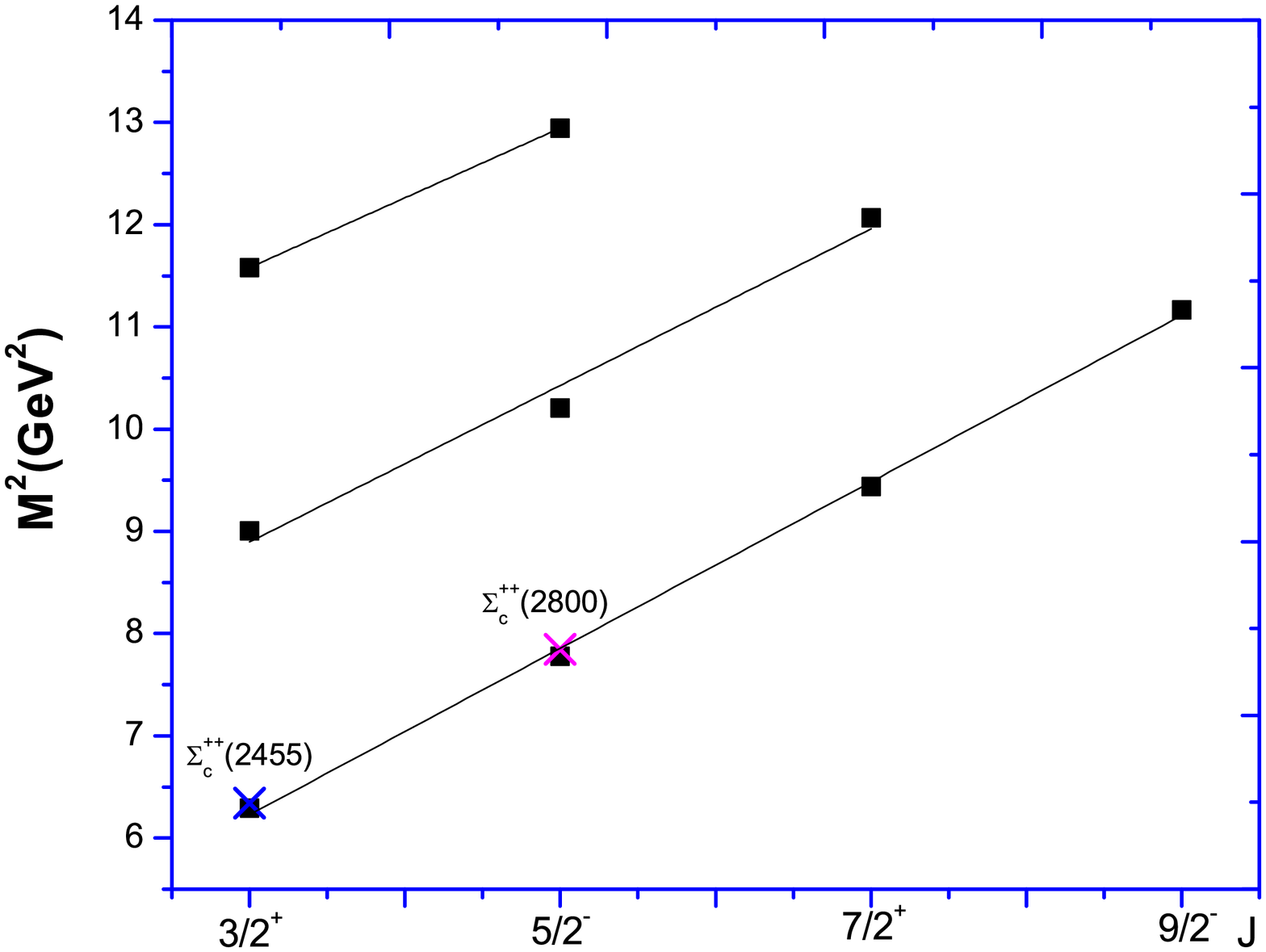}
\end{minipage}
\caption{\label{fig:epsart} Parent and daughter(J, $M^{2}$) Regge trajectories for $\Sigma_{c}^{0}$(first), $\Sigma_{c}^{+}$ (second) and $\Sigma_{c}^{++}$(third) baryons with unnatural parities. Available experimental data are also given with particle names..}
\end{figure*}

\begin{figure*}
\centering
\begin{minipage}[]{0.40\linewidth}
\includegraphics[scale=0.30]{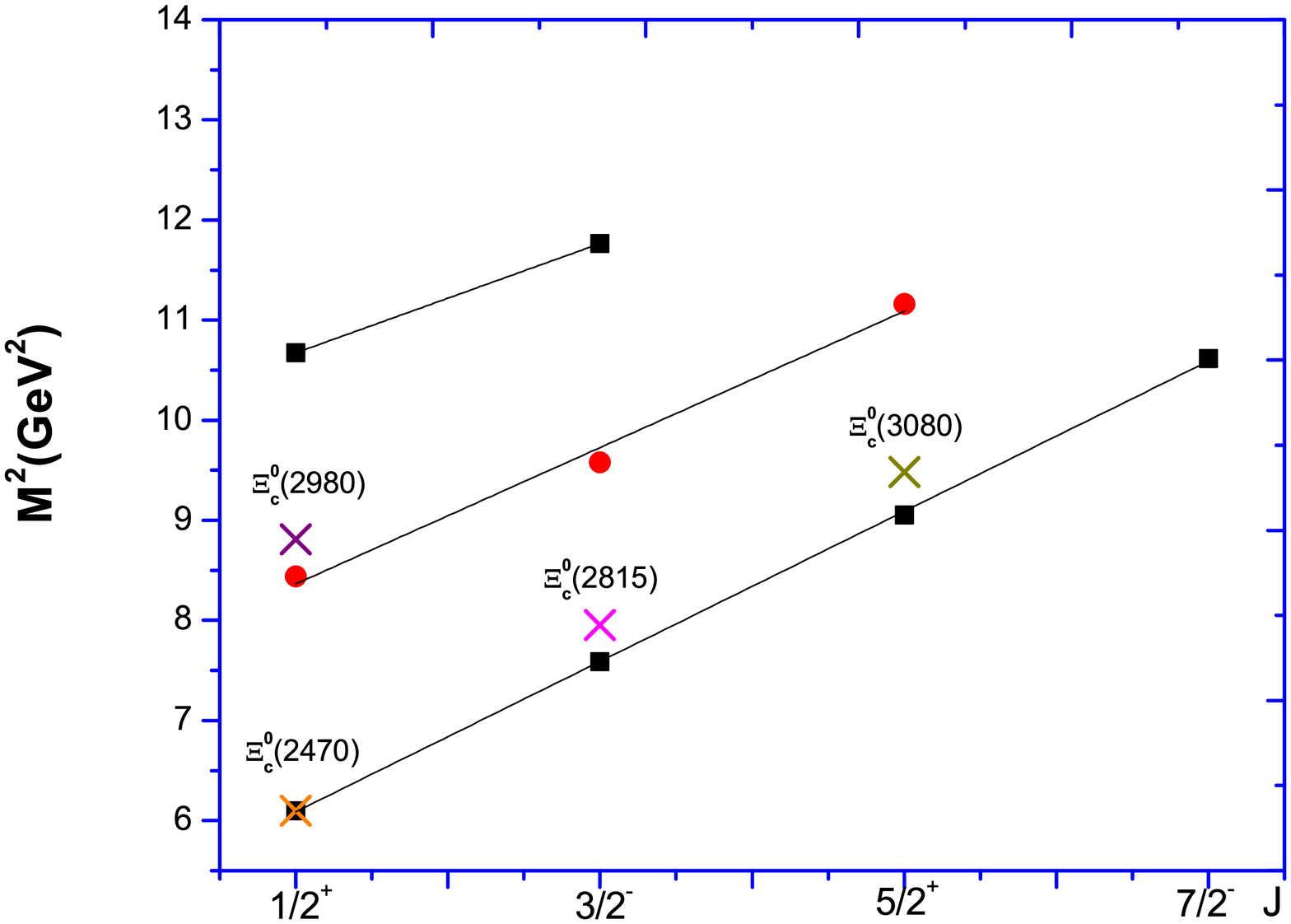}
\label{fig:minipage1}
\end{minipage}
\quad
\begin{minipage}[]{0.40\linewidth}
\includegraphics[scale=0.30]{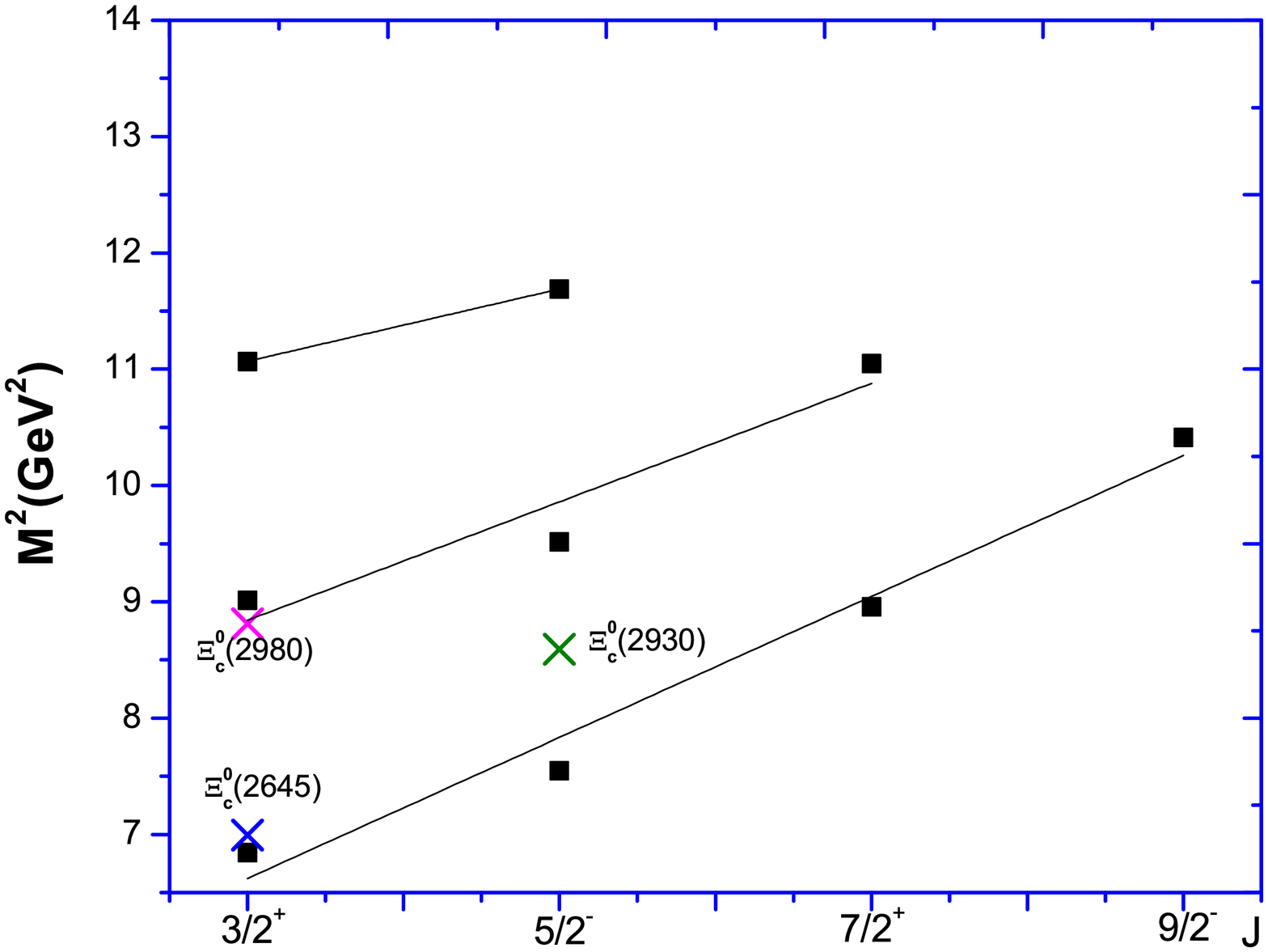}
\label{fig:minipage2}
\end{minipage}
\caption{\label{fig:epsart} Parent and daughter(J, $M^{2}$) Regge trajectories for $\Xi_c^0$ baryons with  natural(first) and unnatural(second) parities. Available experimental data are also given with particle names.}
\end{figure*}

\begin{figure*}
\centering
\begin{minipage}[]{0.40\linewidth}
\includegraphics[scale=0.30]{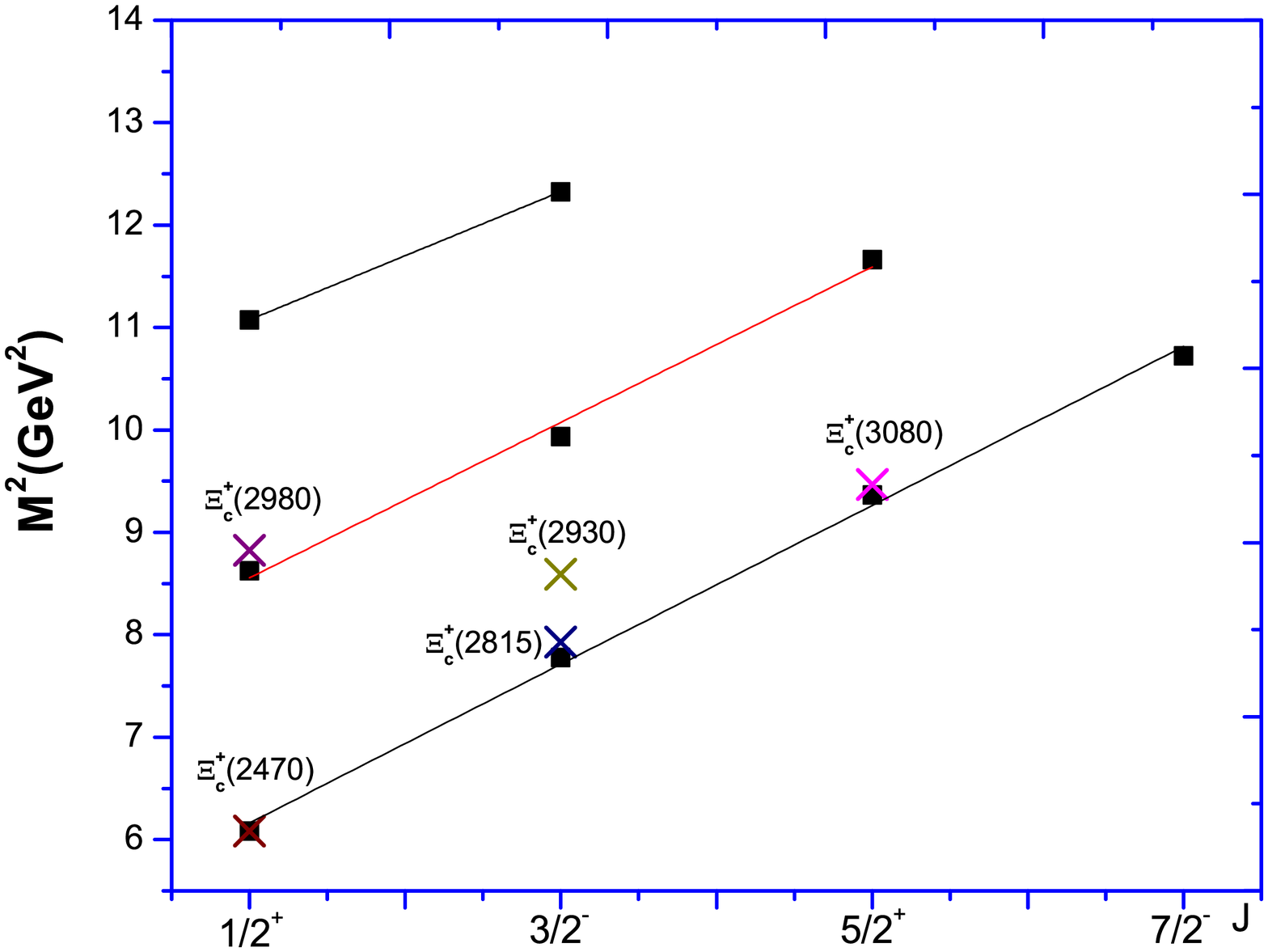}
\label{fig:minipage1}
\end{minipage}
\quad
\begin{minipage}[]{0.40\linewidth}
\includegraphics[scale=0.30]{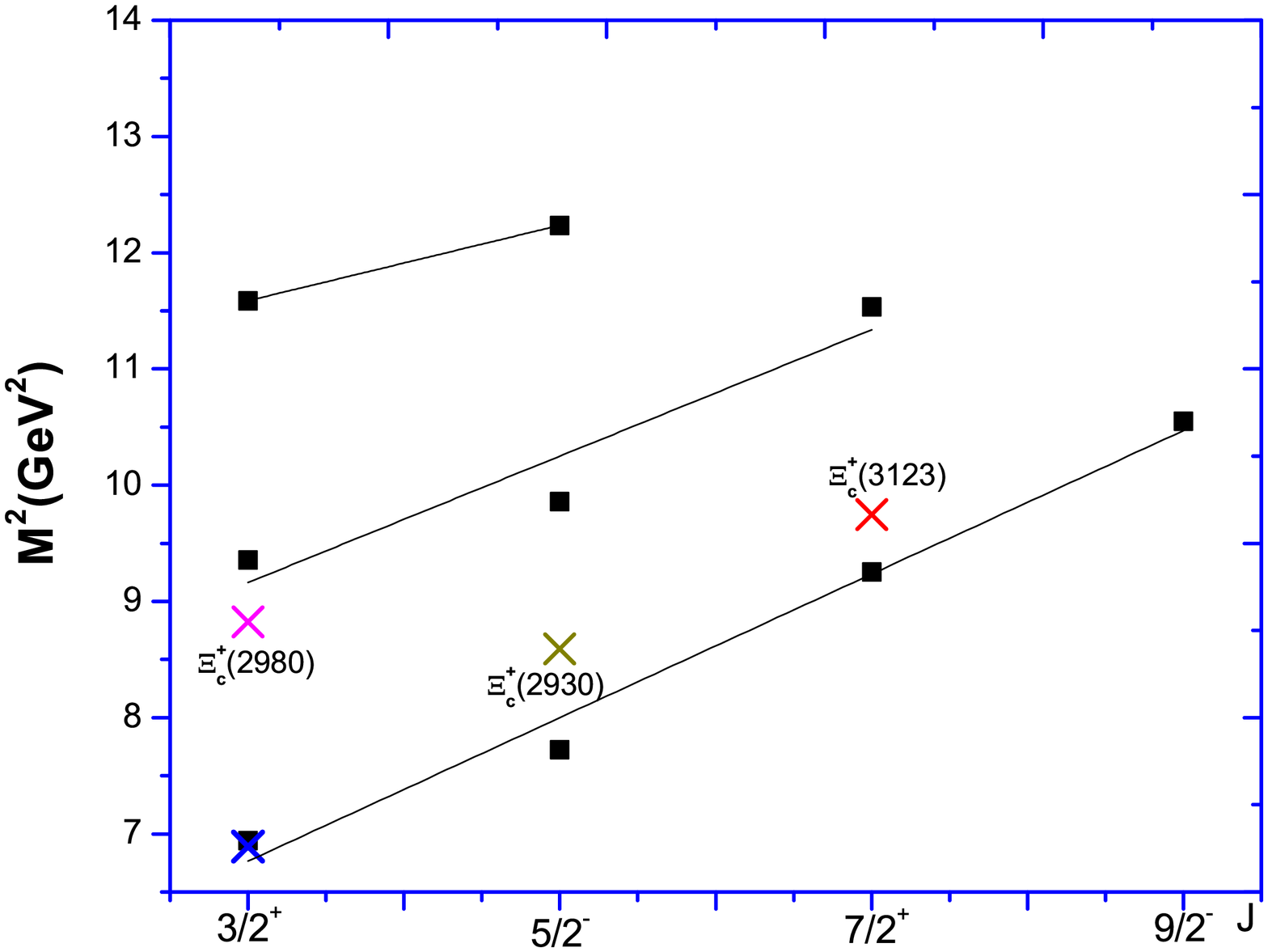}
\label{fig:minipage2}
\end{minipage}
\caption{\label{fig:epsart} Parent and daughter(J, $M^{2}$) Regge trajectories for $\Xi_c^+$ baryons with  natural(first) and unnatural(second) parities. Available experimental data are also given with particle names.}
\end{figure*}

\begin{figure*}
\centering
\begin{minipage}[]{0.40\linewidth}
\includegraphics[scale=0.30]{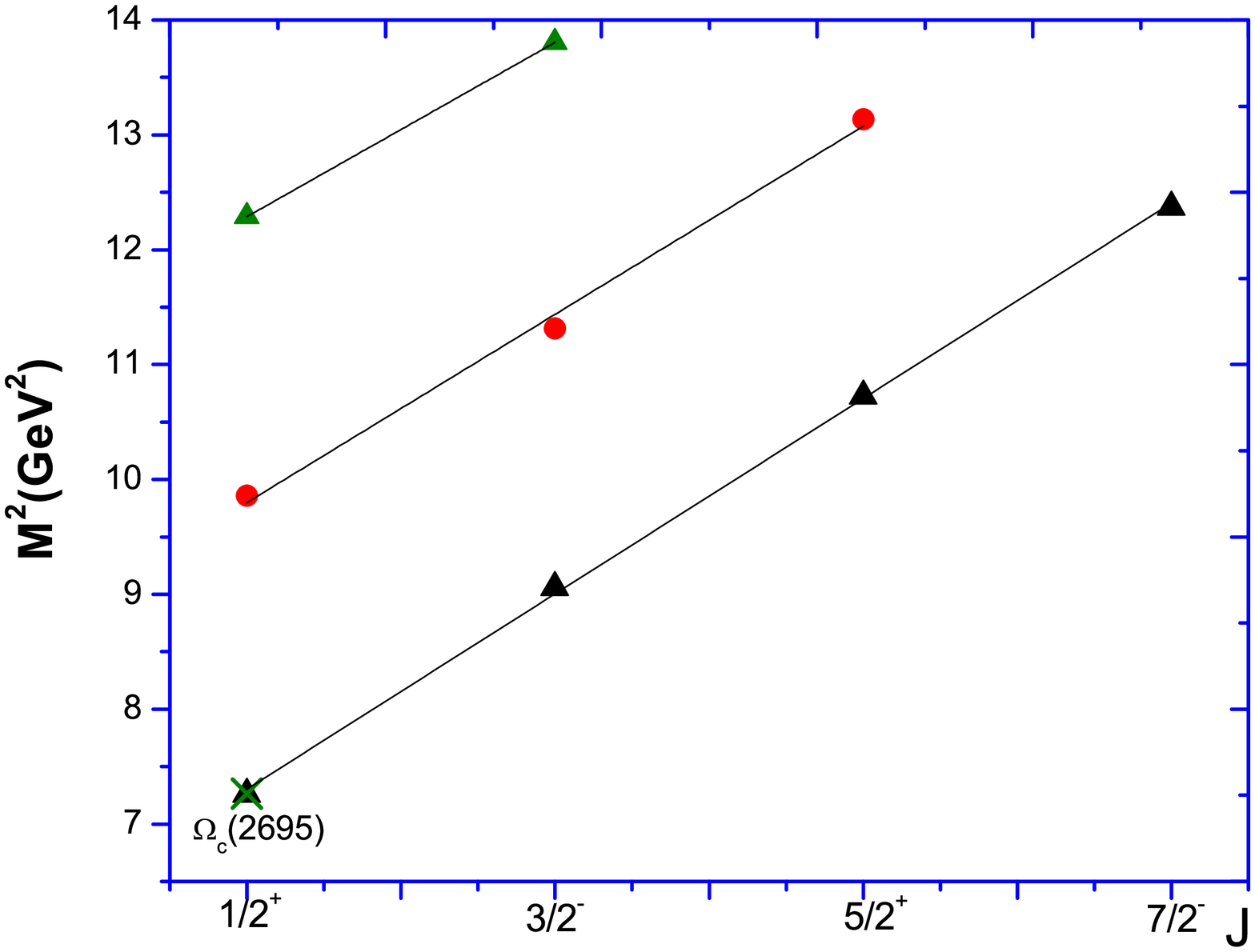}
\label{fig:minipage1}
\end{minipage}
\quad
\begin{minipage}[]{0.40\linewidth}
\includegraphics[scale=0.30]{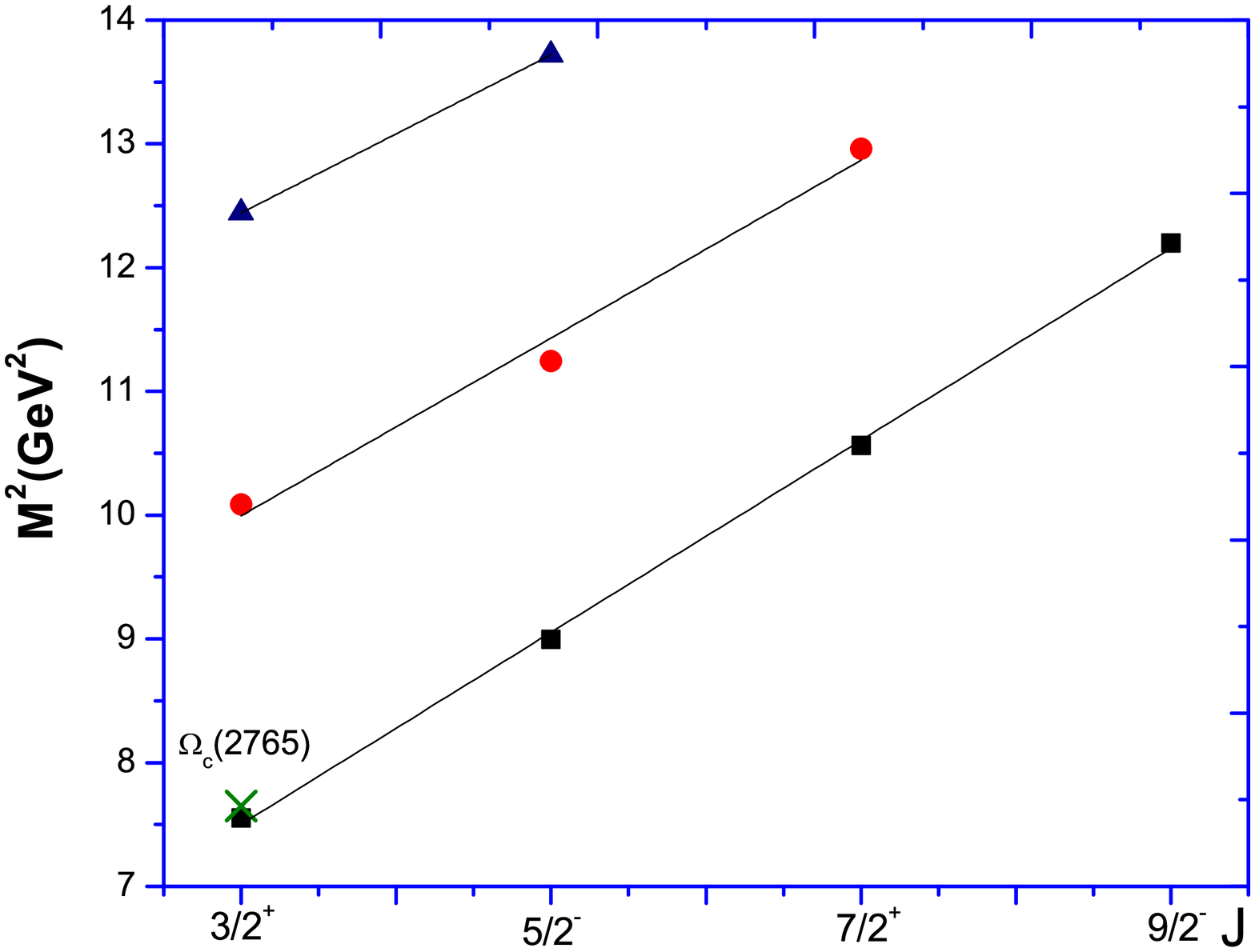}
\label{fig:minipage2}
\end{minipage}
\caption{\label{fig:epsart} Parent and daughter(J, $M^{2}$) Regge trajectories for for $\Omega_c^0$ baryon with natural(first) and unnatural(second) parities. Available experimental data are also given with particle names.}
\end{figure*}

\par $\Xi_{c}(2980)^{+}$ with m = 2971.4 $\pm$ 3.3 MeV and $\Xi_{c}(2980)^{0}$ with m = 2968.0 $\pm$ 2.6 MeV are found by Belle Collaboration \cite{belle}. Although $\Xi_{c}(2980)^{+,0}$ are well established experimentally, its quantum numbers are still unknown. According to our calculations, $\Xi_{c}(2980)$ can be assigned to the 2S state with $J^{P}= \frac{1}{2}^{+}$ or $J^{P}= \frac{3}{2}^{+}$. Similarly, D. Ebert et. al. \cite{ebert2011} also predicted $\Xi_{c}(2980)$ as 2S state.\\

\par $\Xi_{c}(2930)$ is only observed by BABAR \cite{3} and not confirmed yet by any other experimental group. It can be assigned to either 1P or 2S state of $\Xi_{c}(2980)^{0}$ \cite{ebert2011,99}. Our analysis suggests it be the 1P state ($J^{P}= \frac{1}{2}^{-}$).\\

\par The recent experimental paper from Belle \cite{Kato}, reported the developments of $\Xi_c(3055)^{0}$ (m= 3059.0 $\pm$0.5 $\pm$0.6), $\Xi_c(3055)^{+}$ (m= 3055.8 $\pm$0.4 $\pm$0.2) and $\Xi_c(3080)^{+}$ (m= 3079.6 $\pm$0.4 $\pm$0.1) states. We have compared these states with 1D state and they are in good agreement with $\nu$=1.0.\\
 
We can conclude that, the present results are in good agreement with  other theoretical predictions as well as experimental results with few MeV difference [See Table 12- 15]. Fig. [3-4] shows the graph (M $\rightarrow$ $\nu$) for $\Xi_c^{0}$ and $\Xi_c^{+}$. The Regge trajectory for $\Xi_c^{0}$ and $\Xi_c^{+}$ are shown in Fig. (7,11 and 12). In $\Xi_c^{0}$ baryon, 1S ($\frac{1}{2}^{+}$, $\frac{3}{2}^{+}$), 1P($\frac{1}{2}^{-}$), predicted 2S ($\frac{3}{2}^{+}$) states are matched with experimental masses whereas 1D state shows difference($\approx$ 50 MeV). In $\Xi_c^{+}$ baryon, 1S ($\frac{1}{2}^{+}$, $\frac{3}{2}^{+}$), 1P($\frac{1}{2}^{-}$, $\frac{1}{2}^{-}$), predicted 2S ($\frac{1}{2}^{+}$) and 1D state ($\frac{1}{2}^{+}$) are very close with experimental masses. We found $\Xi_c(2930)$ is close to $\Xi_{c}^{+}$ with $J^{P}$=$\frac{5}{2}^{-}$ than $\Xi_{c}^{0,+}$  with $J^{P}$=$\frac{3}{2}^{-}$.
\subsection{$\Omega_c^{0}$}

$\Omega_{c}^0$ is made up of two strange and one charm quark ($\textit{ssc}$). The WA62 collaboration \cite{WA62} reported the $\Omega_{c}^0$ ($J^P= \frac{1}{2}^{+}$) and BaBar reported $\Omega_{c}^0$ ($J^P= \frac{3}{2}^{+}$) primarily. We have calculated mass spectra for radial (1S-5S)[Table 16] and orbital excited states (1P-5P, 1D-2D and 1F) [Table 17]. As we do not have any experimental data for excited states (both radial and orbital), we have compared our predictions with D. Ebert et.al., based on quark-diquark model.	The comparisons with other predictions are also mentioned in Table 16-17. Moreover, Ref. \cite{4} has reported m($\Omega_{c}$)= 3.016(32)(37) GeV (for $J^{P}=\frac{1}{2}^-$) and Ref. \cite{92} has reported m($\Omega_{c}^{*}$)= 3.08 $\pm$ 0.12 GeV (for $J^{P}=\frac{3}{2}^-$) in their papers which are also close to us. Our results are close at the potential index 0.9 for S states as well as for P-states. Various Lattice QCD results obtained for $1^{2}S_{\frac{1}{2}}$ and $1^{4}S_{\frac{3}{2}}$ are listed below.
\begin{center}
$m_{\Omega_{c}}({1}/{2}^{+})$ = 2.648(09)(19) \cite{4}\\
$m_{\Omega_{c}}({1}/{2}^{+})$ = 2.673(5)(12) \cite{pacs}\\
$m_{\Omega_{c}}({1}/{2}^{+})$ = 2.679(37)(20) \cite{brown}\\
$m_{\Omega_{c}}({1}/{2}^{+})$ = 2.629(22) \cite{alex}\\
$m_{\Omega_{c}}({1}/{2}^{+})$ = 2.783(13) \cite{can}\\
{\rule{\linewidth}{0.15mm} }
$m_{\Omega_{c}}({3}/{2}^{+})$ =2.709(11)(21) \cite{4}\\
$m_{\Omega_{c}}({3}/{2}^{+})$ = 2.738(5)(12) \cite{brown}\\
$m_{\Omega_{c}}({3}/{2}^{+})$ = 2.755(37)(24) \cite{pacs}\\
$m_{\Omega_{c}}({3}/{2}^{+})$ = 2.709(26) \cite{alex}\\
$m_{\Omega_{c}}({3}/{2}^{+})$ = 2.837(18) \cite{can}\\
\end{center}
\begin{figure}
\includegraphics[scale=0.30]{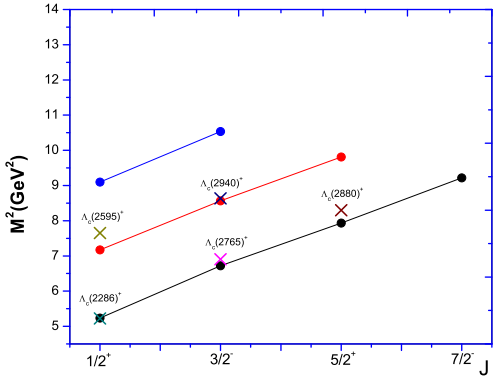}
\caption{Parent and daughter(J, $M^{2}$) Regge trajectories for $\Lambda_c^+$ baryon with natural parity. Available experimental data are also given with particle names.  .}
\end{figure}
\par Mass variation with potential index is shown in Fig. 5 for S and P states. These results also follow the same trend as followed by charm $\Lambda$, $\Sigma$ and $\Xi$ baryons. D and F state results are also compared with ref. \cite{ebert2011} and they are in close agreement at potential index $\nu$=0.9. The Regge, therefore shown in Fig. (6,13) in ($n_r, M^2$) and ($J, M^2$)  plane are almost linear parallel and at equidistance.

\begin{table}[h]
\caption{Our predicted baryonic states are compared with experimental unknown($J^P$) excited states.}
\label{tab:table1} 
\begin{tabular}{ccc}
\hline
Names & M(GeV)\cite{olive}& Baryon State \\
\hline
$\Lambda_{c}^{+}(2765)$ &2766.6$\pm$2.4&$(2^2S_\frac{1}{2})$\\
$\Lambda_{c}^{+}(2940)$&2939.3$^{+1.4}_{-1.5}$& $(2^2P_\frac{1}{2})$,$(2^2P_\frac{3}{2})$ \\
~~~$\Sigma_{c}^{++}(2800)$ & $2801^{+4}_{-6}$&$(1^2P_\frac{3}{2})$,$(1^4P_\frac{5}{2})$\\
~$\Sigma_{c}^{+}(2800)$ & $2792^{+14}_{-5}$&$(1^2P_\frac{3}{2})$,$(1^4P_\frac{5}{2})$\\
$\Sigma_{c}^{0}(2800)$ & $2806 \pm8\pm10$&$(1^2P_\frac{3}{2})$,$(1^4P_\frac{5}{2})$\\
$\Xi_{c}(2930)$ & 2931$\pm$3$\pm$5 &$(1^2P_\frac{3}{2})$,$(1^2P_\frac{5}{2})$\\
$\Xi_{c}^{+}(2980)$ & 2971.4$\pm$3.3 &$(2^2S_\frac{1}{2})$,$(2^4S_\frac{3}{2})$ \\
$\Xi_{c}^{0}(2980)$ & 2968.0$\pm$2.6&$(2^2S_\frac{1}{2})$,$(2^4S_\frac{3}{2})$\\
$\Xi_{c}^{+}(3055)$ & 3054.2$\pm$1.2$\pm$0.5 &$(1^2D_\frac{3}{2})$, $(1^4D_\frac{3}{2})$\\
$\Xi_{c}^{+}(3080)$ & 3077.0$\pm$0.4 & $(1^2D_\frac{5}{2})$, $(1^4D_\frac{5}{2})$\\
$\Xi_{c}^{0}(3080)$ & 3079.9$\pm$1.4&$(1^2D_\frac{3}{2})$, $(1^4D_\frac{3}{2})$ \\
$\Xi_{c}^{+}(3123)$ & 3122.9$\pm$1.3$\pm$0.3& $(1^2D_\frac{7}{2})$\\
\hline
\end{tabular}
\end{table}

\begin{figure*}
\centering
\begin{minipage}[]{0.30\linewidth}
\includegraphics[scale=0.25]{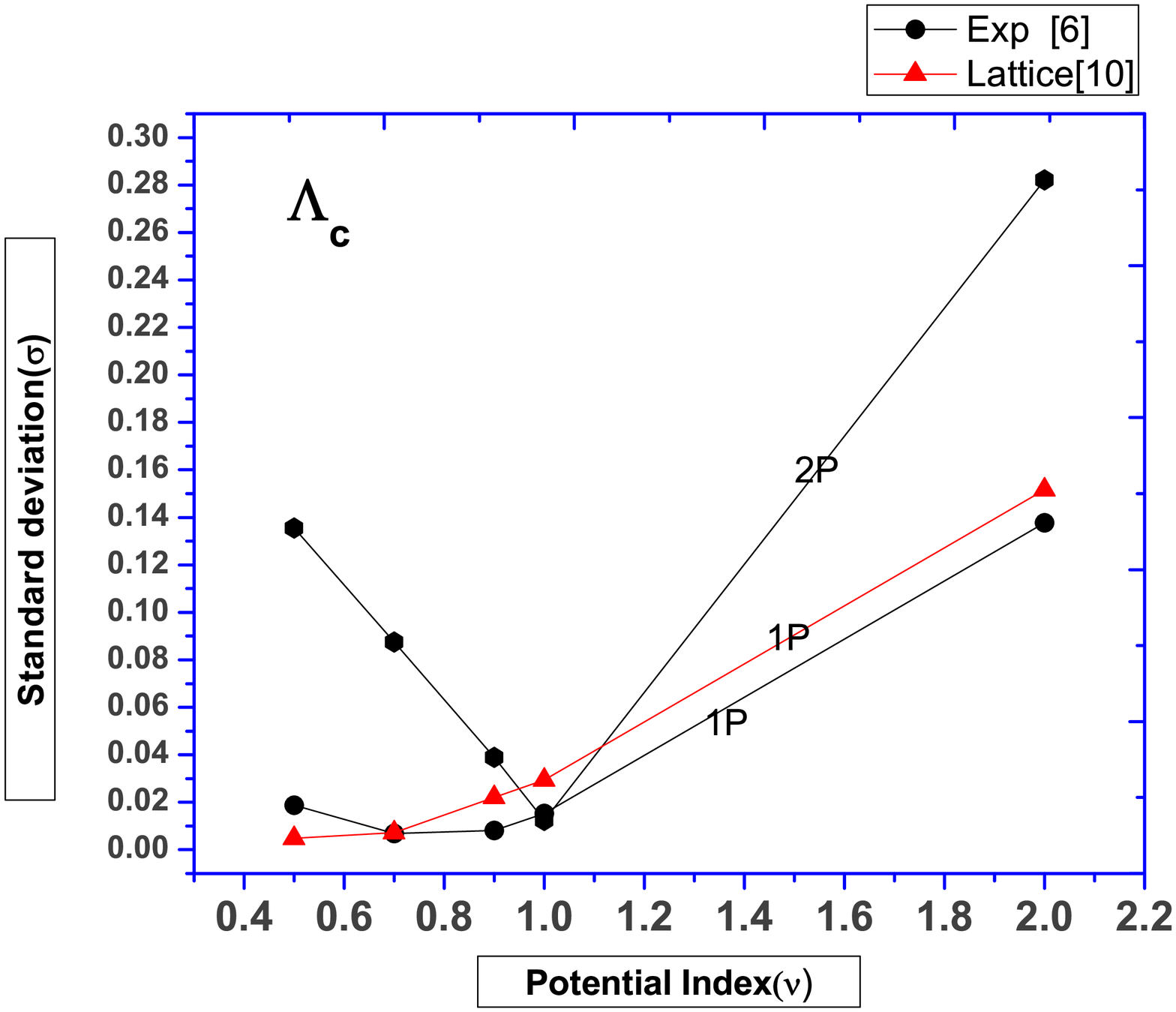}
\label{fig:minipage1}
\end{minipage}
\quad
\begin{minipage}[]{0.30\linewidth}
\includegraphics[scale=0.25]{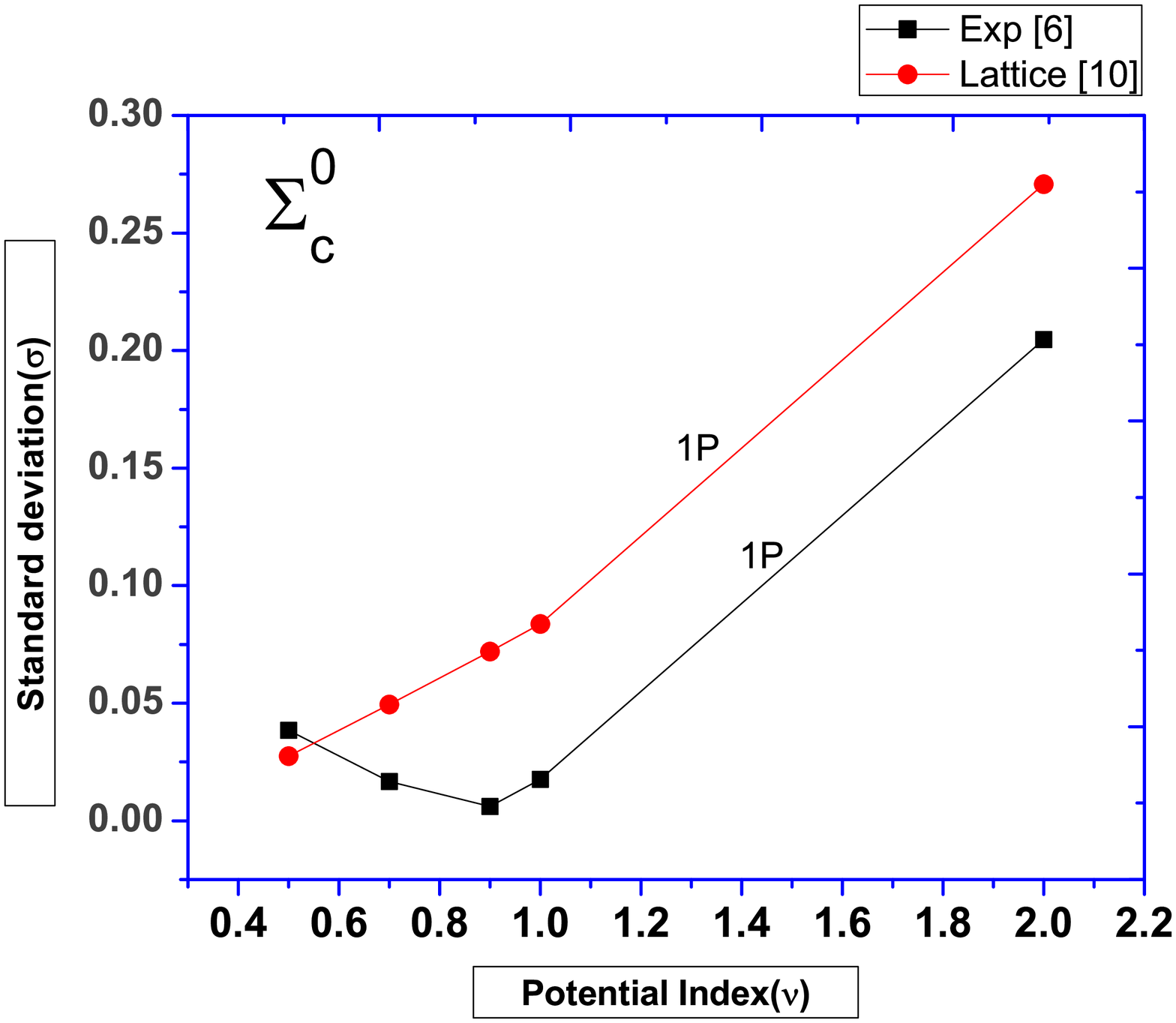}
\label{fig:minipage2}
\end{minipage}
\quad
\begin{minipage}[]{0.30\linewidth}
\includegraphics[scale=0.25]{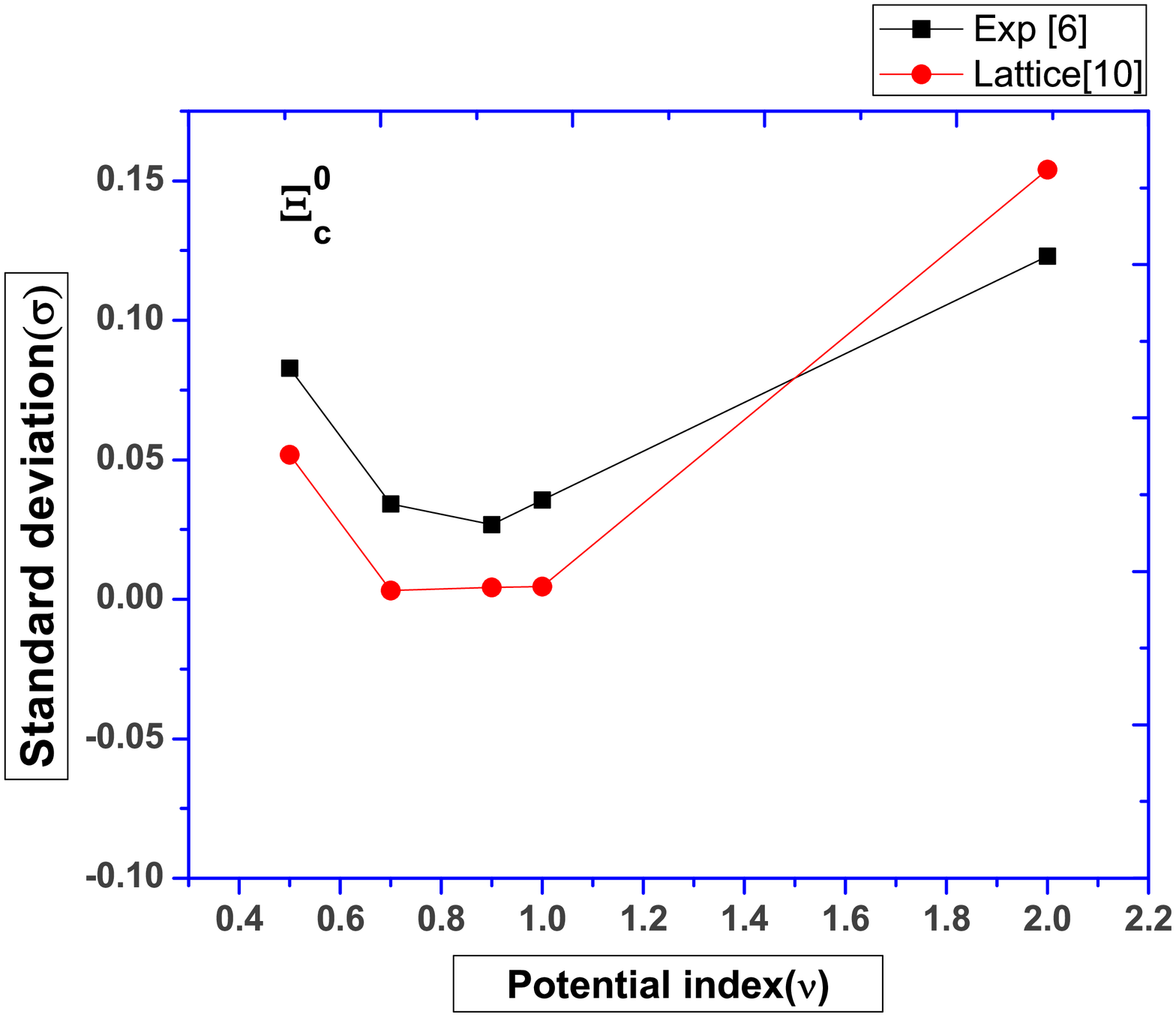}
\label{fig:minipage2}
\end{minipage}
\caption{\label{fig:epsart} Variation of Standard deviation with potential index for P states of $\Lambda_c^+$, $\Sigma_c^0$ and $\Xi_c^0 $.}
\end{figure*}
\section{Conclusion}

The mass spectra of $\Lambda_c$, $\Sigma_c$, $\Xi_c$ and $\Omega_c$ baryons are calculated in hypercentral constituent quark model(hCQM) with coulomb plus power potential along with first order correction(B) and without correction(A). 
As we don't know the exact three body interaction of quarks within hypercentral approximation, we have introduce the power index, $\nu$ and calculated the mass spectra for $\nu$ from 0.5 to 2.0. We have obtained our mass spectra by considering the five values (0.5,0.7,0.9,1.0 and 2.0) of the $\nu$ in each case of baryons. So that, we compute the root mean square (rms) deviations of the predicted masses for each choices of $\nu$ as

\[
\sigma(\nu)=\sqrt{\frac{1}{N} \sum_{n=1}^{N} \left[ \frac{M_{CPP_{\nu}}(n^{2S+1}L_{J})-M_{exp}(n^{2S+1}L_{J})}{M_{exp}(n^{2S+1}L_{J})} \right]^{2}}
\]

The graphs of standard deviation $(\sigma)$ versus potential index $\nu$ are shown in Fig. (15) for $\Lambda_c$, $\Sigma_c$ and $\Xi_c$ baryons. We have taken experimental results from Ref. \cite{olive} and Lattice QCD results from Ref. \cite{4}, to calculate rms deviations for these baryons. The figure shows the distinct minima at potential index near 1.0 in both, experimental as well as Lattice QCD calculations for all these baryons. Thus, we can say the mass spectra of $\Lambda_c$, $\Sigma_c$, $\Xi_c$ and $\Omega_c$ baryons are close to known experimental observations and other theoretical predictions at potential index slightly below 1.0. The semi-electronic decays are also calculated using the formalism given in \cite{sven} for $\Xi_c$ and $\Omega_c$ baryons at potential index 1.0 and the present results, specially decay rates obtained using first order correction(B) are close to the results of Ref \cite{sven}. The higher excited orbital and radial states mass calculation allow us to construct the Regge trajectories in (n, $M^{2}$)[see Figs. (6-8)] and (J, $M^{2}$) [see Figs. (9-14)] plane. Obtained masses are plotted in accordance with quantum number as well as with natural and unnatural parities. This study produces the information about the hadron dynamics and it also important for the hadron production and high energy scattering \cite{ebert2011}. These graphical representation is useful in assigning $J^{P}$ value to experimental unkown states. We are not performing any fitting for the calculated mass(in case of (n,$M^2$) plane, but still we are getting almost linear, parallel and equidistance lines in each case of baryons. \\

\par We have listed various experimentally known($J^{P}$ value unkown) singly charmed baryonic states in Table (19). We have compared these states with our calculated results in Table [4,5,9-15]. From that, we have concluded their baryonic states and also assigned $J^P$ value to these unknown experimental states. After successful implementation of this scheme to the singly charmed baryons, we would like to extend this scheme to calculate the decay rates of these baryons as well as to calculate the mass spectra and the decay rates of bottom baryons. We can also find other dynamics of Regge trajectories like slope and intercepts easily. Finally, we would like to conclude that there are not much theoretical calculations which provide the spectra starting from S to F states. So, we have compare our results to the D. Ebert calculations \cite{ebert2011}. This study will help to experimentalist as well as the theoretician to understand the dynamics of the singly charmed baryons.

\section*{Acknowledgement}
One of the author Z. Shah would like to thank Dr. Nilmani Mathur for his valuable suggestions at CCP 2015 at IIT, Guwahati. A. K. Rai acknowledges the financial support extended by DST, India  under SERB fast track scheme SR/FTP /PS-152/2012.

\end{document}